\documentclass[12pt,draftcls,onecolumn]{IEEEtran}
\usepackage{mathptmx} 
\usepackage{latexsym,bm}
\usepackage{amsmath}
\usepackage{amsfonts}
 \usepackage{amssymb,amsmath}
 \usepackage{graphicx}
\usepackage{epstopdf}
\usepackage{epsfig}       
 \usepackage{amsfonts}
 \usepackage{amsthm}
\usepackage{mathrsfs}
\usepackage{indentfirst}
\usepackage{graphicx}      
\usepackage{epsfig} 
\usepackage{epstopdf}
\usepackage{times}
\usepackage{bbm}
\usepackage[numbers,sort&compress]{natbib}
\usepackage{cleveref}
\usepackage{color}
\usepackage[justification=centering]{caption}
\usepackage{subfigure}

\allowdisplaybreaks
 \newtheorem{theorem}{Theorem}[section]
\newtheorem{corollary}{Corollary}[section]
\newtheorem{lemma}{Lemma}[section]
\newtheorem{remark}{Remark}[section]

\newtheorem{assumption}{Assumption}[section]

\renewcommand{\thelemma}{\arabic{section}.\arabic{lemma}}

\ifCLASSINFOpdf
\else
\fi
\begin{document}

\def\ba{\begin{array}}
\def\ea{\end{array}}
\def\ban{\begin{eqnarray*}}
\def\ean{\end{eqnarray*}}
\def\bd{\begin{description}}
\def\ed{\end{description}}
\def\be{\begin{equation}}
\def\ee{\end{equation}}
\def\bna{\begin{eqnarray}}
\def\ena{\end{eqnarray}}
\allowdisplaybreaks
\title{Graphon Particle Systems, Part I: Spatio-Temporal Approximation and Law of Large Numbers}
%
%
%

\author{Yan~Chen,
        Tao~Li,~\IEEEmembership{Senior Member,~IEEE}, and Xiaofeng Zong,~\IEEEmembership{Member,~IEEE}
\thanks{
Corresponding author: Tao Li. }
\thanks{Yan Chen is   with the State Key Laboratory of Mathematical Sciences, Academy of Mathematics and Systems Science, Chinese Academy of Sciences, Beijing 100190, China (e-mail: yanchen2026@amss.ac.cn).}
\thanks{Tao Li is with the Key Laboratory of Management, Decision and Information Systems, Institute of Systems Science, Academy of Mathematics and Systems Science, Chinese Academy of Sciences,  Beijing 100190, China, and also with School of
Mathematical Sciences, University of Chinese Academy of Sciences, Beijing 100149, China (email: litao@amss.ac.cn).}
\thanks{Xiaofeng Zong is with the School of Artificial Intelligence and Automation, China University of Geosciences, Wuhan 430074, China, and Hubei Key Laboratory of Advanced Control and Intelligent Automation for Complex Systems, Wuhan 430074, China. (e-mail: zongxf@cug.edu.cn).}
}

%
%

\markboth{Journal of \LaTeX\ Class Files, June~2024}%
{Shell \MakeLowercase{\textit{et al.}}: Bare Demo of IEEEtran.cls for IEEE Journals}
%



\maketitle

\begin{abstract}
We study a class of graphon particle systems with  time-varying random coefficients. In a graphon particle system, the interactions among particles are characterized by the coupled mean field terms through an underlying graphon and the randomness of the coefficients comes from    exogenous stochastic processes. By constructing two-level approximated sequences converging in 2-Wasserstein distance,
we prove the existence and uniqueness of the solution to the system. Besides, by constructing two-level approximated functions converging to
 the graphon mean field terms,  we establish the law of large numbers, which reveals that if the number of particles tends to infinity and the discretization step tends to
zero, then the discrete-time interacting particle system over a large-scale network  converges to the graphon particle system. As a byproduct,
we discover that  the graphon particle system can describe the limiting
 dynamics  of the distributed stochastic gradient descent algorithm over the large-scale network and  prove that if the gradients of the local cost functions are Lipschitz continuous, then the graphon particle system can be regarded as the spatio-temporal approximation of the discrete-time distributed stochastic gradient descent algorithm as the number of network nodes tends to infinity and the algorithm step size tends to
zero.
\end{abstract}


%
\IEEEpeerreviewmaketitle

\section{Introduction}
\hspace*{12pt}
\label{sec:introduction}
\IEEEPARstart{M}{any} systems in statistical physics, biological systems and neuroscience are weakly interacting particles, in which  the  interaction strength  among each particle and its neighbours is inversely proportional to the number, such as  continuous ferromagnetism model (\cite{GARTNERJ}), Kuramoto oscillator  (\cite{Gkogkas}-\cite{CM2019}),  Cucker-Smale ensemble (\cite{Ha}), FitzHugh-Nagumo neuron  (\cite{Baladron}-\cite{LuE}) and so on.
  Weakly interacting particle systems can be divided into homogeneous and heterogeneous ones. In a homogeneous  system, each particle interacts with the other particles with the same strength. For homogeneous  systems, Vlasov proposed the concept of  mean field interaction originally in 1938,
 reprinted in \cite{A. A. Vlasov}. Mean field interaction means
that the overall system acts over a given particle through the empirical measure
of the system. This interaction can be
represented by the mean field term. Mckean (\cite{McKean}) introduced the McKean-Vlasov equation to describe the behaviors of the limiting homogeneous weakly interacting particle systems as the number of particles tends to infinity.

\subsection{Related Work}
In a heterogeneous system, the interactions among particles depend on the particle labels and the interaction strength among particles depends on the weights of the edges of the adjacency network. To model the  heterogeneous  interactions among a continuum of particles,  Lov\'{a}sz and Szegedy  (\cite{Lov}) proposed the graphon theory. A graphon, defined by a symmetric measurable function $A: [0,1]\times [0,1] \to [0,1],\ (p,q)\mapsto A(p,q)$, represents  the limit for the sequence  of adjacent networks  as the   number of particles increases to infinity.  Recently,  the heterogeneous weakly interacting particle systems over the graphons have been extensively studied (\cite{Caines}-\cite{M2014}).

One important line of research concerns graphon mean field games. By investigating the limits of noncooperative dynamic games of heterogeneous weakly interacting particle systems, Huang and Caines (\cite{Caines}) proposed graphon mean field game theory, which provides a continuum framework for studying strategic interactions over large heterogeneous networks. This framework has been further studied in \cite{Gao1}-\cite{Fabian} and \cite{ATM2018}. Based on reinforcement learning algorithms, Cui et al. (\cite{Fabian1}-\cite{Fabian}) designed algorithms to approximate Nash equilibria for discrete-time graphon mean field games.
These works establish connections between large finite-player games and their graphon  limits.

Another closely related line of research concerns the dynamics of heterogeneous weakly interacting particle systems over graphons, also called graphon particle systems. Bayraktar et al. (\cite{ERHAN.BAYRAKTAR}-\cite{ERHAN.BAYRAKTAR22}) studied the well-posedness and large-population approximation of graphon particle systems. In particular, they established laws of large numbers for interacting particle systems associated with dense and not-so-dense step graphons, under which the empirical distribution of the states in the interacting particle system converges in probability to the integral of the state distributions of the corresponding graphon particle system. Bet et al. (\cite{Bet G20}) further studied graphon particle systems over exchangeable step graphons and established a law of large numbers for the corresponding state distributions.
These results provide a rigorous connection between continuous-time interacting particle systems over large-scale networks and their graphon limits. However, the graphon particle systems considered in these works mainly have deterministic and time-invariant coefficients. Moreover, their laws of large numbers primarily concern spatial approximation, in which continuous-time interacting particle systems over large-scale networks converge to the corresponding graphon particle systems as the number of particles tends to infinity.

%

\subsection{ Motivations and Models }

Notice that in the Kuramoto oscillator model, for  each oscillator, there is a stochastic process  representing its intrinsic frequency in the phase equation, which is a  random coefficient. Following this idea,  the  mean-field systems with a single population (\cite{Amari}), multiple populations (\cite{Baladron}), and the single-population  mean-field system with random coefficients (\cite{Lu}) all come  down to the following graphon particle system with time-varying  random coefficients. Let $[0,1]$ be the set of a continuum of particles, each element of which represents a particle. The connecting structure among particles is given by the graphon $A$.  The dynamic equation of the graphon particle system is given  by
\begin{eqnarray}\label{randmckvla}
dz_{p}(t) &=& \bigg[\int_{[0,1]} A(p,q)
\left(\int_{ \mathbb{R}^{n}}F(t,p, q,z,z_{p}(t))\mu_{t,q}(dz)\right)dq +G\left(t,p,\eta_{p}(t),z_{p}(t)\right)\bigg]dt\notag\\
&&+H(t,p,\eta_{p}(t),z_{p}(t))dw_{p}(t),\ \forall \ p \in [0,1],
\end{eqnarray}
 where $z_{p}(t) \in \mathbb{R}^{n}$ is the state of particle $p$ at time $t$. Let $(\Omega, \mathcal{F}, P)$ be a complete probability space with a  family of non-decreasing $\sigma$-algebras $\{\mathcal{F}_{t},\ t \geqslant 0\}\subseteq \mathcal{F}$. Given $q \in  [0,1]$, $\mu_{t,q}$  is the distribution of $z_q(t)$. Here, $\int_{[0,1]}A(p,q)
\left(\int_{ \mathbb{R}^{n}}F(t,p, q,z,z_{p}(t))\mu_{t,q}(dz)\right)dq$ is the coupled mean field term based on the graphon $A$, and  $G:[0,\infty)\times [0,1] \times \mathbb{R}^{n} \times
   \mathbb{R}^{n}\to  \mathbb{R}^{n}$, $F:[0,\infty)\times [0,1]
\times [0,1] \times \mathbb{R}^{n} \times \mathbb{R}^{n} \to \mathbb{R}^{n}$ and  $H:[0,\infty)\times [0,1] \times \mathbb{R}^{n} \times
   \mathbb{R}^{n}\to  \mathbb{R}^{n\times n}$  are the functions satisfying some appropriate conditions.
The process $\{(w_{p}(t),\mathcal{F}_{t}), t \geqslant 0,   p\in [0,1]\}$ is a family of independent $n$-dimensional standard Brownian motions  and $\{(\eta_{p}(t),\mathcal{F}_{t}), t \geqslant 0,  p\in [0,1]\}$ is a family of independent $n$-dimensional stochastic processes. The processes $\{w_{p}(t), t \geqslant 0, p \in [0,1]\}$ and $\{\eta_{p}(t), t \geqslant 0, p \in [0,1]\}$ are mutually independent. The initial states $\{z_{p}(0),  \ p \in [0,1]\}$ are adapted to $\mathcal{F}_{0}$, mutually independent and independent of $\{w_{p}(t),  t \geqslant 0, p \in [0,1]\}$ and $\{\eta_{p}(t),  t \geqslant 0, p \in [0,1]\}$.

\begin{remark}
The model of a continuum of independent Brownian motions $\{(w_p(t), \mathcal{F }_t),\ t \geqslant   0,\ p
\in [0,1] \}$ has also been used in \cite{ERHAN.BAYRAKTAR,Bayraktar22,Bet G20,ACL2022}.
Now, we give a method to construct a  continuum of    independent $n$-dimensional standard Brownian motions needed in our work. Given $\nu$ as a Wiener measure on $(\mathcal{C}([0,\infty),\mathbb{R}^n),  \mathscr{B}(\mathcal{C}([0,\infty),\mathbb{R}^n)))$, by   Lemma \ref{infimeas}, there exist independent random elements $w_{p},\ p\in [0,1]$, each of which is with distribution  $\nu$ and is valued in $(\mathcal{C}([0,\infty),\mathbb{R}^n),\mathscr{B}(\mathcal{C}([0,\infty),\mathbb{R}^n)))$, where $\mathscr{B}(\mathcal{C}([0,\infty),\mathbb{R}^n))$ is the $\sigma$-algebra generated by the metric $\rho$  defined by
$
\rho\left(y_{1}, y_2\right) \triangleq \sum_{k=1}^{\infty} \frac{1}{2^k} \max _{0 \leqslant l \leqslant k} (\left\|y_{1}(l)- y_2(l)\right\| \wedge 1 ),\ \forall \ y_{1}, \ y_{2}\in   \mathcal{C}([0,\infty), \mathbb{R}^n),
$
and $\mathcal{C}([0,\infty),\mathbb{R}^n)$ is the space of all continuous, $\mathbb{R}^n$-valued functions on $[0,\infty).$
Then, by Definition 2.2.1 in  \cite{Boksendal} and Remark 4.22 in \cite{Karatzas},    we know that $w_{p},\ p\in [0,1]$ is   a  continuum of  independent $n$-dimensional standard Brownian motions.  The explicit construction of the underlying probability space $(\Omega,\mathcal{F},P)$ can be found in the proof of  Lemma \ref{infimeas}.

\end{remark}

We will give some special examples of the  system (\ref{randmckvla}).

 \textbf{Example 1}:
If   $G\left(t,p,x,y\right)$, $F(t,p, q,z,y)$, $H(t,p,  x,y)$, $\eta_{p}(t)$, $w_{p}(t)$, $A(p,q)$,  $\mu_{t,p}(dz)$ and the distribution of $z_{p}(0)$ do not depend  on the label $p$ in (\ref{randmckvla}),  and are denoted by  $G\left(t,x,y\right)$, $F(t, q,z,y)$, $H(t,x,y)$, $\eta(t)$,
 $w(t)$, $A_{q}$, $\mu_{t}(dz)$ and $\mu_{0}$, respectively, then the system (\ref{randmckvla}) degenerates to
\begin{align}
dz(t)=&\bigg[
\int_{[0,1]}A_{q}
\left(\int_{ \mathbb{R}^{n}}F(t, q,z,z(t))\mu_{t}(dz)\right)dq +G\left(t,\eta(t),z(t)\right)\bigg]dt+H(t,\eta(t),z(t))
dw(t)\notag
\end{align}
in the sense of weak solution, which is the classical Mckean-Vlasov equation (\cite{GARTNERJ}).

 \textbf{Example 2}: The graphon particle system (\ref{randmckvla})  describes not only the models in  \cite{Gkogkas}, \cite{Baladron} and \cite{Amari} but also the dynamics of the consensus-based distributed optimization algorithm  over the graphon. Consider the following optimization problem over a graphon. Let $[0,1]$ be the set of a continuum of nodes, each element of which corresponds to a node. The connecting structure among nodes is given by the graphon $A$. Any node $p \in [0,1]$ has a private local cost function $V(p,x):[0,1] \times \mathbb{R}^{n}\to \mathbb{R}$, which is strongly convex and continuously differentiable w.r.t. $x\in \mathbb{R}^{n}$ and is  integrable w.r.t. $p \in [0,1]$.  The objective of all nodes is to cooperatively solve the optimization problem \begin{align}
\min \limits_{z \in \mathbb{R}^{n}}V(z)\triangleq\int_{[0,1]}V(p,z)dp.\label{globalgoal}
\end{align}
Denote the unique minimizer of $V(z)$ by $z^{*}$. We have proposed the following distributed stochastic gradient descent (SGD) algorithm in \cite{Ycc}.  Given the initial states $\left\{z_{p}(0),\ p \in [0,1]\right\}$, for any node $ p \in [0,1]$,
\begin{align}
dz_p(t)=&\alpha_{1}(t)\int_{ [0,1]}A(p,q)\left( \int_{\mathbb{R}^n}(z-z_p(t))\mu_{t,q}(dz)\right)dqdt
-\alpha_{2}(t)\nabla_{z} V(p,z_p(t))dt\notag\\
& -\alpha_{2}(t)\Sigma_{1} dw_p(t),\label{Mckeanvlasov1}
\end{align}
where $z_p(t)\in \mathbb{R}^n$ is the state of node $p$ at time $t$, representing its local estimate of $z^{*}$,  $\nabla_{z} V(p,z_p(t))\\ \in\mathbb{R}^n$ is the gradient value of the local cost function  at $z_{p}(t)$ and  $\int_{ [0,1]}A(p,q) ( \int_{\mathbb{R}^n}(z-z_p(t))\\ \mu_{t,q}(dz) )dq$ is the coupled mean field term based on the graphon $A$. The initial states $\{z_{p}(0), \ p \in [0,1]\}$ are  adapted to $\mathcal{F}_{0}$, mutually independent and independent of $\{w_{p}(t),  t \geqslant 0, p \in [0,1]\}$.
 The terms $\alpha_{1}(t)$ and
$\alpha_{2}(t)$ are time-varying algorithm gains and $\Sigma_{1} \in \mathbb{R}^{n\times n}.$
Note that the system (\ref{Mckeanvlasov1}) is a special case of (\ref{randmckvla}) with $G=-\alpha_{2}(t)\nabla_{y} V(p,y)$, $F=\alpha_{1}(t)(z-y)$ and $H=-\alpha_{2}(t)\Sigma_{1}$.

For the system (\ref{randmckvla}), there are some fundamental problems worth studying. Firstly, the existence  and   uniqueness of the solution is the theoretical basis. Secondly,  the discrete-time distributed SGD algorithm  over the network  with finite nodes has been extensively studied (\cite{A. Nedic}-\cite{Li T}). A natural question is  whether there is an intrinsic connection between  the  algorithm  and the graphon particle system  (\ref{Mckeanvlasov1}). In this paper, we prove that the system (\ref{Mckeanvlasov1}) is the limit of the  discrete-time distributed SGD algorithm over the large-scale network  in \cite{A. Nedic}-\cite{Li T} as the number of nodes tends to infinity and the  algorithm step size tends to zero. The third one  is the asymptotic property. Especially, for the algorithm (\ref{Mckeanvlasov1}), people expect to figure out whether the states $\{z_{p}(t),\ p \in [0,1],\ t \geqslant 0\}$ of the system (\ref{Mckeanvlasov1}) converge to the minimizer of the global cost function under some proper assumptions.  For the above motivations, the first and second problems are studied in this paper. The third one is investigated in the companion paper \cite{Li}.

\subsection{Contributions}
We prove the existence and uniqueness of the solution to the system (\ref{randmckvla}).  Existing works (\cite{ERHAN.BAYRAKTAR}-\cite{Bet G20})  have been restricted to the cases with time-invariant and deterministic coefficients, while, the system (\ref{randmckvla}) has time-varying random  coefficients  due to  the stochastic processes $\{\eta_{p}(t), t \geqslant 0, p \in [0,1]\}$.  The key in proving the existence and uniqueness of the solution lies in proving the measurability of the map $p \mapsto \mathcal{L}(z_{p}(t))$ in the presence of such coefficients to ensure that the term $\int_{[0,1]}A(p,q)
(\int_{ \mathbb{R}^{n}}F(t,p, q,z,z_{p}(t)) \mu_{t,q}(dz))dq$ is well-defined, where $\mathcal{L}(z_{p}(t))$ is the distribution of $z_{p}(t)$.
To establish this measurability,  we employ a two-level approximation argument similar to   \cite{ERHAN.BAYRAKTAR}.
At the first level, we construct an
approximating sequence
$\{\{z_p^k(t),\,t\in[0,T],\,p\in[0,1]\},\,k\in\mathbb{N}\}$
whose laws converge to the law of $z_p$. At the second level, for
each $k$, we construct a sequence $\{z^{k,l}_{p}(t),  l\in \mathbb{N}\}$ to approximate $z_p^k(t)$. For each $l$, the
process $z_p^{k,l}$ can be expressed as a measurable function of the previous iterate $z_p^{k-1}$, the exogenous process
$\eta_p$, and the Brownian motion $w_p$. This representation yields
the measurability of the map $p\mapsto\mathcal{L}(z_p^{k,l},\eta_p,w_p)$.
The main technical difficulty is to derive the measurability of the map
$p\mapsto\mathcal{L}(z_p(t))$ from that of
$p\mapsto\mathcal{L}(z_p^{k,l},\eta_p,w_p)$ by taking the two
 limits successively.  Unlike the case with time-invariant and deterministic coefficients in \cite{ERHAN.BAYRAKTAR}-\cite{Bet G20},   the time-varying random
coefficients make the error between $z_p^{k,l}(t)$ and $z_p^k(t)$
depend simultaneously on the temporal variations of the coefficients, the increments of the exogenous and state processes, and
the time variation of the measure flow.
To address this difficulty, we derive uniform
moment and temporal estimates for the state iterates and combine them
with the temporal regularity of the coefficients, the exogenous
processes, and the measure flow to control the error. We thereby prove that
$z_p^{k,l}(t)$ converges to $z_p^k(t)$ in probability, and further establish  the measurability  of the map
$p\mapsto\mathcal{L}(z_p^k,\eta_p,w_p)$.  Taking the state marginal
and  noting that the limit of a sequence of measurable maps is measurable, we prove the measurability of the map $p \mapsto \mathcal{L}(z_{p}(t))$.

We prove the law of large numbers, which reveals that the discrete-time interacting particle system  over the large-scale network  spatio-temporally approximates the graphon particle system (\ref{randmckvla}) as the number of particles tends to infinity and the discretization step  tends to zero.
 All the existing laws of large numbers (\cite{ERHAN.BAYRAKTAR}-\cite{Bet G20}) for the graphon particle systems are established in the space dimension.
 Compared with the existing results, we develop the law of large numbers not only in space but also in time dimensions.
  That is, the states  in the continuous-time approximation of the discrete-time interacting particle system converge to those of the graphon particle system (\ref{randmckvla})  in mean square, and the mean 1-Wasserstein distance between the empirical distribution and the integral of state distributions on the node set vanishes. Specially, we prove  that  if the gradients of the local cost functions are Lipschitz continuous, then the dynamics of the discrete-time distributed SGD algorithm converges to the  graphon particle system (\ref{Mckeanvlasov1})
as the number of network nodes tends to infinity and the  algorithm step size tends to zero.



The rest of the paper is organized as follows.  In Section II,  the existence and uniqueness of the solution to the graphon particle system (\ref{randmckvla}) is presented. In Section III, the laws of large numbers for the systems (\ref{randmckvla}) and (\ref{Mckeanvlasov1}) are given. In Section IV, the conclusions are given.

The following notations will be used throughout this paper.  Denote the $n$-dimensional Euclidean space by $\mathbb{R}^{n}$ and the Euclidean norm by $\left\| \cdot \right\|$. For a given matrix $A \in \mathbb{R}^{n\times n}$, $\operatorname{Tr}(A)$ denotes its trace.  Denote $\mathbb{N}$ as the set of nonnegative integers.
 For a number $x \in \mathbb{R}$, denote the greatest integer less than or equal to $x$ and the smallest integer greater than or equal to $x$ as $\lfloor x\rfloor$  and  $\lceil x\rceil$, respectively.
Let $(\Omega, \mathcal{F}, P)$ be a probability space.
Denote the space of continuous functions from $[0,T]$ to $\mathbb{R}^{n}$ by $\mathcal{C}_{T}^{n}$, endowed with the uniform norm
$\|\cdot\|_{*, T}$, that is, $\|x(\cdot)\|_{*, T}\triangleq\sup_{t \in [0,T]}\left\| x(t) \right\|$, $x(\cdot) \in \mathcal{C}_{T}^{n}$,
and denote $\|x\|_{*, t}=\sup_{s\in [0,t]} \| x(s) \|,  t\in [0,T]$.
Denote  $\mathscr{B}(\mathcal{C}_{T}^{n})$  as  the Borel algebra  induced by the norm $\|\cdot\|_{*, T}$.
For any $B\in \mathscr{B}(\mathcal{C}_{T}^{n})$,
if the map $X(\omega):\Omega \mapsto \mathcal{C}_{T}^{n}$ satisfies $X^{-1}(B) \in \mathcal{F}$, then $X(\omega)$ is a random element in $\mathcal{C}_{T}^{n}$.
 For a given random vector $X \in \mathbb{R}^{n}$,
 denote  its  mathematical expectation  and   distribution   by $E[X]$
 and $\mathcal{L}(X)$, respectively.
Denote the sets of  probability measures on
$\mathbb{R}^{n}$ and  $\mathcal{C}_{T}^{n}$ by $\mathscr{P}(\mathbb{R}^{n})$ and $\mathscr{P}(\mathcal{C}_{T}^{n})$, respectively.
Denote the 2-Wasserstein distance on  $\mathscr{P}(\mathbb{R}^{n})$ as
\begin{align}
 W_{2}(\mu, \nu)  = \Big(\inf\limits_{\gamma \in \Pi(\mu,\nu)}\int_{\mathbb{R}^{n} \times \mathbb{R}^{n}} \|x-y\|^2 \gamma(dx,dy)\Big)^{\frac{1}{2}},
\label{defofWassersteindistanceofed}
\end{align}
where $\mu,\ \nu \in \mathscr{P}(\mathbb{R}^{n})$,  $\Pi(\mu,\nu)$ is the set of all couplings of $\mu$ and $\nu$ and a coupling $\gamma$ is a joint probability measure on $\mathbb{R}^{n}\times \mathbb{R}^{n}$ whose marginal distributions are $\mu$ and $\nu$.
Let $p \in [1, \infty)$ and denote the $p$-Wasserstein distance on $\mathscr{P}(\mathcal{C}_{T}^{n})$ as
\begin{align}
  W_{p, t}(\mu, \nu)  =\Big(\inf\limits_{\gamma \in \Pi(\mu,\nu)}\int_{\mathcal{C}_{T}^{n} \times \mathcal{C}_{T}^{n}} \|x-y \|_{*, t}^p \gamma(dx,dy)\Big)^{\frac{1}{p}},
\label{defofWassersteindistance}
\end{align}
where $t \in[0, T] $ and $\mu,$ $\nu \in \mathscr{P}(\mathcal{C}_{T}^{n})$.
Denote the Wasserstein space of order $p$ on $\mathcal{C}_{T}^{n} $ as
$\mathscr{P}_{p}(\mathcal{C}_{T}^{n} )=\{\mu \in \mathscr{P}
(\mathcal{C}_{T}^{n} ):
\int_{\mathcal{C}_{T}^{n} } \| \theta\|_{*,T}^{p} \mu(d \theta)
 <\infty \}$.  Especially, the $1$-Wasserstein distance $W_{1, T}$  can  also be written as
\begin{align}\label{w1wassersteinlip}
  W_{1, T}(\mu, \nu)  =  \sup\limits_{f \in \mathcal{C}_{L}}  \int_{\mathcal{C}_{T}^{n}} f(z) (\mu(d z)
 - \nu(d z)),
\end{align}
where $\mu,\ \nu\in \mathscr{P}_{1}(\mathcal{C}_{T}^{n})$ and $\mathcal{C}_{L}$ is the set of Lipschitz continuous  functions $f:\mathcal{C}_{T}^{n}\to \mathbb{R} $ with Lipschitz constants less  than  or equal to  $1$.
 For notational convenience,  $\{z_{p}(t),\ 0\leqslant t\leqslant T\}\in \mathcal{C}_{T}^{n}$ is denoted by $z_{p}$.
 For any two measurable spaces $(F_{1},\ \mathscr{B}(F_{1}))$ and $(F_{2},\ \mathscr{B}(F_{2}))$,  the measurable map $f: F_{1} \to F_{2}$ and   finite measure $\mu$ on $\mathscr{B}(F_{1})$ (i.e.  $\mu(F_{1})< \infty$),
where $\mathscr{B}(F_{1})$ and $\mathscr{B}(F_{2})$ are the $\sigma$-algebras on $F_{1}$ and $F_{2}$ respectively,  the image measure of
$\mu$ under the map $f$ is given by $\mu \circ f^{-1}(A)=\mu\left(f^{-1}(A)\right), \ \forall \ A \in \mathscr{B}(F_{2}).$
For a given measurable space $(F, \mathscr{G})$ and $x \in F$, where $\mathscr{G}$ is a $\sigma$-algebra on $F$, the Dirac measure $\delta_x$ at $x$ is defined by $\delta_x(A)=1$ if $x \in A$  and $\delta_x(A)=0$ otherwise, $\forall \ A \in \mathscr{G}$.  For a graphon $G$, denote
 $
 \|G\|_{\infty \to 1}=
\sup_{g \in \mathcal{E}}
\int_{[0,1]}\|\int_{[0,1]}G(u,v) g(v)dv\|du,$ where $\mathcal{E}=\{g\in L^{\infty}([0,1], \mathbb{R}^n)\ | \ \text{ess sup}\|g\|\leqslant 1\}$ and
$L^{\infty}([0,1],  \mathbb{R}^n)=\{f\mid f:  [0,1]\rightarrow  \mathbb{R}^{n},  f \ \text{is} \ \text{measurable}\\  \text{and}  \text{ bounded} \ \text{almost} \ \text{everywhere}\}$.  $C_{r}$ inequality is given by $\left\|\sum_{i=1}^N a_i\right\|^r  \leq \sum_{i=1}^N\left\|a_i\right\|^r,\  0<r<1$ and $\left\|\sum_{i=1}^N a_i\right\|^r\leq N^{r-1}  \sum_{i=1}^N \|a_i \|^r$, $r \geq 1,$ $\ a_{i} \in \mathbb{R}^n,\ i=1,\ldots,N$.

\section{The Existence and Uniqueness}
In this section, we will prove the existence and uniqueness of the solution to the graphon particle system (\ref{randmckvla}) on any given interval $[0,T]$.

\vskip  1mm
To prove  the existence and uniqueness, we consider the following  space of probability measures
\begin{align}
\mathcal{M}\triangleq &\Big\{ \nu=\left\{\nu_p: p \in [0,1]\right\} \in\left[\mathscr{P}_{2}\left(\mathcal{C}_{T}^{n}\right)\right]^{[0,1]} \Big|  \ \text{the map} \ [0,1]\ni p \mapsto \nu_p \in \mathscr{P}_{2} (\mathcal{C}_{T}^{n})\ \text{is measurable,}  \notag \\
& \sup _{p \in [0,1]} \int_{\mathcal{C}_{T}^{n}}\|x\|_{*, T}^{2}  \nu_p(d x) <\infty,\ \text{and for any}\ \epsilon>0,\ \text{there exists} \ \delta>0, \  \text{such that} \notag\\
 & \sup_{  |t_{1}- t_{2}|<\delta,\ p \in [0,1] }W_{2}(\nu_{t_{1},p},\nu_{t_{2},p})<\epsilon
  \Big\}.\notag
\end{align}
\vskip -0.5mm
Denote $W_{2,\mathcal{M},t}(\mu,\nu)=\sup_{p \in [0,1]}W_{2,t}(\mu_{p}, \nu_{p}), \forall \ \mu,\ \nu \in \mathcal{M}, \ t \in [0,T] $. We give the following assumptions on the graphon particle system (\ref{randmckvla}) so as to guarantee the uniqueness and existence of the exact solution, and the   convergence of the approximate solutions.

\vskip 1.5mm

\begin{assumption}\label{assumption01}
 Graphon $A(p,q)$ is continuous w.r.t. $(p,q) \in [0,1]\times [0,1]$.
\end{assumption}
\vskip 1.5mm

\begin{assumption}\label{assumption001}
There exist  $\zeta \geqslant 0$ and $\upsilon_{0} \geqslant 0$ such that $\sup _{p \in [0,1]} E\big[ \|z_{p}(0) \|^{2+\upsilon_{0}}\big] \leqslant \zeta$; the map  $ [0,1] \ni p\mapsto \mathcal{L}(z_{p}(0))=\mu_{0,p}\in \mathscr{P}(\mathbb{R}^{n})$ is measurable; 
 for any $\epsilon >0$, there exists $\delta>0$, such that if   $|p_1-p_2|<\delta$, then $  W_{2}(\mu_{0,p_{1}}, \mu_{0,p_{2}})<\epsilon,\ \forall \ p_1,\ p_2 \in [0,1]$.
\end{assumption}
\vskip 1.5mm

\begin{assumption}\label{assumption1}
There exist positive constants $\sigma_{i},\ i=1,2,\cdots,6$,  $C_{1}$ and $C_{2}$, such that the following conditions hold.



(i)  $ \|G(t,p,x, y)\| +\|H(t,p,x,y)\| \leqslant C_{1} (1+\|x\| +\|y\|  ), \ \forall \ x, \ y \in \mathbb{R}^{n}, \ t \in [0,T]$,\ $p \in [0,1]$;
$\|G(t,p,x_{1},y_{1})-G(t,p,x_{2},y_{2})\|^{2}+\|H(t,p,x_{1},y_{1})-H(t,p,x_{2},y_{2})\|^{2}\leqslant \sigma_{1} ( \|x_{1} -x_{2}\|^{2}+\|y_{1}-y_{2}\|^{2}),$ $ \ \forall \  x_{1}, \ x_{2},\ y_{1}, \ y_{2} \in \mathbb{R}^{n}, \ t  \in [0,T], \ p \in [0,1]$; for any $\epsilon>0,$  there exists $\delta>0$ such that if $|p_{1}-p_{2}|<\delta$, then $\|G(t,p_{1},x,y)-G(t,p_{2},x,y)\|^2+\|H(t,p_{1},x,y)-H(t,p_{2},x,y)\|^2<\epsilon (\sigma_{2} \|x\|^{2}+\sigma_{2}\|y
\|^{2}+\sigma_{3})$,  $\forall \ p_{1},\ p_{2} \in [0,1], \ t \in [0,T], \ x, \ y \in \mathbb{R}^{n}$;
for any $\epsilon >0$, there exists $\delta>0$ such that if $|t_{1}-t_{2}|<\delta$, then $\|H(t_{1},p,x,y)-H(t_{2},p,x,y)\|^{2} +\|G(t_{1},p,x,y) -G(t_{2},p,x,y)\|^{2} \leqslant \epsilon (\sigma_{2} \|x\|^{2}+\sigma_{2}\|y
\|^{2}+\sigma_{3}), \ \forall \ t_{1},  \ t_{2} \in [0,T],\ x, \ y \in \mathbb{R}^{n}, \ p \in [0,1]$.

(ii)
$\|F(t,p,q,z_{1},y_{1})-F(t,p,q,z_{2},y_{2})\| \leqslant \sigma_{4}(\|z_{1}-z_{2}\|+\|y_{1}-y_{2}\|), \ \forall \ z_{1}, \ z_{2}, \ y_{1}, \ y_{2}, \ \in \mathbb{R}^{n},\ t \ \in [0,T],  \ p, \ q \in [0,1]$;
 for any $\epsilon>0,$ there exists $\delta>0$ such that if $|p_{1}-p_{2}|+|q_{1}-q_{2}|<\delta$, then $\|F(t,p_{1},q_{1},z,y)-F(t,p_{2},q_{2},z,y)\|^{2}<\epsilon  (\sigma_{5}\|z\|^{2}+\sigma_{5}\|y\|^{2}+\sigma_{6})$,
$\forall \ p_{1},\ p_{2}, \ q_{1}, \ q_{2} \in [0,1], \ t \in [0,T],\ z,\ y \in \mathbb{R}^{n}$;
for any $\epsilon >0$, there exists $\delta>0$ such that if $|t_{1}-t_{2}|<\delta$, then $\big\|F(t_{1},p,q,z,y)-F(t_{2},p,q,z,
y)\big\|^{2} \leqslant \epsilon (\sigma_{5}\|z\|^{2}+\sigma_{5}\|y\|^{2}+\sigma_{6})
,\ \forall \ t_{1}, \ t_{2}\in [0,T], \ z, \ y\in \mathbb{R}^{n}, \ p, \ q \in [0,1]$;
$\|F(t,p,q,z,y)\|  \leqslant C_{2}(1+\|z\|+\|y\|),\ \forall \ z, \ y \in \mathbb{R}^{n}, \ t\in [0,T], \ p, \ q \in [0,1]$.

(iii) The map $[0,1] \ni p\mapsto \mathcal{L}(\eta_{p}(t))\in \mathscr{P}(\mathbb{R}^{n})$ is measurable, $t \geqslant 0$;
 for $p\in [0,1]$, $\eta_{p}$ is a random element in $ \mathcal{C}_{T}^{n}$;
   there exists $\upsilon_{1} \geqslant 0,\ r \geqslant 0$ such that $\sup\limits_{t \in [0,T], \ p \in [0,1]}E\big[\|\eta_{p}(t) \|^{2+\upsilon_{1}} \big]  \leqslant r$; $\eta_{p}(t)$ is uniformly continuous  w.r.t. $t$ in mean square, that is, for  any $\epsilon >0$, there exists $\delta>0$, such that if   $|t_1-t_2|<\delta$, then $ E\left[\left\| \eta_{p}(t_{1})-\eta_{p}(t_{2}) \right\|^2\right]<\epsilon,\ \forall \ t_1,\ t_2 \in [0,T],\ p\in [0,1]$;
  for any $\epsilon>0$, there exists $\delta>0$, such that, if $|p_{1}-p_{2}|<\delta$, then $W_{2,T}(\mathcal{L}(\eta_{p_{1}}), \mathcal{L}(\eta_{p_{2}}))<\epsilon$.

\end{assumption}
\vskip 1.5mm

The following theorem shows the existence and uniqueness of the solution to the system (\ref{randmckvla}).
\vskip 1.5mm
\begin{theorem}\label{existencegeneral}
If Assumptions  \ref{assumption001}-\ref{assumption1}  hold, then there exists a unique solution $ \{z_{p},\ \mu_{p},\ p\in[0,1] \}$  to the system (\ref{randmckvla}) on $[0,T]$,  satisfying that  $\sup_{p \in [0,1]}E\big[\sup_{t\in[0,T]}\|z_{p}(t)\|^{2+\upsilon}\big]<\infty$  and the map $[0,1] \ni p \mapsto \mu_{p}\in \mathscr{P}_{2}\left(\mathcal{C}_{T}^{n}\right)$ is measurable, where $\mu_{p}=\mathcal{L}(z_{p})\in \mathscr{P}_{2}\left(\mathcal{C}_{T}^{n}\right)$ and $\upsilon=\min\{\upsilon_{0},\ \upsilon_{1}\}$.
\end{theorem}
\vskip 1.5mm
\begin{proof}
See Appendix A for the proof.
\end{proof}

\vskip 1.5mm

\begin{remark}
It is known that dealing with the states with  a continuum of independent Brownian motions poses
technical challenges on the measurability issue of the mapping $p\mapsto z_{p}$. One way to avoid this question is
converting the system of a continuum of states to HJB and FPK equations (\cite{Caines}). Another way is to construct the underlying probability space  directly.
  In fact, the theory developed in \cite{Sun2006} grants the existence of a Fubini extension
of the   product space,
carrying a collection of essentially pairwise independent
(e.p.i.) Brownian motions   with sufficient joint measurability (in the extension), which has been used in graphon games (\cite{ACL2022}) to ensure the measurability of the mapping.
 Dunyak and Caines in \cite{SAC2022} constructed a $Q$-space noise   without the independence when examining the linear discrete-time dynamical control system.

In this work, we do not need the measurability of the mappings $p \mapsto z_p$ and $p \mapsto w_p$.
Since (\ref{randmckvla}) only involves the integral  with respect to $\mu_{p}$ rather than $z_{p}$, it suffices  that the mapping $p \mapsto \mu_{p}$ is measurable, as established in Theorem \ref{existencegeneral}.  Related discussions can also be found in \cite{ERHAN.BAYRAKTAR} and \cite{Lacker}.
\end{remark}

\vskip 1.5mm

The following lemma shows that the solution $ \{z_{p}, \ \mu_{p}, \ p\in[0,1] \}$  to the system (\ref{randmckvla}) on $[0,T]$ is uniformly continuous. This will be used in Section III.
\begin{lemma}\label{labelconti}
	 If Assumptions  \ref{assumption01}-\ref{assumption1} hold, then $ \{ \mu_{p}, \ p\in[0,1] \}$ in the solution $ \{z_{p}, \ \mu_{p}, \ p\in[0,1] \}$  to the system (\ref{randmckvla}) on $[0,T]$ are uniformly continuous w.r.t. $p$, that is, for any $\epsilon >0$, there exists $\delta>0$, such that if $|p_1-p_2|<\delta$, then
		$ W_{2,T}^2\left(\mu_{p_1},\mu_{p_2}\right)<\epsilon, \ \forall \ p_1,\ p_2 \in [0,1]$, and $
		W_{2}^2\left(\mu_{t,p_1},\mu_{t,p_2}\right)<\epsilon, \ \forall \ p_1,\ p_2 \in [0,1],\ t \in [0,T].
		$
\end{lemma}

\begin{proof}
See Appendix B for the proof.
\end{proof}

%
\section{ Spatio-Temporal  Approximation and  Law of Large Numbers}
\subsection{Spatio-Temporal Approximation of Graphon Particle System }
In this subsection, we prove that the graphon particle system (\ref{randmckvla}) is the spatio-temporal approximation of a discrete-time interacting particle system over the large-scale network.

\vskip 1.5mm
Consider the spatial discretization of the graphon particle system (\ref{randmckvla}). For any given positive integer $N$,  define  a step graphon  $A^{N}:[0,1]\times [0,1] \to [0,1]$ as  $A^{N}(0,0)=A(0,0),\  A^{N}(p,q)=A\big(\frac{\lceil Np \rceil}{N},\frac{\lceil Nq \rceil}{N}\big)=A\left(\frac{i}{N},\frac{j}{N}\right), \ p \in \big(\frac{i-1}{N},\frac{i}{N}\big], \ q \in (\frac{j-1}{N},\frac{j}{N}]$,
$i, \ j=1,2,\ldots,N$.
Define $z_{N,p}(t)=z_{\frac{i}{N}}(t)$, $\eta^{N}_{p}(t)=\eta_{\frac{i}{N}}(t),$ $w^{N}_{p}(t)=w_{\frac{i}{N}}(t),\ p \in (\frac{i-1}{N},\frac{i}{N}]$, $i=1,2,\ldots,N$.
 Let $\mu_{t}^{N}(dz,dq)$ be the distribution on $\mathbb{R}^{n} \times [0,1]$ satisfying the following conditions.  (i) The marginal distribution $\mu_{t}^{N}(\cdot,dq)$ is always the uniform distribution on $[0,1]$, that is, $\mu_{t}^{N}(\cdot,dq)=dq$, $\forall \ t \geqslant 0$. (ii) For any $j=1,2,\ldots,N$, given $q \in \big(\frac{j-1}{N},\frac{j}{N}\big]$, the conditional distribution $\mu_{t}^{N}\big(dz|q \big)=\delta_{z_{N,\frac{j}{N}}(t)}(dz)$. This together with (\ref{randmckvla}) leads to the following system
\begin{align}\label{stepgraphonbig}
dz_{N,p}(t)=&\bigg[\int_{[0,1]}A^{N}(p,q)
\bigg(\int_{ \mathbb{R}^{n}}F\big(t,p, q,z,z_{N,p}(t)\big)\mu^{N}_{t,q}(dz)\bigg)dq\notag\\
&+G\big(t,p,\eta^{N}_{p}(t),z_{N,p}(t)\big)\bigg]dt +H\big(t,p,\eta^{N}_{p}(t),
z_{N,p}(t)\big)dw^{N}_{p}(t),\ \forall \ p \in (0,1].
\end{align}
 Take $p=\frac{i}{N},\ i=1,2,\ldots,N$ in (\ref{stepgraphonbig}) and
 denote $z^{N}_{i}(t)=z_{N,\frac{i}{N}}(t)$  and $a_{N,ij}=A^{N}(\frac{i}{N},\frac{j}{N}), \ i, \ j=1,2,\ldots,N$.
 From the  definition of the conditional distribution, we have   $\mu^{N}_{t}(dz,dq)=\mu_{t}^{N}\big(dz|q \big)dq$.  Let $z_{i}^{N}(0)=z_{\frac{i}{N}}(0),  \  i=1,2,\ldots,N.$
 Then, we have the $N$-particle system
 \begin{align}
dz^{N}_{i}(t)
=&\Bigg[
\sum_{j=1}^{N}\int_{\big(\frac{j-1}{N},\frac{j}{N}\big]}a_{N,ij}
\bigg(\int_{ \mathbb{R}^{n}}F\Big(t,\frac{i}{N}, \frac{j}{N},z,z^{N}_{i}(t)\Big) \mu^{N}_{t}(dz,dq)\bigg)
+G\Big(t,\frac{i}{N},\eta_{\frac{i}{N}}(t),z^{N}_{i}(t)\Big)\Bigg]dt
 \notag\\
& +H\Big(t,\frac{i}{N},\eta_{\frac{i}{N}}(t),
z^{N}_{i}(t)\Big)dw_{\frac{i}{N}}(t)\notag\\
=& \Bigg[
\sum_{j=1}^{N}\int_{\big(\frac{j-1}{N},\frac{j}{N}\big]}a_{N,ij}
\bigg(\int_{ \mathbb{R}^{n}}F\Big(t,\frac{i}{N}, \frac{j}{N},z,z^{N}_{i}(t)\Big)\mu^{N}_{t}(dz|q)\bigg)dq
+G\Big(t,\frac{i}{N},\eta_{\frac{i}{N}}(t),z^{N}_{i}(t)\Big)\Bigg]dt
\notag\\
&+H\Big(t,\frac{i}{N},\eta_{\frac{i}{N}}(t),
z^{N}_{i}(t)\Big)dw_{\frac{i}{N}}(t)\notag\\
=& \Bigg[
\sum_{j=1}^{N}\int_{\big(\frac{j-1}{N},\frac{j}{N}\big]}a_{N,ij}
\bigg(\int_{ \mathbb{R}^{n}}F\Big(t,\frac{i}{N}, \frac{j}{N},z,z^{N}_{i}(t)\Big)\delta_{z_{j}^{N}(t)}(dz)\bigg)dq
+G\Big(t,\frac{i}{N},\eta_{\frac{i}{N}}(t),z^{N}_{i}(t)\Big)\Bigg]dt
\notag\\
&+H\Big(t,\frac{i}{N},\eta_{\frac{i}{N}}(t),
z^{N}_{i}(t)\Big)dw_{\frac{i}{N}}(t)\notag\\
=& \Bigg[G\Big(t,\frac{i}{N},\eta_{\frac{i}{N}}(t),z^{N}_{i}(t)\Big)
+\frac{1}{N}\sum_{j=1}^{N}a_{N,ij}
F\Big(t,\frac{i}{N},\frac{j}{N},
z_{j}^{N}(t),z^{N}_{i}(t)\Big)\Bigg]dt \notag\\
&+H\Big(t,\frac{i}{N},
\eta_{\frac{i}{N}}(t),
z^{N}_{i}(t)\Big)dw_{\frac{i}{N}}(t),\  i=1,2,\ldots,N. \label{finiteparticles}
\end{align}

We then consider the time discretization of the above stochastic differential equation. For any given positive integer $k$ and a  sequence $\{t_{m}=\frac{mT}{k},\ m=0,1, \ldots, k-1\}$ of the time interval $[0,T]$, $\frac{T}{k}$ is the  discretization step. By (7.1) in \cite{X. R. Mao}, the Euler-Maruyama approximate solutions to (\ref{finiteparticles}) satisfy that
\begin{align}
z^{N,k}_{i}(t_{m+1})
=&z^{N,k}_{i}\left(t_{m}\right)
+\frac{T}{k}
\bigg[G\Big(t_{m},\frac{i}{N},\eta_{\frac{i}{N}}(t_{m}),
z^{N,k}_{i} (t_{m} )\Big)\notag\\
&+\frac{1}{N}\sum_{j=1}^{N}a_{N,ij}
F\Big(t_{m},\frac{i}{N}, \frac{j}{N},z^{N,k}_{j}(t_{m}), z^{N,k}_{i}(t_{m})\Big)\bigg]\notag\\
&+
H\Big(t_{m},\frac{i}{N},\eta_{\frac{i}{N}}(t_{m}),
z^{N,k}_{i} (t_{m} )
\Big) \big(w_{\frac{i}{N}}(t_{m+1})
-w_{\frac{i}{N}}(t_{m})\big),\label{finiteparticlesdiscret1}
\end{align}
 $m=0,1,\ldots,k-1, \ i=1,2, \ldots,N$, where  $\big\{z_{i}^{N,k}(0)=z_{\frac{i}{N}}(0),\ i=1,2,\ldots,N\big\},\ k=1,2,\ldots$

\vskip  0.1cm
Now, we  establish the connection between the systems (\ref{randmckvla}) and (\ref{finiteparticlesdiscret1})  in both time and space dimensions.
\vskip 0.1cm

At first, we construct  the continuous-time approximation $\{\{z_{i}^{N,k}(t),\ t \in [0,T],\ i=1, 2, \ldots, N\}, \  k=1,2,\ldots\}$ of the system (\ref{finiteparticlesdiscret1}), defined as follows. For any $k=1,2,\ldots$,  $t \in \left(t_{m},t_{m+1}\right]$, $ m=0,1,\ldots,k-1$ and $ i=1,2,\ldots,N,$
\begin{align}
z^{N,k}_{i}(t)
=&z^{N,k}_{i} (t_{m} )
+\int_{t_{m}}^{t}
\Bigg[G\Big(t_{m},\frac{i}{N},\eta_{\frac{i}{N}}(t_{m}),
z^{N,k}_{i} (t_{m} )\Big)\notag\\
&
+\frac{1}{N}\sum_{j=1}^{N}a_{N,ij}
F\Big(t_{m},\frac{i}{N}, \frac{j}{N},z^{N,k}_{j} (t_{m} ),z^{N,k}_{i}
 (t_{m} )\Big)\Bigg]ds\notag\\
&+\int_{t_{m}}^{t}
H\Big(t_{m},\frac{i}{N},\eta_{\frac{i}{N}}(t_{m}),
z^{N,k}_{i} (t_{m} )
\Big)dw_{\frac{i}{N}}(s).\label{finiteparticlesdiscret}
\end{align}
Note that   $ \{z_{i}^{N,k}(t),\ t \in [0,T],\ i=1,2,  \ldots, N\},\ k=1,2,\ldots $  are also the Euler-Maruyama approximate solutions to the system (\ref{finiteparticles}).
 \vskip 0.1cm

 We give the existence and uniqueness of the solution to the system (\ref{finiteparticles}).
\vskip 0.1cm
\begin{lemma}\label{existenceoffinite0}
If Assumptions  \ref{assumption001}-\ref{assumption1}  hold, then the system (\ref{finiteparticles}) has a unique solution  $ \{z_{i}^{N}(t),\ t \in [0,T], \ i=1,2,\ldots,N \}$ satisfying that $\sup_{i \in \left\{1,\ldots,N\right\}}E\big[\sup_{t \in [0,T]} \|z_{i}^{N}(t) \|^{2}\big]<\infty.$
\end{lemma}
\vskip 1.5mm

The proof of the above lemma is similar to that of Theorem 5.2.1 in  \cite{Boksendal} and is therefore omitted.

Denote $\bar{z}_{p}^{N}(t)=z_{0}(t)\delta_{p}(\{0\})+\sum\limits_{i=1}^{N}z_{i}^{N}(t)
\delta_{p}
\big(\big(\frac{i-1}{N},
\frac{i}{N}\big]\big)$ and $\hat{z}_{p}^{N,k}(t) =z_{0}(t)\delta_{p}(\{0\})
+\sum_{i=1}^{N}z_{i}^{N,k}(t)$ $\delta_{p}\big(\big(\frac{i-1}{N},
\frac{i}{N}\big]\big),\ t \in [0,T], \ p \in [0,1]$.
\vskip 1.5mm

For the time dimension, we establish the connection between the systems (\ref{finiteparticles}) and (\ref{finiteparticlesdiscret1}) by the following lemma.
\vskip 0.2cm
\begin{lemma}\label{dsffinite}
For any given positive integer $N \geqslant 1$, if  Assumptions \ref{assumption001}-\ref{assumption1} hold, then the Euler-Maruyama approximate solutions (\ref{finiteparticlesdiscret})  converge to the solution to the continuous-time interacting particle system (\ref{finiteparticles}) in the sense that
 \begin{align}
\lim_{k\to \infty}\int_{[0,1]}E\bigg[\sup_{t \in [0,T]} \big\|\hat{z}_{p}^{N,k}(t)-\bar{z}^{N}_{p}(t) \big\|^{2}\bigg]dp=0.\notag
\end{align}
\end{lemma}

\vskip 1.5mm
The proof of the above lemma is similar to that of Theorem 7.3 in  \cite{X. R. Mao} and is therefore omitted.

\vskip 1.5mm
For the space dimension,  we will establish the approximation relation between the systems  (\ref{randmckvla})  and   (\ref{finiteparticles}).
The difficulty lies in the analysis of  the  difference between
 the step graphon $A^{N}$ and the  graphon $A$ coupled with  $F(t,p, q,z,z_{p}(t))$. To solve this, we construct two-level approximated functions converging to the graphon mean field terms. On the first level, we
approximate $F(t,p, q,z,z_{p}(t))$ with an indicator function of a bounded set and on the second level, we construct a polynomial of $z$ and $z_{p}(t)$ to approximate the part of $F(t,p, q,z,z_{p}(t))$ inside the bounded set, with
   the approximation error decreasing as the bounded set  expands. Then,  we prove that the coupling term inside the bounded set can be controlled by the approximation error of the polynomial and the difference between the step graphon $A^{N}$  and the graphon $A$. For the coupling term outside the bounded set,  using  H{\"o}lder inequality and Chebyshev inequality,  we prove that it vanishes  with the expansion of the bounded set.
   Then,  by the continuity of the graphon, we prove that the coupling term vanishes with the increase of the number of particles and the expansion of the bounded set, which in turn gives the following lemma.

\vskip 1.5mm
\begin{lemma}\label{difflisandao}
If Assumption  \ref{assumption01}, Assumption  \ref{assumption001}  and   Assumption \ref{assumption1}  with $\upsilon=\min\{\upsilon_{0}, \upsilon_{1}\}>0$ hold, then the continuous-time interacting particle system (\ref{finiteparticles}) over the large-scale network approximates the graphon particle system  (\ref{randmckvla}) in the sense that
\begin{align}
\lim_{N\to \infty}E\bigg[W_{1,T}\bigg(\frac{1}{N}\sum_{i=1}^{N}\delta_{z_{i}^{N}},\int_{[0,1]}
\mu_{p}dp\bigg)\bigg]=0.\label{convergewasserstein1}
\end{align}
Especially, if  $\{z_{p}(0), \ p\in[0,1]\}$ and  $\{\eta_{p}, \ p\in [0,1]\}$ are deterministic   and all the above assumptions are also satisfied, then we have
\begin{align}
\lim_{N\to \infty}\int_{[0,1]}E\bigg[\sup_{t \in [0,T]} \big\|\bar{z}_{p}^{N}(t)-z_{p}(t) \big\|^{2}\bigg]dp=0.\label{NTOINFZ}
\end{align}
\end{lemma}
\vskip 1.5mm
\begin{proof}
See Appendix B for the proof.
\end{proof}
\vskip 1.5mm

By Lemmas \ref{dsffinite}-\ref{difflisandao}, we give  the following law of large numbers.
\vskip 1.5mm
\begin{theorem}\label{finitetographon}
If Assumption  \ref{assumption01}, Assumption  \ref{assumption001} and   Assumption \ref{assumption1}  with $\upsilon=\min\{\upsilon_{0},  \upsilon_{1}\}>0$ hold,  then the continuous-time approximation (\ref{finiteparticlesdiscret}) of the discrete-time interacting particle system (\ref{finiteparticlesdiscret1}) over the large-scale network approximates the graphon particle system (\ref{randmckvla}) in the sense that
\begin{align}
&\lim_{N \to \infty}\lim_{k\to \infty} E\bigg[W_{1,T}\bigg(\frac{1}{N}\sum_{i=1}^{N}\delta_{z _{i}^{N,k}},\int_{[0,1]}
\mu_{p}dp\bigg)\bigg]=0. \label{NTOINFTYwass}
\end{align}
Especially, if  $\{z_{p}(0),\  p\in[0,1]\}$ and  $\{\eta_{p}, \ p\in [0,1]\}$ are deterministic  and all the above assumptions are also satisfied, then we have
\begin{align}
&\lim_{N \to \infty}\lim_{k\to \infty}\int_{[0,1]}E\bigg[\sup_{t \in [0,T]} \left\|\hat{z}_{p}^{N,k}(t)-z_{p}(t) \right\|^{2}\bigg]dp=0.\label{NTOINFTYGRS}
\end{align}

\end{theorem}
\vskip 2mm
\begin{proof}
See Appendix B for the proof.
\end{proof}

\vskip 2mm

\begin{remark}
Assumption \ref{assumption01}  is stronger than
Condition 2.2 (b) of \cite{ERHAN.BAYRAKTAR}. The former
requires the graphon to be continuous, whereas the latter
requires it to be piecewise continuous.
Indeed,  Theorem~ \ref{finitetographon}   can be further extended to the graphon particle systems with  piecewise continuous graphons.
 For such  graphons,
Lemma \ref{labelconti} can be reformulated by replacing global
continuity with the continuity on the interior of each subinterval, which
remains sufficient to establish (\ref{lianxu}). The piecewise
continuity condition is also sufficient to establish
(\ref{fidsts1}). Consequently, the proof of Lemma
\ref{difflisandao} remains valid with only minor modifications, and
hence Theorem \ref{finitetographon} still
holds for piecewise continuous graphons.
\end{remark}
\vskip 2mm
\begin{remark}
 In   \cite{ERHAN.BAYRAKTAR}, the convergence rates with respect to $N$ are obtained under globally Lipschitz continuity conditions on the functions $F,\ G$ and $H$.  In this paper, we consider the general case with time-varying and random coefficients, which are only uniformly    continuous w.r.t.   $p$.
 It is expected that the convergence rate with respect to $N$ can also be established if the globally Lipschitz  continuity conditions are assumed, which is an interesting topic for future investigation.
\end{remark}

\subsection{Spatio-Temporal Approximation of SGD Algorithm }

In this subsection, we will prove that a special graphon particle system (\ref{Mckeanvlasov1}) can be regarded as the spatio-temporal approximation of the distributed SGD algorithm  over the large-scale network.

\vskip 1.5mm
We give assumptions on the graphon particle system (\ref{Mckeanvlasov1}).

\begin{assumption}\label{assumption2}
 There exists a constant $\kappa>0$, such that $\left\|\nabla_{z} V(p,z_{1})-\nabla_{z} V(p,z_{2})\right\| \leqslant \kappa \| z_{1}- z_{2}\|$, $\forall \ z_{1}, \ z_{2} \in \mathbb{R}^{n},  \ p \in [0,1]$; there exist constants $\sigma_{v}>0$ and $C_{v}>0$ such that $\|\nabla_{z} V(p,z)\| \leqslant \sigma_{v}\|z\|+C_{v}$,  $\forall \ z \in \mathbb{R}^{n},\ p \in [0,1]$.
\end{assumption}

\begin{assumption}\label{assumption205}
 For any $\epsilon>0$, there exists $\delta>0$, such that if $|p_{1}-p_{2}|<\delta$, then $$\left\|V(p_{1},z)-V(p_{2},z)\right\|+\left\|\nabla_{z}V(p_{1},z)-\nabla_{z}V(p_{2},z)\right\|<\epsilon(L_{v}\|z\|+ L_{v}),\ \forall\ p_{1}, \ p_{2} \in [0,1],\ z \in \mathbb{R}^{n}.$$
\end{assumption}

\begin{assumption}\label{assumptiondescent02}
 There exist constants $\zeta_{2} >0$ and $\upsilon_{1}\geqslant 0$, such that $\sup_{p \in [0,1]}E\big[\|z_{p}(0)\|^{2+\upsilon_{1}}\big] \leqslant \zeta_{2}$, the map $ [0,1] \ni p\mapsto \mathcal{L}(z_{p}(0))\in \mathscr{P}(\mathbb{R}^{n})$ is measurable and   for any $\epsilon >0$, there exists $\delta>0$, such that if   $|p_1-p_2|<\delta$, then $  W_{2}(\mathcal{L}(z_{p_{1}}(0)), \mathcal{L}(z_{p_{2}}(0)))<\epsilon,\ \forall \ p_1,\ p_2 \in [0,1]$.
\end{assumption}

\vskip 1.5mm
\begin{assumption}\label{assumptiondescent1}
The time-varying algorithm gains satisfy that $\alpha_{1}(t)>0$, $\alpha_{2}(t)>0$  and $\alpha_{1}(t),$ $\alpha_{2}(t)$ are continuous w.r.t. $t$.
\end{assumption}

\vskip 1.5mm

By the spatial and temporal discretization, we can show how (\ref{globalgoal}) and (\ref{Mckeanvlasov1}) are related to the distributed optimization over the network with finite nodes.

\vskip 1.5mm
For any given positive integer $N$, we define $V^{N}(p,z)=V(\frac{i}{N},z), \ p \in \left(\frac{i-1}{N},\frac{i}{N}\right]$, $i=1,2,\ldots,N$ and denote
 $v_{N,i}(z)$ $=V^{N}(\frac{i}{N},z)$, $i=1,2,\ldots,N$. Then one obtains the distributed optimization problem over the network with $N$ nodes, that is,
\setlength{\abovedisplayskip}{3pt}   
\setlength{\belowdisplayskip}{3pt}   
\begin{align}\label{finitedistr}
\min \limits_{z \in \mathbb{R}^{n}}\int_{[0,1]}V^{N}(p,z)dp=\min \limits_{z \in \mathbb{R}^{n}}\frac{1}{  N}\sum\limits_{i=1}^{N}v_{N,i}(z).
\end{align}
By Assumptions \ref{assumption205}, we have
$\lim_{N\to \infty} \int_{[0,1]}V^{N}(p,z)dp= \int_{[0,1]}V(p,z)dp,  \ z \in \mathbb{R}^{n}$. By Assumption \ref{assumption01}  and Lemma 8.11 in \cite{L. Lovasz}, we have
 $\lim_{N\to \infty}\|A^{N}-A\|_{\infty \to 1}^{2}=0.$
 Therefore, the distributed optimization problem (\ref{globalgoal}) over the graphon $A$ is the limit of
   the distributed optimization problem (\ref{finitedistr}) over the  graph  with the weighted adjacency matrix $(a_{N,ij})_{i,j=1,\cdots,N}$ as the  number of nodes $N$ goes to infinity.

\vskip 1.5mm

Similar to the proof of (\ref{finiteparticles})-(\ref{finiteparticlesdiscret1}), we have the following discrete-time interacting particle system over the large-scale network. For any $\ k=1,2,\ldots,$ $m=0,1,\ldots,k-1$ and $ i=1,2,\ldots,N,$
\begin{align}
z^{N,k}_{i}(t_{m+1})
=&z^{N,k}_{i}(t_{m})+\frac{\alpha_{1}(t_{m})T}{Nk} \sum\limits_{j=1}^{N}
a_{N,ij}\big(z^{N,k}_{j}(t_{m})-z^{N,k}_{i}(t_{m})\big)\notag\\
&-\frac{\alpha_{2}(t_{m})T}{k}\nabla_{z} v_{N,i}\big(z^{N,k}_{i}(t_{m})\big) -\alpha_{2}(t_{m})\xi^{N,k}_{i}(t_{m}),\label{finitestochgrad}
\end{align}
where $ \xi^{N,k}_{i}(t_{m})=\Sigma_{1} \big(w_{\frac{i}{N}}(t_{m+1})-w_{\frac{i}{N}}(t_{m})\big)$ is an $n$-dimensional  martingale difference sequence with zero mean and covariance matrix
$\frac{T}{k}\Sigma_{1} \Sigma_{1}^{T}$. It can be verified that (\ref{finitestochgrad}) is just the distributed SGD algorithm  over the network with finite nodes in \cite{A. Nedic}-\cite{Li T}.

\begin{remark}
 Generally, the true gradient value is approximated by a noisy estimate computed on a mini-batch of data. This noise arises due to the randomness in sampling the mini-batch. Under certain assumptions (e.g., small learning rate, independent samples), the cumulative effect of this noise over many iterations can be modeled as a diffusion process   (\ref{Mckeanvlasov1}), where the noise term resembles the increments of a Brownian motion (\cite{Wojtowytsch}).
\end{remark}

\vskip 1.5mm
The distributed SGD algorithm (\ref{finitestochgrad}) can be written as
\begin{align}
z^{N,k}_{i}(t_{m+1})
=&z^{N,k}_{i}(t_{m})
+\frac{\alpha_{1}(t_{m})T}{Nk}\sum_{j=1}^{N}A^{N}\Big(\frac{i}{N},\frac{j}{N}\Big)
 \big(z^{N,k}_{j}(t_{m})-z^{N,k}_{i}(t_{m})\big)
 \notag\\
& -\frac{\alpha_{2}(t_{m})T}{k} \nabla_{z} V\Big(\frac{i}{N},z^{N,k}_{i}\left(t_{m}\right)\Big) -\alpha_{2}(t_{m})
\xi^{N,k}_{i}(t_{m}),\label{finisgradient}
\end{align}
where $ m=0,1,\ldots,k-1, \ i=1,2,\ldots,N, \ k=1,2,\ldots$
\vskip 1.5mm

The continuous-time approximation of the system (\ref{finisgradient}) is given as follows. Given the initial states $\big\{z_{i}^{N,k}(0)=z_{\frac{i}{N}}(0),\ i=1,2,\ldots,N,\ k=1,2,\ldots\big\},$ for any $\ k=1,2,\ldots$,  $t \in \left(t_{m},t_{m+1}\right]$, $m=0,1,\ldots,k-1$ and $ i=1,2,\ldots,N,$
\begin{align}
z^{N,k}_{i}(t)=&z^{N,k}_{i}\left(t_{m}\right)
+\int_{t_{m}}^{t}
\bigg[\frac{\alpha_{1}(t_{m})}{N}\sum_{j=1}^{N}A^{N}\Big(\frac{i}{N},
\frac{j}{N}\Big)
\big(z^{N,k}_{j}\left(t_{m}\right)-z^{N,k}_{i}\left(t_{m}\right)\big)\notag\\
&
-\alpha_{2}(t_{m})\nabla_{z} V\Big(\frac{i}{N},z^{N,k}_{i}\left(t_{m}\right)\Big)\bigg]ds-\int_{t_{m}}^{t}\alpha_{2}(t_{m})\Sigma_{1}
dw_{\frac{i}{N}}(s).\label{finiteparticlegradient}
\end{align}

Denote  $\frac{T}{k}$ as the   algorithm step size and $\hat{z}_{p}^{N,k}(t)=z_{0}(t)\delta_{p}\left(\{0\}\right)+
\sum_{i=1}^{N}
z_{i}^{N,k}(t)
\delta_{p} ( (\frac{i-1}{N},
\frac{i}{N} ] ), \ t \in [0,T], \ p \in [0,1]$.
\vskip 1.5mm

Next, we show that the step graphon particle system  $\{\hat{z}_{p}^{N,k}(t),\ t \in [0,T],
\   p \in [0,1]\}$  converges to the graphon particle system (\ref{Mckeanvlasov1}).
As the graphon particle system (\ref{Mckeanvlasov1}) and the interacting particle system  (\ref{finiteparticlegradient}) are the special cases of the systems
  (\ref{randmckvla}) and   (\ref{finiteparticlesdiscret}), respectively, by Theorem \ref{existencegeneral} and Theorem \ref{finitetographon}, we have the following corollaries.
\vskip 1.5mm
\begin{corollary}\label{exopti}
If Assumptions \ref{assumption2}, \ref{assumptiondescent02}-\ref{assumptiondescent1} hold, then the graphon particle system (\ref{Mckeanvlasov1}) has a unique solution $ \{z_{p}, \ \mu_{p},\ p\in[0,1] \}$  on $[0,T]$  satisfying that the map $[0,1]  \ni  p\mapsto \mu_{p}\in \mathscr{P}_{2}(\mathcal{C}_{T}^{n})$ is measurable and  $$\sup\limits_{p \in [0,1]} E\bigg[\sup_{t\in[0,T]}  \|z_{p}(t)\|^{2+\upsilon_{1}}\bigg]  <\infty,$$ where $\mu_{p}= \mathcal{L}(z_{p})\in  \mathscr{P}_{2}(\mathcal{C}_{T}^{n})$.
\end{corollary}
\vskip 1.5mm
\begin{corollary}\label{largop}
If Assumption  \ref{assumption01},  Assumption  \ref{assumption2}, \ref{assumption205}, Assumption \ref{assumptiondescent02} with $\upsilon_{1}>0$ and Assumption  \ref{assumptiondescent1} hold, then the continuous-time approximation (\ref{finiteparticlegradient})  of the SGD algorithm (\ref{finisgradient}) over the large-scale network approximates the graphon particle system (\ref{Mckeanvlasov1}) in the sense that
 $$\lim_{N \to \infty}\lim_{k\to \infty} E\bigg[W_{1,T}\bigg(\frac{1}{N}\sum_{i=1}^{N}\delta_{ z _{i}^{N,k}},\int_{[0,1]}
\mu_{p}dp\bigg)\bigg] =0.$$
Especially, if  $\{z_{p}(0),\ p\in[0,1]\}$  are deterministic and all the above assumptions are also satisfied, then we have
$$\lim_{N\to \infty}\lim_{k\to \infty}\int_{[0,1]}E\Bigg[\sup\limits_{t \in [0,T]}\|\hat{z}_{p}^{N,k}(t)-z_{p}(t)\|^{2}\Bigg]dp=0.$$
\end{corollary}

Corollary \ref{largop} has shown the relation between the graphon particle system (\ref{Mckeanvlasov1})    and  the discrete-time SGD algorithm  (\ref{finisgradient}).
For the distributed optimization problem (\ref{globalgoal})  and the SGD algorithm (\ref{Mckeanvlasov1}) over the graphon, people are also concerned with the convergence of the algorithm, which will be investigated in the companion paper \cite{Li}.

\section{Conclusions}
We study a class of graphon particle systems with time-varying random coefficients and  prove the existence and uniqueness  of solutions under some suitable conditions. We then  prove the law of large numbers in both time and space dimensions, that is, this class of graphon particle systems is the limit of the discrete-time interacting particle system over the large-scale network  as the number of particles tends to   infinity and the discretization step tends to   zero.  Moreover, we prove that the limiting dynamics of the  distributed SGD algorithm over the large-scale network is a graphon particle system.

\appendices

\setcounter{equation}{0}
\renewcommand{\theequation}{A.\arabic{equation}}
\section{}


\emph{Proof of Theorem \ref{existencegeneral}:} For any given $t\in [0,T]$,  denote the map $\pi_{t}:\mathcal{C}_{T}^{n} \to \mathbb{R}^{n}$,  $ \theta \mapsto \theta(t)$.
Define the map $\mathcal{M} \ni \mu \mapsto \Phi(\mu) \in \left[\mathscr{P}\left(\mathcal{C}_{T}^{n}\right)\right]^{[0,1]}$ as $\Phi(\mu)=\left\{\mathcal{L}(z_{p}), \ p \in [0,1]\right\}$, where $z_{p}$ is the solution to the equation
\begin{align}\label{integralform}
z_{p}(t)-z_{p}(0)
=&\int_{0}^{t}\bigg[\int_{[0,1]\times \mathbb{R}^{n}} A(p,q)F(s,p, q,z,z_{p}(s))\mu_{q} \circ \pi_{s}^{-1}(dz)dq \notag\\
&+ G(s,p,\eta_{p}(s),z_{p}(s))\bigg]ds+ \int_{0}^{t} H(s,p,\eta_{p}(s),z_{p}(s))dw_{p}(s).
\end{align}

We will prove the existence and uniqueness of the solution to the system (\ref{randmckvla}) by  that of the fixed point of the map  $\Phi$. What's more, we need to prove
the measurability of the map $[0,1]\ni q  \mapsto\mu_{q}\in \mathscr{P}\left(\mathcal{C}_{T}^{n}\right)$ to guarantee that the term $\int_{[0,1]}A(p,q)\big(\int_{ \mathbb{R}^{n}}F(t,p, q,z, z_{p}(t))\mu_{t,q}(dz)\big)dq$  in (\ref{randmckvla}) is well-defined, that is, the map $[0,1] \ni q \mapsto \int_{ \mathbb{R}^{n}}A(p,q)F(t,p, q,z, z_{p}(t))\mu_{t,q}(dz)\in \mathbb{R}^{n}$ is measurable. So, we prove the existence and uniqueness  of the fixed point of the map  $\Phi$ in $\mathcal{M}$.

The proof  can be divided into the following three steps. (i) The map $\Phi$ is well-defined.  (ii)  The fixed point of the map $\Phi$ in $\mathcal{M}$ exists. (iii)  The fixed point of the map $\Phi$ in $\mathcal{M}$ is unique.

At first, we will prove that the map $\Phi$ is well-defined, that is, for any given $\mu \in \mathcal{M}$,   (\ref{integralform}) has a unique strong solution and $\Phi(\mu) \in \mathcal{M}$.
Now, we show the existence.
Let $z_{p}^{0}(t)=z_{p}(0)$ and $z_{p}^{k}(0)=z_{p}(0),\ \forall \ t \in [0,T], \ k \in \mathbb{N}$. For any $k =1,2,\ldots$, let
\begin{align}\label{iteexist}
 z_{p}^{k}(t)
=&z_{p}^{k-1}(0)+M_{1p}^{k-1}(t)+M_{2p}^{k-1}(t)+M_{3p}^{k-1}(t),
\end{align}
where
\begin{align*}
M_{1p}^{k-1}(t)=&\int_{0}^{t}   H (s,p,\eta_{p}(s),z^{k-1}_{p}(s) )dw_{p}(s),\\
 M_{2p}^{k-1}(t)  =& \int_{0}^{t}  G (s,p,\eta_{p}(s),z^{k-1}_{p}(s) )  ds,\\
 M_{3p}^{k-1}(t) =&  \int_{0}^{t}  \int_{[0,1]\times \mathbb{R}^{n}}  A(p,q) F(s,p, q,z,z_{p}^{k-1}(s))\mu_{s,q}(dz)dq  ds.
\end{align*}
 Denote  $L_{p}^k(t) =  E [\sup_{s^{\prime} \in [0,t]} \|z_{p}^{k}(s^{\prime}) \|^{2+\upsilon} ]$ and $\widetilde{L}= \sup_{q \in {[0,1]}}
\int_{\mathcal{C}_{T}^{n}}\|z\|_{*,T}^{2}  \mu_{q}(dz)$.
We will prove
  $$  \sup_{k \in \mathbb{N},\ p \in [0,1]}  E \bigg[\sup_{s \in [0,t]} \|z_{p}^{k}(s) \|^{2+\upsilon} \bigg]\leqslant m_{1}(t),\ t\in [0,T]$$
  by induction, where
$m_{1}(t)=\big(P_{2}\widetilde{L}^{\frac{2
+\upsilon}{2}} +P_3\big)e^{P_{1}t}, $
$P_{1}=P_{2}+ 12 ^{1+\upsilon} C_{1}^{2+\upsilon} ( H_{\upsilon}T^{\frac{\upsilon}{2}}+ T^{1+\upsilon}),$
 $P_{2}=  12 ^{1+\upsilon}C_{2}^{2+\upsilon}T^{1+\upsilon}(1+T),$    $P_3=P_{2}+4^{1+\upsilon}  (\zeta^{\frac{2+\upsilon}{2+\upsilon_{0}}})    +12^{1+\upsilon}C_{1}^{2+\upsilon} ( H_{\upsilon}T^{\frac{\upsilon}{2}}+ T^{1+\upsilon})(1+r^{\frac{2+\upsilon}{2+\upsilon_{1}}}).$
For $k=0$, by Assumption \ref{assumption001}, we know that the conclusion holds. Now,  suppose that   $$\sup_{ p \in [0,1]}E \bigg[\sup_{s \in [0,t]}   \|z_{p}^{m}(s) \|^{2+\upsilon} \bigg]\leqslant  m_{1}(t)$$
 holds for   $k=m$, $m=0,1,2,...$
  By  Assumption \ref{assumption1} (i), Lyapunov inequality and  $C_{r}$ inequality, we have
\begin{align}
&E\Big[\textstyle{ \int_{0}^{T}} \big\| H\left(s,p,\eta_{p}(s),z^{m}_{p}(s)\right)\big\|^{2+\upsilon} ds \Big]\notag\\
\leqslant  &C_{1}^{2+\upsilon} 3^{1+\upsilon}\textstyle{  \int_{0}^{T}} E\left[ 1+\|\eta_{p}(s)\|^{2+\upsilon} +\| z^{m}_{p}(s)\|^{2+\upsilon} \right]ds \notag\\ \leqslant &\textstyle{TC_{1}^{2+\upsilon} 3^{1+\upsilon}  \big(1+\hspace{-2pt}r^{\frac{2+\upsilon}{2+\upsilon_{1}}}+\hspace{-2pt}E\big[ \sup_{t \in [0,T]}\|z^{m}_{p}(t)\|^{2+\upsilon} \big]\big)<\infty.}\notag
\end{align}
Then, by Theorem 1.7.2 in  \cite{X. R. Mao}, we have
\begin{align}
&E\bigg[\sup_{s \in [0,t]} \|M_{1p}^{m}(s) \|^{2+\upsilon}  \bigg] \notag\\
\leqslant &
H_{\upsilon}T^{\frac{\upsilon}{2}}  \int_{0}^{t}   E\big[\|H(s,p,\eta_{p}(s),z^{m}_{p}(s)
)\|^{2+\upsilon}\big]ds \notag\\
\leqslant & 3^{1+\upsilon} C_{1}^{2+\upsilon}\left(H_{\upsilon}T^{\frac{\upsilon}{2}}  \right) \bigg(T(1+r^{\frac{2+\upsilon}{2+\upsilon_{1}}}) +\int_{0}^{t} L_{p}^m(s)ds\bigg),\label{estiHse}
\end{align}
where $H_{\upsilon}=
\left(\frac{(2+\upsilon)^{3}}{2(1+\upsilon)}\right)^{\frac{2+\upsilon}{2}}$.
 By  H{\"o}lder inequality, Assumption \ref{assumption1} (ii),  and  $C_{r}$ inequality, we have
\begin{align}
&E\bigg[\sup_{s \in [0,t]}\|M_{3p}^{m}(s)\|^{2+\upsilon}\bigg]\notag\\
\leqslant & \textstyle{ t^{1+\upsilon}\int_{0}^{t}}   E\Big[ \big\|\textstyle{\int_{[0,1]\times \mathbb{R}^{n}}}  A(p,q)  F(s,p, q,z,z_{p}^{m}(s))\mu_{s,q}(dz)dq \big\|^{2+\upsilon}\Big]  ds\notag\\
\leqslant& T^{1+\upsilon} 3^{1+\upsilon}C_{2}^{2+\upsilon}  \textstyle{\int_{0}^{t}}   \big(1 + E[\|z_{p}^{m}(s)\|^{2+\upsilon}] \big)ds\notag\\
&+  3^{1+\upsilon}T^{1+\upsilon} C_{2}^{2+\upsilon}  \textstyle{\int_{0}^{t}}
 \big( \textstyle{\int_{[0,1]\times \mathbb{R}^{n}}}   \|z\|^{2}\mu_{s,q}(dz)dq\big)^{\frac{2+\upsilon}{2}} ds\notag\\
\leqslant &3^{1+\upsilon}T^{1+\upsilon}C_{2}^{2+\upsilon}\big(T\big(1+\widetilde{L}^{\frac{2+\upsilon}{2}}\big)+\textstyle{\int_{0}^{t}} L_{p}^m(s)ds
\big).\notag
\end{align}
This together with Assumption \ref{assumption001}, Assumption \ref{assumption1} (i)-(ii), (\ref{iteexist})-(\ref{estiHse}) and H{\"o}lder inequality gives
\begin{align}
&E\bigg[\sup_{s \in [0,t]}\|z_{p}^{m+1}(s)\|^{2+\upsilon}\bigg]\notag\\
\leqslant &
 4^{1+\upsilon}\bigg(E\bigg[\sup_{s \in [0,t]}\big\|M_{2p}^{m}(s)\big\|^{2+\upsilon}+\sup_{s \in [0,t]}\|M_{3p}^{m}(s)\|^{2+\upsilon}\bigg]\notag\\
&+ E\big[\|z_{p}(0)\|^{2+\upsilon}\big]+E\bigg[\sup_{s \in [0,t]}  \|M_{1p}^{m}(s) \|^{2+\upsilon}\bigg]\bigg)\notag\\
\leqslant & P_{1}\int_{0}^{t}L_{p}^m(s)ds+P_3+P_{2}\widetilde{L}^{\frac{2
+\upsilon}{2}}\leq m_{1}(t).\label{boudesan}
\end{align}
Then,  the conclusion holds for $m+1$. Therefore, there exists $M_{1}=m_{1}(T)$ such that
\begin{align}
\sup\limits_{k \in \mathbb{N},\ p \in [0,1]}E \Bigg[\sup_{t \in [0,T]} \big\|z_{p}^{k}(t) \big\|^{2+\upsilon}\Bigg]\leqslant M_{1}.\label{finiteboundesem}
\end{align}
 By (\ref{iteexist}), $z_{p}^{k}(0)=z_{p}(0),\ k \in \mathbb{N}$ and $C_{r}$ inequality, we have
\begin{align}
&E\left[ \big\|z_{p}^{k+1}(t)-z_{p}^{k}(t) \big\|^{2}\right]\notag\\
\leqslant & 3E\left[\left\| \int_{0}^{t} (G(s,p,\eta_{p}(s),z_{p}^{k}(s))
-G(s,p,\eta_{p}(s),z_{p}^{k-1}(s)) )ds \right\|^{2}\right]\notag\\
&
+3E\left[\left\|\int_{0}^{t}\int_{[0,1]\times \mathbb{R}^{n}} A(p,q)\big(F(s,p, q,z,z_{p}^{k}(s)) -F(s,p,q,z,z_{p}^{k-1}(s))\big)\mu_{s,q}(dz)dqds\right\|^{2}\right]
\notag\\
& +3E\left[\left\|\int_{0}^{t}(H(s,p,\eta_{p}(s),z_{p}^{k}(s))-H(s,p,\eta_{p}(s),z_{p}^{k-1}(s))\big)dw_{p}(s)\right\|^{2}\right]\notag\\
=&:3E\left[\left\|S_{1p}(t)\right\|^{2}\right]+3E\left[\left\|S_{2p}(t)\right\|^{2}\right]+3    E\left[\left\|S_{3p}(t)\right\|^{2}\right].\label{uniqueness}
\end{align}
For the first term on the r.h.s. of  the above inequality, by Assumption \ref{assumption1} (i) and H{\"o}lder inequality, we have
\begin{align}
E\left[\left\|S_{1p}(t)\right\|^{2}\right]\leqslant & T\int_{0}^{t}E\Big[ \big\|G(s,p,\eta_{p}(s),z_{p}^{k}(s))
-G(s,p,\eta_{p}(s),z_{p}^{k-1}(s))\big \|^{2}\Big]ds\notag\\
\leqslant  & T\sigma_{1}\int_{0}^{t} E\Big[\big\| z_{p}^{k}(s)-z_{p}^{k-1}(s)\big\|^{2}\Big]ds.\label{uniquefirst}
\end{align}
For the second term on the r.h.s. of (\ref{uniqueness}), by Assumption \ref{assumption1} (ii) and H{\"o}lder inequality, we have
\begin{align}
E\left[\left\|S_{2p}(t)\right\|^{2}\right]\leqslant & TE\bigg[\int_{0}^{t}\int_{[0,1]\times \mathbb{R}^{n}}\big\|F(s,p, q,z,z_{p}^{k}(s)) -F(s,p,q, z,z_{p}^{k-1}(s))\big\|^{2}
\mu_{s,q}(dz)dqds\bigg]\notag\\
\leqslant  &
\sigma_{4}^{2}T\int_{0}^{t}E\Big[ \big\|z_{p}^{k}(s)-
z_{p}^{k-1}(s) \big\|^{2}\Big]ds.\label{uniquefirst1}
\end{align}
For the third term on the r.h.s. of  (\ref{uniqueness}), note that the solutions satisfy that $\sup_{t \in [0,T]}E\big[\| z_{p}^{k-1}(t)\|^{2}\big]\\ <\infty$  and $\sup_{t \in [0,T]}    E\big[\| z_{p}^{k}(t)\|^{2}\big]  <\infty$. This together with Assumption \ref{assumption1} (i) leads to
\begin{align}
&E\bigg[\int_{0}^{t} \|H(s,p,\eta_{p}(s),z_{p}^{k}(s))- H(s,p,\eta_{p}(s),  z_{p}^{k-1}(s))\|^{2} ds\bigg]\notag\\
\leqslant & \sigma_{1}  \int_{0}^{t} E\left[\|z_{p}^{k}(s) -z_{p}^{k-1}(s)\|^{2}\right]ds \notag\\
 \leqslant & 2T \sigma_{1}  \bigg( \sup_{t \in [0,T]} E\left[\| z_{p}^{k-1}(t)\|^{2}\right] + \sup_{t \in [0,T]} E\left[\| z_{p}^{k}(t)\|^{2}\right]\bigg)
 <   \infty,\notag
\end{align}
 which together with  It\^{o}  isometry  and Assumption \ref{assumption1} (i) gives
\begin{align}
E\left[\left\|S_{3p}(t)\right\|^{2}\right]\leqslant  \sigma_{1} \int_{0}^{t} E\Big[\big\|z_{p}^{k}(s)-z_{p}^{k-1}(s)\big\|^{2}\Big]ds.\notag
\end{align}
This together with (\ref{uniqueness})-(\ref{uniquefirst1}) gives
\begin{align}
 E\Big[ \|z_{p}^{k+1}(t)-z_{p}^{k}(t) \|^{2}\Big]
 \leqslant   L\int_{0}^{t}  E\big[ \| z_{p}^{k}(s)-z_{p}^{k-1}(s) \|^{2}\big]ds,\label{uniqthird}
\end{align}
where $L=3(T+1)\sigma_{1}+3T\sigma_{4}^{2}$.
By  Cauchy formula  $$  \int_{t_0}^t \int_{t_0}^{t_{k-1}} \dotsi \int_{t_0}^{t_1} f(s) d s d t_1 \dotsi d t_{k-1}=\frac{1}{(k-1) !} \int_{t_0}^t(t-s)^{k-1} f(s) ds  $$ and \eqref{uniqthird}, we have
\begin{align}
E\big[ \|z_{p}^{k+1}(t)-z_{p}^{k}(t) \|^{2}\big]
 \leqslant \frac{L^{k}}{(k-1)!}\int_{0}^{t}(t-s)^{k-1}
E \big[\|z_{p}^{1}(s)-z_{p}^{0}(s)\|^{2} \big]ds. \label{befgron}
\end{align}
 By
$C_{r}$ inequality, Assumption \ref{assumption001}  and (\ref{finiteboundesem}), we know that there exists a constant $L_{1}=2\big(M_{1}+\zeta^{\frac{2}{2+\upsilon_{0}}}\big)$ such that
$\sup_{t \in [0,T]}E\big[ \|z_{p}^{1}(t)-z_{p}^{0}(t) \|^{2}\big]
 \leqslant   2\sup_{t \in [0,T]}E\big[ \|z_{p}^{1}(t) \|^{2}\big]+2\sup_{t \in [0,T]}E\big[ \|z_{p}^{0}(t) \|^{2}\big] \\ \leqslant  L_{1}.$
Note that $L$ and $L_{1}$  are uniform w.r.t.  $p\in [0,1]$ and $t \in [0,T]$. Then, by (\ref{befgron}), we have
\begin{align}
\sup\limits_{p \in [0,1],\ t \in [0,T]}E\big[ \|z_{p}^{k+1}(t)-z_{p}^{k}(t) \|^{2}\big] \leqslant \frac{L^{k}T^{k}L_{1}}{k!}.\notag
\end{align}
Combining this with Assumption \ref{assumption1} (i)-(ii), $C_{r}$ inequality, (\ref{iteexist}), (\ref{finiteboundesem}) and Theorem 3.6 in \cite{Frideman} gives
\begin{align}
&E\Bigg[\bigg(\sup_{t \in [0,T]} \big\|z_{p}^{k+1}(t)-z_{p}^{k}(t) \big\|\bigg)^{2}\Bigg]\notag\\
=&E\bigg[\sup_{t \in [0,T]} \big\|z_{p}^{k+1}(t)-z_{p}^{k}(t) \big\|^{2}\bigg]\notag\\
\leqslant & 3\sum_{i=1}^3E\Bigg[\sup_{t \in [0,T]}\|S_{ip}(t)\|^{2}\Bigg]\notag\\
\leqslant & 3T\int_{0}^{T}E\Big[  \big\|G(s,p,\eta_{p}(s),z^{k}_{p}(s))
-G\ (s,p,\eta_{p}(s),z^{k-1}_{p}(s) ) \big\|^{2}\Big]ds
\notag\\
&+12\int_{0}^{T}E\Big[  \big\|H (s,p,\eta_{p}(s),z^{k}_{p}(s) )
-H (s,p,\eta_{p}(s),z^{k-1}_{p}(s) ) \big\|^{2}\Big]ds\notag\\
&+3T\int_{0}^{T} E\bigg[\int_{[0,1]\times \mathbb{R}^{n}} \big\|F (s,p, q,z,z_{p}^{k}(s) ) -F (s,p, q,z,z_{p}^{k-1}(s) ) \big\|^{2}\mu_{s,q}(dz)dq \bigg]ds\notag\\
\leqslant &\left(L+9\sigma_{1}\right)\int_{0}^{T}E\big[ \|z_{p}^{k}(s)-z_{p}^{k-1}(s) \|^{2}\big]ds\notag\\
\leqslant&\left(L+9\sigma_{1}\right)T\sup \limits_{t \in [0,T]}E\Big[ \|z_{p}^{k}(t)-z_{p}^{k-1}(t) \|^{2}\Big]\notag\\
\leqslant & \left(L+9\sigma_{1}\right)
\frac{L^{k-1}T^{k}L_{1}}{(k-1)!}.\notag
\end{align}
Noting that the term on the r.h.s. of the last inequality is independent of $p \in [0,1]$, we have
\begin{align}
&E\bigg[\sup_{t \in [0,T]} \big\|z_{p}^{k+1}(t)-z_{p}^{k}(t) \big\|^{2}\bigg]
\leqslant \left(L+9\sigma_{1}\right)
\frac{L^{k-1}T^{k}L_{1}}{(k-1)!}, \ \forall \ p \in [0,1].\label{beforecauchy}
\end{align}
By the above inequality and  Chebyshev inequality, we have
\begin{align}
&\sum_{k=2}^{\infty} P\bigg\{\sup_{t \in [0,T]}\big\|z_{p}^{k+1}(t)-z_{p}^{k}(t) \big\|>\frac{1}{k^2}\bigg\}\leqslant (L+9\sigma_{1} ) L_{1}\sum_{k=2}^{\infty} \frac{L^{k-1}T^{k}k^{4}}{(k-1)!},\ \forall \ p \in [0,1]. \label{beforecauchy000}
\end{align}
It can be proved that $\sum_{k=1}^{\infty} \frac{L^{k}T^{k+1}k^{4}}{k!}<\infty$. This together with  Borel-Cantelli Lemma shows that $\lim_{k \to \infty} \big\|z_{p}^{k+1} -z_{p}^{k}  \big\|_{*,T}=0$ a.s. Then, for any $p \in [0,1]$, \ $\left\{z_{p}^{k}, \ k \in \mathbb{N} \right\}$ is a Cauchy sequence in $\left(\mathcal{C}_{T}^{n},\ \|\cdot\|_{*,T}\right)$. As $\left(\mathcal{C}_{T}^{n}, \|\cdot\|_{*,T}\right)$ is complete,  for any $p \in [0,1]$, there exists  $z_{p}\in \mathcal{C}_{T}^{n}$, such that
$\lim_{k \to \infty}\|z_{p}^{k}-z_{p}\|_{*,T}=0$  a.s. By Theorem  4.5.1 in \cite{Kailaizhong} and (\ref{finiteboundesem}), we have
$ \sup_{p \in [0,1]} E\big[ \|z_{p}(t) \|^{2+\upsilon}\big] \\
\leqslant   \sup\limits_{p \in [0,1]} \liminf\limits_{k \to \infty}E\big[ \|z^{k}_{p}(t) \|^{2+\upsilon}\big]
\leqslant  \sup_{p \in [0,1],\ k \in \mathbb{N}}  E\big[ \|z^{k}_{p}(t) \|^{2+\upsilon}\big]\leqslant M_{1},\ \forall \ t \in [0,T].$
Similar to the proof of Theorem 5.2.1  in \cite{Boksendal} and by the above inequality, (\ref{beforecauchy}) and Assumption \ref{assumption1} (i)-(ii),  it can be proved that $\{z_{p},\ p\in [0,1]\}$ satisfies (\ref{integralform}).
Now, we show the uniqueness.
If $\left\{z_{p},\ p \in [0,1]\right\}$ and  $\left\{\widetilde{z}_{p},\ p \in [0,1]\right\}$ are the strong solutions to the system (\ref{integralform}) given $\mu \in \mathcal{M}$, then by $z_{p}(0)=\widetilde{z}_{p}(0)$, (\ref{integralform}) and (\ref{beforecauchy000}) and similar to the proof of (\ref{uniqueness})-(\ref{uniqthird}), we have $E\big[\|z_{p}(t)-\widetilde{z}_{p}(t)\|^{2}\big]=0,\ \forall \ t \in [0,T]$. Then, by the continuity of $z_{p}(\cdot)$ and $\widetilde{z}_{p}(\cdot)$ and similar to the proof of Theorem 5.2.1  in \cite{Boksendal}, we know that, if the strong solution to the system (\ref{integralform}) given $\mu \in \mathcal{M}$ exists, then it must be unique. Denote $\nu_{p}=\mathcal{L}(z_{p})$ and $\nu_{p}^{k}=\mathcal{L}(z_{p}^{k})\in \mathscr{P}(\mathcal{C}_{T}^{n})$. Now, we will prove that $\{\nu_{p},\ p \in [0,1]\} \in \mathcal{M}$.
 By $\mu \in \mathcal{M}$, we have $\sup_{p \in [0,1]}\int_{\mathcal{C}_{T}^{n}} \|\theta\|_{*,T}^{2} \mu_{p}(d \theta) <\infty$. Denote
 \begin{align}
 K_{2}=\sup_{p \in [0,1]}\int_{\mathcal{C}_{T}^{n}} \|\theta\|_{*,T}^{2} \mu_{p}(d \theta).\label{boundofmusecond}
 \end{align}
Similar to the proof of  (\ref{boudesan}) and by the above equality, Assumption \ref{assumption1} (i)-(ii), (\ref{integralform}), $C_{r}$ inequality and H{\"o}lder inequality, we have
\begin{align}
&E\bigg[\sup_{t \in [0,T]} \|z_{p}(t) \|^{2+\upsilon}\bigg]\notag\\
 \leqslant &  P_{2}T\bigg( \sup_{p \in [0,1]} \int_{\mathcal{C}_{T}^{n}} \|\theta\|_{*,T}^{2} \mu_{p}(d \theta)\bigg)^{\frac{2+\upsilon}{2}} +  P_{1}T \sup_{t \in [0,T]} E \big[\|z_{p}(t) \|^{2+\upsilon}\big]+ P_3 \notag\\
\leqslant   & P_{1}TM_{1}+P_{2}TK_{2}^{\frac{2+\upsilon}{2}}+P_3.\label{boundeofsolution}
\end{align}
 Noting that the r.h.s. of the above inequality is independent of $p$,  then we have $$\sup _{p \in [0,1]} \int_{\mathcal{C}_{T}^{n}}\|x\|_{*, T}^2 \nu_p(d x)<\infty.$$ Therefore, by Assumption \ref{assumption1} (i)-(ii), (\ref{defofWassersteindistanceofed}) and  (\ref{integralform}), it can be proved that for any $\epsilon>0$, there exists $\delta>0$, such that  $\sup_{  |t_{1}- t_{2}|<\delta,\ p \in [0,1] }W_{2}(\nu_{t_{1},p},\nu_{t_{2},p})<\epsilon$. By
(\ref{integralform}), (\ref{finiteboundesem}) and (\ref{boundeofsolution}), we have $\nu_{p}^{k},\ \nu_{p} \in  \mathscr{P}_{2}(\mathcal{C}_{T}^{n})$. Then, it's sufficient to  prove that the map $[0,1] \ni p \mapsto \nu_{p} \in \mathscr{P}_{2}(\mathcal{C}_{T}^{n})$ is measurable for $\Phi(\mu)=\nu \in \mathcal{M}.$
Noting that
$z_{p}^{k}$ converges to $z_{p}$ as $k \to \infty $ a.s. and by Theorem 7.1.5  in \cite{L. Ambrosio}, we have $\lim_{k \to \infty} W_{2,T}(\nu_{p}^{k}, \nu_{p})=0,\ \forall \  p \in [0,1]$. Then, by Theorem 4.2.2 in \cite{R.M.Dudley}, it's sufficient to prove the measurability of the maps $[0,1] \ni p \mapsto \nu_{p}^{k} \in \mathscr{P}_{2}(\mathcal{C}_{T}^{n}), \ k\geqslant 0 $ for the measurability of the map $[0,1] \ni p \mapsto \nu_{p}\in \mathscr{P}_{2}(\mathcal{C}_{T}^{n})$.
Noting that $(\mathcal{C}_{T}^{n},\|\cdot\|_{\infty})$ is a separable metric space and by Lemma 7.26 in \cite{Bertsekas}, we know that the map $[0,1]\ni  p \mapsto \nu_{p} \in \mathscr{P}(\mathcal{C}_{T}^{n})$ is measurable if and only if for any $B \in \mathscr{B}(\mathcal{C}_{T}^{n})$, the map $[0,1]\ni  p \mapsto \nu_{p}(B) \in [0,1]$ is measurable. Denote $\widetilde{\nu}_{p}^{k}=\mathcal{L}(z_{p}^k,\eta_{p},w_{p}) \in \mathscr{P}(\mathcal{C}_{T}^{3n})$.  Notice that if the map  $[0,1] \ni p \mapsto \widetilde{\nu}_{p}^{k}\in \mathscr{P}(\mathcal{C}_{T}^{3n})$ is measurable, then for any $B\in \mathscr{B}(\mathcal{C}_{T}^{n})$, the map  $[0,1] \ni p \mapsto \nu_{p}^{k}(B)=\widetilde{\nu}_{p}^{k}(B\times \mathcal{C}_{T}^{2n})\in \mathbb{R}$ is measurable.
Then, it's sufficient to prove the measurability of the map $[0,1] \ni p \mapsto \widetilde{\nu}_{p}^{k}\in \mathscr{P}(\mathcal{C}_{T}^{3n})$ for the  measurability of the map
$[0,1] \ni p \mapsto \nu_{p}^{k} \in \mathscr{P}(\mathcal{C}_{T}^{n})$. We will prove that the maps
$[0,1] \ni p \mapsto \widetilde{\nu}_{p}^{k}\in \mathscr{P}(\mathcal{C}_{T}^{3n}), \ k \in \mathbb{N}$ are measurable by induction. By Assumption \ref{assumption001}, the conclusion holds for $k=0$. Suppose that the conclusion holds for $k=\widetilde{k}$, $\widetilde{k}=0,1,2,\ldots$ and we  will prove that it holds for $k=\widetilde{k}+1$.
Similar to the proof of Proposition 2.1 in  \cite{ERHAN.BAYRAKTAR}, it's sufficient to prove that for all
$0 \leqslant t_{1} \leqslant \cdots \leqslant t_{m} \leqslant T$ and any bounded  continuous functions $f_{i}, \ g_{i}, \ h_{i}: \mathbb{R}^{n} \to \mathbb{R},\ i=1,   \ldots, m, $  the map $[0,1] \ni p \mapsto E\big[\prod_{i=1}^{m}f_{i} (z_{p}^{\widetilde{k}+1}(t_{i}) )h_{i}
 (\eta_{p} (t_{i}  ) )g_{i}(w_{p}(t_{i}))\big]\in \mathbb{R}$
is measurable. Denote $s_{l}=\left\lfloor sl\right\rfloor\frac{1}{l},$ where $ l $ is a positive integer. For any given $t \in [0,T]$ and $p \in [0,1]$, consider the following process
\begin{align}\label{discreteform}
z_{p}^{\widetilde{k}+1,l}(t)
=&z_{p}^{\widetilde{k}}(0)+\int_{0}^{t}G\big(s_{l},p,\eta_{p}
\left(s_{l}\right),z_{p}^{\widetilde{k}}\left(s_{l}\right)\big)ds\notag\\
&+\int_{0}^{t}\int_{[0,1]\times \mathbb{R}^{n}} A(p,q)F \big(s_{l},p, q,z,z_{p}^{\widetilde{k}}\left(s_{l}\right) \big)\mu_{s_{l},q}(dz)
dqds\notag\\
&+\int_{0}^{t}H\big(s_{l},p,\eta_{p}\left(s_{l}\right)
,z_{p}^{\widetilde{k}}\left(s_{l}\right)\big)dw_{p}(s).
\end{align}
Then $z_{p}^{\widetilde{k}+1,l}(t)=h(t,p,\eta_{p},z_{p}^{\widetilde{k}},w_{p})$, where
 \begin{align}
 & h(t,p,\eta_{p}, z_{p}^{\widetilde{k}},w_{p})\notag\\
 =& \sum_{j=0}^{ \lfloor tl \rfloor}\frac{1}{l}G\big(l_{j},p,\eta_{p}\big(l_{j}\big),
z_{p}^{\widetilde{k}}\big(l_{j}\big)\big)+z_{p}^{\widetilde{k}}(0) \notag\\
&+\frac{1}{l} \sum_{j=0}^{ \lfloor tl \rfloor}  \int_{[0,1]\times \mathbb{R}^{n}}A(p,q)F(l_{j}, p, q, z, z_{p}^{\widetilde{k}}(l_{j})) \mu_{l_{j},q}(dz)dq\notag\\
& +\sum_{j=0}^{\lfloor tl\rfloor }H\big(l_{j},p,\eta_{p}\big(l_{j}\big),
z_{p}^{\widetilde{k}}\big(l_{j}\big)\big)
\big(w_{p}\big(l_{j+1}\big) -w_{p}\big(l_{j}\big)\big)\notag\\
   &
   + \big(t -t_{l}\big)G\big (t_{l},p,\eta_{p} (t_{l} ), z_{p}^{\widetilde{k}}\big(t_{l}\big)\big)\notag\\
   & +\big(t-t_{l}\big) \int_{[0,1]\times \mathbb{R}^{n}}   A(p,q)  F\big(t_{l}, p, q, z, z_{p}^{\widetilde{k}}\big(t_{l}\big)\big)  \mu_{t_{l},q}(dz)dq\notag\\
    &+  H\big(t_{l},  p,\eta_{p}\big(t_{l}\big),
z_{p}^{\widetilde{k}}\big(t_{l}\big)\big)
\big(w_{p}(t) -w_{p}\big(t_{l}\big)\big) \notag
 \end{align}
 and $l_{j}=\frac{j}{l}$.
Then, by Assumption \ref{assumption1} (i)-(ii),  $h(t,p,x,z,y)$ is measurable w.r.t. $p$ and continuous w.r.t. $(x,z,y)$, that is, $h$ is measurable. By $$E\bigg[\prod_{i=1}^{m} f_{i}\big(z_{p}^{\widetilde{k}+1,l}(t_{i})\big)h_{i} (\eta_{p}
 (t_{i}  ) ) g_{i}(w_{p}(t_{i}))\bigg]\\ =
E\bigg[\prod_{i=1}^{m}  h_{i} (\eta_{p}
 (t_{i} ) )  f_{i}\big(h(t_{i}, p,\eta_{p},z_{p}^{\widetilde{k}},w_{p}) \big) g_{i} (w_{p}(t_{i}) ) \bigg],$$ the measurability of the map $[0,1] \ni p \mapsto \mathcal{L} (z_{p}^{\widetilde{k}},\eta_{p},w_{p} )$ and  (5.3.1) in \cite{L. Ambrosio}, we know that the map
\begin{align}
[0,1] \ni p \mapsto E\bigg[\prod_{i=1}^{m}f_{i}\big(z_{p}^{\widetilde{k}+1,l}(t_{i})\big)h_{i} (\eta_{p}
 (t_{i} ) )g_{i}(w_{p}(t_{i}))\bigg]\in \mathbb{R}\label{dicremeasu}
\end{align}
is measurable. Now, we will prove that $z_{p}^{\widetilde{k}+1,l}(t)$ converges to $z_{p}^{\widetilde{k}+1} (t)$ as $l \to \infty$ in probability, that is, for any $\epsilon, \ \epsilon_{0}>0$, we will prove that there exists $l_{1}>0$, such that if $l \geqslant l_{1}$, then $P\left\{\big\|z_{p}^{\widetilde{k}+1,l}(t)-z_{p}^{\widetilde{k}+1}(t)\big\|>\epsilon\right\}<\epsilon_{0}$.
By  Chebyshev inequality, we have $$P\left\{\big\|z_{p}^{\widetilde{k}+1,l}(t)-z_{p}^{\widetilde{k}+1}(t) \big \|>\epsilon\right\}
\leqslant  E\left[ \|z_{p}^{\widetilde{k}+1,l}(t)-z_{p}^{\widetilde{k}+1}(t) \|^{2}\right] \epsilon^{-2}.$$ Then, it's sufficient to prove  $E\big[ \|z_{p}^{\widetilde{k}+1,l}(t)-z_{p}^{\widetilde{k}+1}(t) \|^{2}\big]  <\epsilon_{0} \epsilon^{2} $. By
(\ref{iteexist}), (\ref{discreteform}) and $C_{r}$ inequality, we have
\begin{align}
 & E\big[\|z_{p}^{\widetilde{k}+1}(t)-z_{p}^{\widetilde{k}+1,l}(t)\|^{2}\big]\notag\\
\leqslant & 3E\Bigg[\bigg\|\int_{0}^{t}\big(G\big(s,p,\eta_{p}(s),z^{\widetilde{k}}_{p}(s)\big)
-G\big(s_{l},p,\eta_{p}(s_{l}),z_{p}^{\widetilde{k}}(s_{l})\big)
\big)ds\bigg\|^{2}\Bigg]\notag\\
&+3E\Bigg[\bigg\|\int_{0}^{t}\int_{[0,1]\times \mathbb{R}^{n}} A(p,q)\big(F \big(s,p,q,z,z^{\widetilde{k}}_{p}(s) \big)\mu_{s,q}(dz)\notag\\
&-F (s_{l},p, q,z,z^{\widetilde{k}}_{p} (s_{l} ) ) \mu_{s_{l},q}(dz)\big)dqds\bigg\|^{2}\Bigg] \notag\\
&
+3E\Bigg[\bigg\|\int_{0}^{t}\big(H(s,p,\eta_{p}(s),z^{\widetilde{k}}_{p}(s))
-H(s_{l},p,\eta_{p}\left(s_{l}\right),z_{p}^{\widetilde{k}}\left(s_{l}\right)
)\big)dw_{p}(s)
\bigg\|^{2}\Bigg]\notag\\
&=:3J_{1p\widetilde{k}}^l(t)+3J_{2p\widetilde{k}}^l(t)+3J_{3p\widetilde{k}}^l(t).\label{inexex}
\end{align}
Denote $M_{2}=M_{1}^{\frac{2}{2+\upsilon}}$. By Lyapunov inequality and (\ref{finiteboundesem}), we have $\sup_{k \in \mathbb{N},\ p \in [0,1],\ t \in [0,T]}E\big[\|z_{p}^{k}(t) \|^{2}\big]\\ \leqslant M_{2}$. Denote $l_{r}=\sigma_{2}(M_{2}+r^{\frac{2}{2+\upsilon_{1}}})+\sigma_{3}$. Let $O:=[0,T]\times[0,1]\times \mathbb{N}$. By Assumption \ref{assumption1} (i)-(ii), (\ref{finiteboundesem}) and (\ref{boundofmusecond}), we know that, for any $\epsilon >0,$ there exists
$\epsilon_{1}>0$, such that if $\|t_{1}-t_{2}\|<\epsilon_{1}$, then
\begin{align}
& \sup_{(s, p,\widetilde{k}) \in O} E \Big[\big\|G\big(t_{1},p,\eta_{p}(s_{l}),z_{p}^{\widetilde{k}}(s_{l})\big) -G\big(t_{2},p,\eta_{p}(s_{l}),z_{p}^{\widetilde{k}}(s_{l})\big) \big\|^{2}\Big]\notag\\
&\leqslant \frac{\epsilon^{2}\epsilon_{0} }{54T^{2}l_{r}} \sup_{{(s, p,\widetilde{k}) \in O}} \left(\sigma_{2} E\left[ \big\|z_{p}^{\widetilde{k}}\left(s_{l}\right) \big\|^{2}\right]+\sigma_{3}+\sigma_{2} E \left[\|\eta_{p}\left(s_{l}\right) \|^{2}\right]\right) \notag\\
&
\leqslant  \frac{\epsilon^{2}\epsilon_{0} }{54T^{2}},\label{errorofG}
\end{align}
\begin{align}
& \sup_{{(s, p,\widetilde{k}) \in O}}  E\bigg[
 \int_{[0,1]\times \mathbb{R}^{n}}  \big\|F \big(t_{1},p, q,z,z^{\widetilde{k}}_{p} (s_{l} ) \big)-F \big(t_{2},p, q,z,z^{\widetilde{k}}_{p} (s_{l} ) \big) \big\|^{2}\mu_{s_{l},q}(dz)dq\bigg]\notag\\
\leqslant & \gamma_0 \sup_{{(s, p,\widetilde{k}) \in O}} \left(\sigma_{5}E\Big[\big\|z^{\widetilde{k}}_{p} (s_{l} ) \big\|^{2}\Big]+\sigma_{5}\int_{[0,1]\times \mathbb{R}^{n}} \|z\|^{2}\mu_{s_{l},q}(dz)dq+\sigma_{6}\right)\notag\\
\leqslant&
\frac{\epsilon^{2}\epsilon_{0}}{81T^{2}}, \label{errorofF}
\end{align}
and
\begin{align}
 &\sup_{{(s, p,\widetilde{k}) \in O}} E\bigg[ \Big\|H \big(t_{1},p,\eta_{p}(s),z^{\widetilde{k}}_{p}(s) \big)-H \big(t_{2},p,\eta_{p} (s ),
z_{p}^{\widetilde{k}} (s ) \big)\Big\|^{2}\bigg] \notag\\
\leqslant& \frac{\epsilon_{0}\epsilon^{2}}{216Tl_{r}}\sup_{{(s, p,\widetilde{k}) \in O}} \left(\sigma_{2} E\Big[\big\|z_{p}^{\widetilde{k}}(s) \big\|^{2}\Big]+\sigma_{2} E\Big[\|\eta_{p} (s ) \|^{2}\Big] +\sigma_{3} \right)\notag\\
\leqslant&
\frac{\epsilon_{0}\epsilon^{2}}{216T},\label{errorofH}
\end{align}
where $\gamma_0=\frac{\epsilon^{2}\epsilon_{0}}{81T^{2}
 (\sigma_{5}M_{2}+\sigma_{5}K_{2}+\sigma_{6} )}$.
 By Assumption \ref{assumption1} (iii), there exists $\epsilon_{2}>0$, such that
\begin{align}
\sup_{p \in [0,1], \ |t_{1}-t_{2}|<\epsilon_{2} }E\big[\|\eta_{p}(t_{1})-\eta_{p}(t_{2})\|^{2}\big]< \min \Big\{
\frac{\epsilon^{2}\epsilon_{0}}{216\sigma_{1}T^{2}},\ \frac{\epsilon^{2}\epsilon_{0}}{216\sigma_{1}T} \Big\}. \label{etaptep}
\end{align}
  By $\mu\in \mathcal{M}$, there exists $\epsilon_{3}>0$ such that
  \begin{align}
  \sup_{p \in [0,1],\ |t_{1}-t_{2}|<\epsilon_{3} }W_{2}\big(\mu_{t_{1},p},\mu_{t_{2},p}\big) \leqslant  \frac{\epsilon \sqrt{\epsilon_{0}}}{9T\sigma_{4}\sqrt{n}}. \label{w2zptep}
  \end{align}
 By Assumption
\ref{assumption1} (i)-(ii), (\ref{iteexist}) and (\ref{finiteboundesem}), we know that there exists $\epsilon_{4}>0$, such that
\begin{align}
&\sup_{p \in [0,1],\ \widetilde{k}\in \mathbb{N}, \ |t_{1}-t_{2}|<\epsilon_{4} }E\Big[\big\|z_{p}^{\widetilde{k}}(t_{1})-z_{p}^{\widetilde{k}}\left(t_{2}\right)\big\|^{2}\Big] \leqslant    \min\bigg\{ \frac{\epsilon^{2}\epsilon_{0}}{81T\sigma_{4}^{2}},\
 \frac{\epsilon^{2}\epsilon_{0}}{216\sigma_{1}T^{2}},\
 \frac{\epsilon^{2}\epsilon_{0}}{216\sigma_{1}T}\bigg\}. \label{zaptep}
\end{align}
Denote $\epsilon_{5}=\min\left\{\epsilon_{1},\ \epsilon_{2},\ \epsilon_{3}, \ \epsilon_{4}\right\}$.
 By $\lim_{l\to\infty}\sup_{s\in [0,T]} |s_{l}-s|=0$, we know that there exists $l_{1}>0$, such that if $l \geqslant l_{1}$, then $\sup_{s\in [0,T]} \left\|s_{l}-s\right\|<\epsilon_{5}$.
  Therefore,  for the first term on the r.h.s. of
(\ref{inexex}), by Assumption \ref{assumption1} (i), $C_{r}$ inequality, H{\"o}lder inequality, (\ref{errorofG}), (\ref{etaptep}) and (\ref{zaptep}), we have
\begin{align}
&3J_{1p\widetilde{k}}^l(t)\notag\\
\leqslant &6T \int_{0}^{T} E\Big[\big\|G\big(s,p,\eta_{p}(s),z^{\widetilde{k}}_{p}(s)\big)
-G\big(s,p,\eta_{p}(s_{l}),z_{p}^{\widetilde{k}}(s_{l})\big)\big\|^{2}\Big]ds\notag\\
&+6T\int_{0}^{T} E\Big[ \big\|G \big(s,p,\eta_{p} (s_{l} ),z_{p}^{\widetilde{k}} (s_{l} ) \big)
-G \big(s_{l},p,\eta_{p} (s_{l} ),z_{p}^{\widetilde{k}} (s_{l} ) \big) \big\|^{2}\Big] ds\notag\\
\leqslant& \frac{\epsilon^{2}\epsilon_{0} }{9}+
6T\sigma_{1}\int_{0}^{T}\Big(E\big[ \|\eta_{p}(s)-
\eta_{p}\left(s_{l}\right) \|^{2}\big] +E\left[ \big\|z^{\widetilde{k}}_{p}(s)
-z_{p}^{\widetilde{k}}\left(s_{l}\right) \big\|^{2}\right]\Big)ds\notag\\
\leqslant & \frac{\epsilon^{2}\epsilon_{0} }{3}, \ \forall \ l \geqslant l_{1}.\label{firsrtepec}
\end{align}
 By Remarks 6.5-6.6 in \cite{Villani}, we have $$W_{2}(\mu, \nu) \geqslant \sup_{f:\ f\ \text{is\ 1-Lipschitz} }\bigg|\int_{\mathbb{R}^{n}}f(z) \mu(d z)-\int_{\mathbb{R}^{n}} f(z) \nu(d z)\bigg|,$$
where $ \mu, \ \nu \in \mathscr{P}(\mathbb{R}^{n})$. Then
for the second term on the r.h.s. of
(\ref{inexex}), by Assumption  \ref{assumption1} (ii), $C_{r}$ inequality, H{\"o}lder inequality, (\ref{defofWassersteindistance}),  (\ref{errorofF}) and (\ref{w2zptep})-(\ref{zaptep}), we have
\begin{align}
&3J_{2p\widetilde{k}}^l(t)\notag\\
\leqslant & 9E\Bigg[\bigg\|\int_{0}^{t}\int_{[0,1]\times \mathbb{R}^{n}} A(p,q)F\big(s,p, q,z,z^{\widetilde{k}}_{p}(s)\big)\big(\mu_{s,q}(dz)-\mu_{s_{l},q}(dz)\big)dqds\bigg\|^{2}\Bigg]
\notag\\
  &+9E\Bigg[\bigg\|\int_{0}^{t}
  \int_{[0,1]\times \mathbb{R}^{n}} A(p,q) \big(F(s,p,q,z,z^{\widetilde{k}}_{p}(s))-F(s,p, q,z,z^{\widetilde{k}}_{p}(s_{l})) \big)\mu_{s_{l},q}(dz)dqds\bigg\|^{2}\Bigg]\notag\\
 &+9E\Bigg[\bigg\|\int_{0}^{t}
 \int_{[0,1]\times \mathbb{R}^{n}} A(p,q)\big(F(s,p, q,z,z^{\widetilde{k}}_{p}\left(s_{l}\right))-F(s_{l},p, q,z,z^{\widetilde{k}}_{p}\left(s_{l}\right))\big)\mu_{s_{l},q}(dz)dqds
 \bigg\|^{2}\Bigg]\notag\\
\leqslant & 9Tn\sigma_{4}^{2}\int_{0}^{T}\int_{[0,1]}W^{2}_{2}
\left(\mu_{s,q},\mu_{s_{l},q}\right)dqds\notag\\
&+9T\int_{0}^{T}E\bigg[
 \int_{[0,1]\times \mathbb{R}^{n}}\big\|F(s,p, q,z,z^{\widetilde{k}}_{p}(s))-F(s,p, q,z,z^{\widetilde{k}}_{p}\left(s_{l}\right))\big\|^{2}
 \mu_{s_{l},q}(dz)dq\bigg]ds\notag\\
 &+9T\int_{0}^{T}E\bigg[
 \int_{[0,1]\times \mathbb{R}^{n}}\big\|F(s,p, q,z,z^{\widetilde{k}}_{p}\left(s_{l}\right))-F(s_{l},p, q,z,z^{\widetilde{k}}_{p}\left(s_{l}\right))\big\|^{2}\mu_{s_{l},q}(dz)dq\bigg]ds\notag\\
\leqslant & 9T\sigma_{4}^{2}\int_{0}^{T}E\Big[ \big\|z^{\widetilde{k}}_{p}(s)-z^{\widetilde{k}}_{p}
\left(s_{l}\right) \big\|^{2}\Big]ds +\frac{2\epsilon^{2}\epsilon_{0}}{9}\notag\\
\leqslant &   \frac{\epsilon^{2}\epsilon_{0} }{3}, \ \forall \ l \geqslant l_{1}.\label{secontepec}
\end{align}
 For the third term on the r.h.s. of
(\ref{inexex}), by Assumption \ref{assumption1} (i), Theorem 3.6 in \cite{Frideman},  $C_{r}$ inequality, (\ref{errorofH}), (\ref{etaptep}) and (\ref{zaptep}), we have
\begin{align}
3J_{3p\widetilde{k}}^l(t)
\leqslant &12\int_{0}^{T}E\Big[ \big\|H (s,p,\eta_{p}(s),z^{\widetilde{k}}_{p}(s) ) -H \big(s_{l},p,\eta_{p}\left(s_{l}\right),z_{p}^{\widetilde{k}}\left(s_{l}\right) \big)\big\|^{2}\Big]ds\notag\\
\leqslant & 24\int_{0}^{T}E\Big[ \big\|H\big(s,p,\eta_{p}(s),z^{\widetilde{k}}_{p}(s)\big)
-H\big(s_{l},p,\eta_{p}\left(s\right)
,z_{p}^{\widetilde{k}}\left(s\right)\big) \big\|^{2}\Big]ds \notag\\
&+24\int_{0}^{T}E\Big[ \big\|H\big(s_{l},p,\eta_{p}\left(s\right)
,z_{p}^{\widetilde{k}}\left(s\right)\big) -H\big(s_{l},p,
\eta_{p}\left(s_{l}\right),z_{p}^{\widetilde{k}}\left(s_{l}\right)\big) \big\|^{2}\Big]ds\notag\\
\leqslant &  \frac{\epsilon_{0}\epsilon^{2}}{9}+ 24\sigma_{1}\int_{0}^{T}\Big(E\big[ \|\eta_{p}\left(s\right)-\eta_{p}\left(s_{l}\right) \|^{2}\big]
+E\left[ \big\|z_{p}^{\widetilde{k}}\left(s\right)-z_{p}^{\widetilde{k}}\left(s_{l}\right) \big\|^{2}\right]\Big)ds\notag\\
\leqslant  & \frac{\epsilon^{2}\epsilon_{0}}{3}, \ \forall \ l \geqslant l_{1}.\notag
\end{align}
This together with (\ref{inexex}) and (\ref{firsrtepec})-(\ref{secontepec}) leads to $E\big[ \|z_{p}^{\widetilde{k}+1}(t)\notag -z_{p}^{\widetilde{k}+1,l}(t) \|^{2}\big]
<\epsilon^{2}\epsilon_{0}$. Therefore, for any $\epsilon,\ \epsilon_{0}>0$, there exists  $l_{1}>0$, such that if $l\geqslant l_{1}$, then
$P\big\{ \|z_{p}^{\widetilde{k}+1,l}(t)-z_{p}^{\widetilde{k}+1}(t) \|>\epsilon\big\}
\leqslant  E\big[ \|z_{p}^{\widetilde{k}+1,l}(t)-z_{p}^{\widetilde{k}+1}(t) \|^{2}\big]
 \epsilon^{-2} <\epsilon_{0}$,
 that is, $z_{p}^{\widetilde{k}+1,l}(t)$ converges to $z_{p}^{\widetilde{k}+1}(t)$ in probability. Then, noting that $f_{i}, \ g_{i}$ and $ h_{i}$  are bounded and continuous, we know that $$\lim_{l\to \infty}\Bigg|E \bigg[\prod_{i=1}^{m}f_{i}(z_{p}^{\widetilde{k}+1,l}(t_{i}))h_{i}
 \left(\eta_{p}
\left(t_{i}\right)\right)g_{i}\left(w_{p}(t_{i})\right) \bigg]-E\bigg[\prod_{i=1}^{m} f_{i}(z_{p}^{\widetilde{k}+1}(t_{i}))  h_{i}(\eta_{p}
(t_{i}))g_{i}(w_{p}(t_{i}))\bigg]\Bigg|=0.$$
This  together with Theorem 4.2.2 in \cite{R.M.Dudley}  and (\ref{dicremeasu}) gives that the map
$$
[0,1] \ni p \mapsto E\bigg[\prod_{i=1}^{m}f_{i} (z_{p}^{\widetilde{k}+1}(t_{i}) ) h_{i} (\eta_{p}
 (t_{i} ) )  g_{i} (w_{p}(t_{i}) )\bigg] \in \mathbb{R}$$ is measurable. Then, for any
$ k \in \mathbb{N}$,  the map $[0,1] \ni p \mapsto \widetilde{\nu}_{p}^{k} \in  \mathscr{P}(\mathcal{C}_{T}^{3n})$ is measurable. In conclusion, the map $\Phi$ is well-defined.

Second, we will prove the existence of the fixed point of the map $\Phi$ in $\mathcal{M}$. Let $z_{p}^{0}(t)=z_{p}(0)$, $\forall \ t \in [0,T]$,
$\widetilde{\mu}=\{\mathcal{L}\left(z_{p}^{0}\right), \ p \in [0,1]\}$ and $\Phi_{0}(\widetilde{\mu})=\widetilde{\mu}$. For any $k \in \mathbb{N}$, define the following iterative sequence
\begin{align}
 &z_{p}^{k+1}(t)\notag\\
=&z_{p}^{k+1}(0)+\int_{0}^{t}
\bigg[G\left(s,p,\eta_{p}(s),z^{k+1}_{p}(s)\right)\notag\\
&+\int_{[0,1]\times \mathbb{R}^{n}}A(p,q) F\left(s,p, q,z,z_{p}^{k+1}(s)\right)
\Phi_{k,q}(\widetilde{\mu})\circ \pi_{s}^{-1}(dz)dq\bigg]ds\notag\\
&+\int_{0}^{t}H\left(s,p,\eta_{p}(s),z^{k+1}_{p}(s)\right)dw_{p}(s),\label{contracmu}
\end{align}
where $\Phi_{k,q}(\widetilde{\mu})=\mathcal{L}(z_{q}^{k})$. Denote $\Phi_{k}(\widetilde{\mu})= \{\Phi_{k,p}(\widetilde{\mu}), \ p \in [0,1] \}$ and
  $L_{2}=3T\sigma_{1}+6T\sigma_{4}^{2}+12\sigma_{1}$.
Similar to the proof of (\ref{beforecauchy}) and by (\ref{contracmu}), Assumption \ref{assumption1} (i)-(ii), $C_{r}$ inequality, H{\"o}lder inequality and Theorem 3.6 in \cite{Frideman}, we have
\begin{align}
&E\Bigg[\bigg(\sup_{ s  \in [0,t]} \|z_{p}^{k+1}(s )-z_{p}^{k}(s ) \|\bigg)^{2}\Bigg]\notag\\
   \leqslant & (3T+12)\sigma_{1}\int_{0}^{t}E\big[ \|z_{p}^{k+1}(s )-z_{p}^{k}(s )
   \|^{2}\big]ds\notag\\
   &+6TE\Bigg[\int_{0}^{t} \int_{[0,1]}\bigg\| \int_{ \mathbb{R}^{n}}F\left(s,p, q,z,z^{k+1}_{p}(s)\right) (\Phi_{k,q}(\widetilde{\mu})\circ \pi_{s}^{-1}(dz)\notag\\ &-\Phi_{k-1,q}(\widetilde{\mu})\circ \pi_{s}^{-1}(dz) )\bigg\|^{2}dqds\Bigg]\notag\\
&+6T\int_{0}^{t}  \int_{[0,1]\times \mathbb{R}^{n}}E\Big[\big\|F (s,p, q,z,z^{k+1}_{p}(s) ) -F (s,p, q,z,z^{k}_{p}(s) )\big\|^{2}\Big]\Phi_{k-1,q}(\widetilde{\mu})\circ \pi_{s}^{-1}(dz)dqds\notag\\
  \leqslant & L_{2}\int_{0}^{t}E\Big[ \big\|z^{k+1}_{p}(s)
  -z^{k}_{p}(s) \big\|^{2}\Big]ds +6Tn\sigma_{4}^{2}\int_{0}^{t} \sup \limits_{p\in [0,1]}W_{2,s}^{2}(\Phi_{k,p}(\widetilde{\mu}),\Phi_{k-1,p}(\widetilde{\mu}))ds\notag\\
  \leqslant &  L_{2}
  \int_{0}^{t}E\left[\left(  \big\|z^{k+1}_{p} -z_{p}^{k}  \big\|_{*,s}\right)^{2}\right]ds +6Tn\sigma_{4}^{2}\int_{0}^{t} \sup \limits_{p\in [0,1]}W_{2,s}^{2}(\Phi_{k,p}(\widetilde{\mu}),\Phi_{k-1,p}(\widetilde{\mu}))ds.\label{conrfixed01}
\end{align}
This together with Gr{\"o}nwall's inequality gives
$$E\Big[ \big( \|z_{p}^{k+1} -  z_{p}^{k}  \|_{*,t} \big)^{2}\Big]
 \leqslant  6Tn\sigma_{4}^{2}e^{L_{2}T}
\int_{0}^{t}  \sup\limits _{p\in [0,1]} W_{2,s}^{2}(\Phi_{k,p}(\widetilde{\mu}), \Phi_{k-1,p}(\widetilde{\mu})) ds.$$
Then, by (\ref{defofWassersteindistance}),   we have
$$W_{2,\mathcal{M},T}^{2}  (\Phi_{k+1}(\widetilde{\mu}), \Phi_{k}(\widetilde{\mu}))
\leqslant   6T   n\sigma_{4}^{2} e^{L_{2}T}
\int_{0}^{T} W_{2,\mathcal{M},s}^{2} (\Phi_{k}(\widetilde{\mu}), \Phi_{k-1}
(\widetilde{\mu}))ds.$$
Then, by  Cauchy formula, we have
\begin{align}
&W_{2,\mathcal{M},T}^{2}(\Phi_{k+1}(\widetilde{\mu}),
\Phi_{k}(\widetilde{\mu}))
\leqslant \big(6Tn\sigma_{4}^{2}e^{L_{2}T}\big)^{k}\frac{T^{k}W_{2,\mathcal{M},T}^{2}\left(
\Phi_{1}(\widetilde{\mu}),\widetilde{\mu}\right)}{k!}.\label{nearcauch}
\end{align}
Combining Assumption \ref{assumption001}, (\ref{boundeofsolution}), $C_{r}$ inequality and $\widetilde{\mu},\ \Phi_{1}(\widetilde{\mu})   \in \mathcal{M}$ gives
\begin{align}
W_{2,\mathcal{M},T}^{2}(\Phi_{1}(\widetilde{\mu}),\widetilde{\mu})
\leqslant & \sup \limits_{p \in [0,1]}E\bigg[\sup_{t \in [0,T]} \big\|z_{p}^{1}(t)-z_{p}^{0}(t) \big\|^{2}\bigg]\notag\\
\leqslant & 2\sup_{p\in [0,1]}E\bigg[\sup_{t \in [0,T]} \|z_{p}^{1}(t) \|^{2}\bigg]
+2\sup_{p\in [0,1]}E\bigg[\sup_{t \in [0,T]}  \|z_{p}^{0}(t) \|^{2}\bigg] < \infty.\notag
\end{align}
Then, by
(\ref{nearcauch}), we know that $\{\Phi_{k}(\widetilde{\mu}), \ k \in \mathbb{N}\}$  is a Cauchy sequence in $\left[\mathscr{P}_{2}(\mathcal{C}_{T}^{n})\right]^{[0,1]}$. Nothing that the space
$\left(\mathscr{P}_{2}(\mathcal{C}_{T}^{n}), W_{2,T}\right)$ is complete,  there exists $\mu \in \left[\mathscr{P}_{2}(\mathcal{C}_{T}^{n})\right]^{[0,1]}$  such that $$\lim_{k\to \infty}W_{2,\mathcal{M},T}^{2}(\Phi_{k}(\widetilde{\mu}),\mu) =0$$  and $\sup _{p \in [0,1]} \int_{\mathcal{C}_{T}^{n}}\|x\|_{*, T}^2 \mu_p(d x)<\infty$.
 By $\Phi_{k}(\widetilde{\mu})\in \mathcal{M}$, we know that the map $[0,1] \ni p \mapsto \Phi_{k,p}(\widetilde{\mu})$ is measurable. Then, by Theorem 4.2.2 in \cite{R.M.Dudley}, we know that the map $[0,1] \ni p \mapsto \mu_{p}$ is measurable. This together with the triangle inequality of the 2-Wasserstein distance gives $\mu \in \mathcal{M}$.
By $W_{2,\mathcal{M},T}^{2}(\Phi(\mu),\mu)=\lim_{k \to \infty}W_{2,\mathcal{M},T}^{2}(\Phi_{k+1}(\widetilde{\mu}),\Phi_{k}(\widetilde{\mu}))=0$, we know that $\mu$ is the fixed point of the map $\Phi$ in $\mathcal{M}$.

At last, we  prove the uniqueness of the fixed point of the map $\Phi$ in $\mathcal{M}$.
%
 Suppose that $z_{p}^{\mu}(0)=z_{p}^{\nu}(0)=z_{p}(0)$, and $\{z_{p}^{\mu},\  \mu_{p}=\mathcal{L}(z_{p}^{\mu})\}$ and $\{z_{p}^{\nu}, \ \nu_{p}=\mathcal{L}(z_{p}^{\nu})\}$  are the solutions to (\ref{randmckvla}).
Then, similar to the proof of (\ref{conrfixed01})-(\ref{nearcauch}),
 we have $W_{2,\mathcal{M},T}^{2}( \mu, \nu )=W_{2,\mathcal{M},T}^{2}(\Phi(\mu),\Phi(\nu))=0,$
which  shows the uniqueness of the fixed point of the map $\Phi$ in $\mathcal{M}$.

Combining the above three steps, we know that there exists a unique solution   to the system (\ref{randmckvla}). $\hfill\blacksquare$

\setcounter{equation}{0}
\renewcommand{\theequation}{B.\arabic{equation}}
\setcounter{lemma}{0}
\renewcommand{\thelemma}{B.\arabic{lemma}}
\section{}

%

\emph{Proof of Lemma \ref{labelconti}:} The fact that the system (\ref{randmckvla}) with Brownian motions $\{w_{p},\ p\in [0,1]\}$ and $\{\eta_{p},\ p\in [0,1]\}$ is used to emphasise the independence of  $\{\widetilde{z}_{p},\ p\in [0,1]\}$, which is not relevant with this proof, so  we work with the following equivalent system with a single Brownian motion $\{(B(t),\ \mathcal{F}_{t}),\ t \geqslant 0\}$ here, that is,
\begin{align}\label{equivalentsystesigb}
d \widetilde{z}_{p}(t) =& \bigg[\int_{[0,1]}A(p,q)
\left(\int_{ \mathbb{R}^{n}}F(t,p, q,z,\widetilde{z}_{p}(t))\mu_{t,q}(dz)\right)dq +G\left(t,p,\widetilde{\eta}_{p}(t),\widetilde{z}_{p}(t)\right)\bigg]dt\notag\\ &+H(t,p,\widetilde{\eta}_{p}(t),\widetilde{z}_{p}(t))dB(t),\ \forall \ p \in [0,1],
\end{align}
where $\mathcal{L}(\widetilde{z}_{p}(0))=\mathcal{L}(z_{p}(0))$, $\widetilde{\eta}_{p}$ is a random element in   $\mathcal{C}_{T}^n$ and  $\mathcal{L}(\widetilde{\eta}_{p})=\mathcal{L}(\eta_{p})$.
Note that the distributions in the solution to the above system are identical to those in the solution to the system (\ref{randmckvla}). We denote the solution to the system (\ref{equivalentsystesigb}) as $ \{\widetilde{z}_{p}, \ \mu_{p}, \ p\in[0,1] \}$ and the solution also satisfies
\begin{align}\label{boundesectinmomenysofb}
 \sup_{p \in [0,1]}E\bigg[\sup_{t\in[0,T]}\|\widetilde{z}_{p}(t)\|^{2+\upsilon}\bigg]<\infty.
\end{align}
For any $p_{1},\ p_{2} \in [0,1]$, by (\ref{equivalentsystesigb}) and $C_{r}$ inequality, we have
\begin{align}\label{equivalentsystesigb1}
&E\left[\left\|\widetilde{z}_{p_{1}}-\widetilde{z}_{p_{2}} \right\|_{*,T}^{2}\right]\notag\\
\leqslant & 4 E\left[\left\|\widetilde{z}_{p_{1}}(0)-\widetilde{z}_{p_{2}}(0) \right\|^{2}\right]
+4E\Bigg[\sup_{t \in [0,T]}\bigg\|\int_{0}^{t} \bigg(\int_{[0,1]}A(p_{1},q)
\left(\int_{ \mathbb{R}^{n}}F(s,p_{1}, q,z,\widetilde{z}_{p_{1}}(s))\mu_{s,q}(dz)\right)dq\notag\\
&-\int_{[0,1]}A(p_{2},q)
\left(\int_{\mathbb{R}^{n}}F(s,p_{2}, q,z,\widetilde{z}_{p_{2}}(s))\mu_{s,q}(dz)\right)dq\bigg)ds\bigg\|^{2}\Bigg]\notag\\
&+4E\Bigg[\sup_{t \in [0,T]}\bigg\|\int_{0}^{t} \left(G\left(s,p_{1},\widetilde{\eta}_{p_{1}}(s),\widetilde{z}_{p_{1}}(s)\right)  -G\left(s,p_{2},\widetilde{\eta}_{p_{2}}(s),\widetilde{z}_{p_{2}}(s)\right ) \right)ds\bigg\|^{2}\Bigg]\notag\\
&
+4E\Bigg[\sup_{t \in [0,T]}\bigg\|\int_{0}^{t} \left(H\left(s,p_{1},\widetilde{\eta}_{p_{1}}(s),\widetilde{z}_{p_{1}}(s)\right)  -H\left(s,p_{2},\widetilde{\eta}_{p_{2}}(s),\widetilde{z}_{p_{2}}(s)\right ) \right)dB(s)\bigg\|^{2}\Bigg].
\end{align}
By   H{\"o}lder inequality and $C_{r}$ inequality, we have
\begin{align}\label{equivalentsystesigb3}
 & 4E\Bigg[\sup_{t \in [0,T]}\left\|\int_{0}^{t} \left(G\left(s,p_{1},\widetilde{\eta}_{p_{1}}(s),\widetilde{z}_{p_{1}}(s)\right)  -G\left(s,p_{2},\widetilde{\eta}_{p_{2}}(s),\widetilde{z}_{p_{2}}(s)\right ) \right)ds\right\|^{2}\Bigg]\notag\\
 \leqslant &4T\int_{0}^{T} E\left[\left\| G\left(s,p_{1},\widetilde{\eta}_{p_{1}}(s),\widetilde{z}_{p_{1}}(s)\right)  -G\left(s,p_{2},\widetilde{\eta}_{p_{2}}(s),\widetilde{z}_{p_{2}}(s)\right ) \right\|^{2}\right]ds\notag\\
  \leqslant &8T\int_{0}^{T} E\left[\left\| G\left(s,p_{1},\widetilde{\eta}_{p_{1}}(s),\widetilde{z}_{p_{1}}(s)\right)  -G\left(s,p_{2},\widetilde{\eta}_{p_{1}}(s),\widetilde{z}_{p_{1}}(s)\right)   \right\|^{2}\right]ds\notag\\
  & + 8T\int_{0}^{T} E\left[\left\| G\left(s,p_{2},\widetilde{\eta}_{p_{1}}(s),\widetilde{z}_{p_{1}}(s)\right)  -G\left(s,p_{2},\widetilde{\eta}_{p_{2}}(s),\widetilde{z}_{p_{2}}(s)\right ) \right\|^{2}\right]ds.
\end{align}
By Assumption \ref{assumption1} (i), (iii) and (\ref{boundesectinmomenysofb}), we know that for any $\epsilon>0$, there exists $\delta_{2}>0$, such that
 \begin{align}\label{equivalentsystesigb4}
  \sup_{|p_{1}-p_{2}|<\delta_{2}}  8T\int_{0}^{T} E\left[\left\| G\left(s,p_{1},\widetilde{\eta}_{p_{1}}(s),\widetilde{z}_{p_{1}}(s)\right)  -G\left(s,p_{2},\widetilde{\eta}_{p_{1}}(s),\widetilde{z}_{p_{1}}(s)\right)   \right\|^{2}\right]ds<\epsilon.
 \end{align}
  By Assumption \ref{assumption1} (i), we have
  \begin{align}\label{equivalentsystesigb5}
   & 8T\int_{0}^{T} E\left[\left\| G\left(s,p_{2},\widetilde{\eta}_{p_{1}}(s),\widetilde{z}_{p_{1}}(s)\right)  -G\left(s,p_{2},\widetilde{\eta}_{p_{2}}(s),\widetilde{z}_{p_{2}}(s)\right ) \right\|^{2}\right]ds\notag\\
   \leqslant &  8T^{2}\sigma_{1} \sup_{|p_{1}-p_{2}|<\delta_{3},\ t\in [0,T]} E\Big[\big\|  \widetilde{\eta}_{p_{1}}(t) - \widetilde{\eta}_{p_{2}}(t) \big\|^{2}\Big] + 8T\sigma_{1}\int_{0}^{T}E\Big[\big\| \widetilde{z}_{p_{1}}(s)  -\widetilde{z}_{p_{2}}(s)\big\|^{2}\Big]ds\notag\\
   \leqslant & 8T^{2}\sigma_{1} \sup_{|p_{1}-p_{2}|<\delta_{3},\ t\in [0,T]} E\Big[\big\|  \widetilde{\eta}_{p_{1}}(t) - \widetilde{\eta}_{p_{2}}(t) \big\|^{2}\Big]  +  8T\sigma_{1}\int_{0}^{T}\sup_{|p_{1}-p_{2}|<\delta_{3}} E\Big[\big\| \widetilde{z}_{p_{1}}  -\widetilde{z}_{p_{2}}\big\|_{*,t}^{2}\Big]dt.
  \end{align}
 By   Theorem 3.6 in \cite{Frideman} and $C_{r}$ inequality, we have
 \begin{align}\label{equivalentsystesigb6}
  &4E\Bigg[\sup_{t \in [0,T]}\bigg\|\int_{0}^{t} \big(H\left(s,p_{1},\widetilde{\eta}_{p_{1}}(s),\widetilde{z}_{p_{1}}(s)\right)  -H\left(s,p_{2},\widetilde{\eta}_{p_{2}}(s),\widetilde{z}_{p_{2}}(s)\right ) \big)dB(s)\bigg\|^{2}\Bigg]\notag\\
 \leqslant &  16\int_{0}^{T}E\left[ \left\|  H\left(s,p_{1},\widetilde{\eta}_{p_{1}}(s),\widetilde{z}_{p_{1}}(s)\right)  -H\left(s,p_{2},\widetilde{\eta}_{p_{2}}(s),\widetilde{z}_{p_{2}}(s)\right )  \right\|^{2}\right]ds\notag\\
 \leqslant &  32\int_{0}^{T}E\left[ \left\| H\left(s,p_{1},\widetilde{\eta}_{p_{1}}(s),\widetilde{z}_{p_{1}}(s)\right)  -H\left(s,p_{2},\widetilde{\eta}_{p_{1}}(s),\widetilde{z}_{p_{1}}(s)\right )   \right\|^{2}\right]ds\notag\\
 &+  32 \int_{0}^{T}E\left[ \left\|  H\left(s,p_{2},\widetilde{\eta}_{p_{1}}(s),\widetilde{z}_{p_{1}}(s)\right)  -H\left(s,p_{2},\widetilde{\eta}_{p_{2}}(s),\widetilde{z}_{p_{2}}(s)\right )   \right\|^{2}\right]ds.
 \end{align}
By Assumption \ref{assumption1} (i), (iii) and (\ref{boundesectinmomenysofb}), we know that for any $\epsilon>0$, there exists $\delta_{4}>0$, such that
\begin{align}\label{equivalentsystesigb7}
  32\sup_{|p_{1}-p_{2}|<\delta_{4}} \int_{0}^{T}E\left[ \left\| H\left(s,p_{1},\widetilde{\eta}_{p_{1}}(s),\widetilde{z}_{p_{1}}(s)\right)  -H\left(s,p_{2},\widetilde{\eta}_{p_{1}}(s),\widetilde{z}_{p_{1}}(s)\right )  \right\|^{2}\right]ds<\epsilon.
\end{align}
  By Assumption \ref{assumption1} (i), we have
  \begin{align}\label{equivalentsystesigb8}
   &32\int_{0}^{T}E\left[\left\|  H\left(s,p_{2},\widetilde{\eta}_{p_{1}}(s),\widetilde{z}_{p_{1}}(s)\right)  -H\left(s,p_{2},\widetilde{\eta}_{p_{2}}(s),\widetilde{z}_{p_{2}}(s)\right )   \right\|^{2}\right]ds\notag\\
   \leqslant & 32 \sigma_{1} T \sup_{|p_{1}-p_{2}|<\delta_{4},\ t\in [0,T]} E\Big[\big\|  \widetilde{\eta}_{p_{1}}(t) - \eta_{p_{2}}(t) \big\|^{2}\Big] \notag\\
   &+  32 \sigma_{1}\int_{0}^{T}\sup_{|p_{1}-p_{2}|<\delta_{4}}E\Big[\big\| \widetilde{z}_{p_{1}}(s)  -\widetilde{z}_{p_{2}}(s)\big\|^{2}\Big]ds\notag\\
   < & 32 \sigma_{1} T \sup_{|p_{1}-p_{2}|<\delta_{4},\ t\in [0,T]} E\Big[\big\|  \widetilde{\eta}_{p_{1}}(t) -  \widetilde{\eta}_{p_{2}}(t) \big\|^{2}\Big] \notag\\
   &+  32 \sigma_{1}\int_{0}^{T}\sup_{|p_{1}-p_{2}|<\delta_{4}}E\Big[\big\| \widetilde{z}_{p_{1}}  -\widetilde{z}_{p_{2}}\big\|_{*,t}^{2}\Big]dt.
  \end{align}
By H{\"o}lder inequality and $C_{r}$ inequality, we have
\begin{align}\label{equivalentsystesigb9}
 & 4E\Bigg[\sup_{t \in [0,T]}\bigg\|\int_{0}^{t} \bigg(\int_{[0,1]}A(p_{1},q)
\left(\int_{ \mathbb{R}^{n}}F(s,p_{1}, q,z,\widetilde{z}_{p_{1}}(s))\mu_{s,q}(dz)\right)dq\notag\\
&-\int_{[0,1]}A(p_{2},q)
\left(\int_{\mathbb{R}^{n}}F(s,p_{2}, q,z,\widetilde{z}_{p_{2}}(s))\mu_{s,q}(dz)\right)dq\bigg)ds\bigg\|^{2}\Bigg]\notag\\
\leqslant & 4T\int_{0}^{T}E\Bigg[ \bigg\|   \int_{[0,1]}A(p_{1},q)
\left(\int_{ \mathbb{R}^{n}}F(t,p_{1}, q,z,\widetilde{z}_{p_{1}}(t))\mu_{t,q}(dz)\right)dq\notag\\
&-\int_{[0,1]}A(p_{2},q)
\left(\int_{\mathbb{R}^{n}}F(t,p_{2}, q,z,\widetilde{z}_{p_{2}}(t))\mu_{t,q}(dz)\right)dq \bigg\|^{2}\Bigg]dt\notag\\
\leqslant & 12T \int_{0}^{T} E\left[\left\|  \int_{ \mathbb{R}^{n}\times[0,1]} (A(p_{1},q)-A(p_{2},q))
F(t,p_{1}, q,z,\widetilde{z}_{p_{1}}(t))\mu_{t,q}(dz)dq \right\|^{2}\right]dt\notag\\
&+ 12T \int_{0}^{T}E\left[\left\|  \int_{ \mathbb{R}^{n}\times[0,1]} A(p_{2},q)
\left(F(t,p_{1}, q,z,\widetilde{z}_{p_{1}}(t))-F(t,p_{1}, q,z,\widetilde{z}_{p_{2}}(t))\right)\mu_{t,q}(dz)dq \right\|^{2}\right]dt\notag\\
&+12T  \int_{0}^{T}E\left[\left\|  \int_{ \mathbb{R}^{n}\times[0,1]} A(p_{2},q)
\left(F(t,p_{1}, q,z,\widetilde{z}_{p_{2}}(t))-F(t,p_{2}, q,z,\widetilde{z}_{p_{2}}(t))\right)\mu_{t,q}(dz)dq \right\|^{2}\right]dt.
\end{align}
By Assumption \ref{assumption01}, we know that for any $\epsilon>0$, there exists $\delta_{5}>0$, such that
\begin{align}
  \sup_{|p_{1}-p_{2}|<\delta_{5},\ q\in [0,1]} |A(p_{1},q)-A(p_{2},q)|^{2}<\frac{\epsilon}{36T^{2}C_{2}^{2}\left(1+2\sup_{p \in[0,1]}E\left[\left\|\widetilde{z}_{p}\right\|_{*,T}^{2}\right]\right)}. \notag
\end{align}
Then, for the $\epsilon$ and $\delta_{5}$ given by the above inequality, by (\ref{boundesectinmomenysofb}), Assumption \ref{assumption1} (ii), H{\"o}lder inequality and $C_{r}$ inequality, we know that if $|p_{1}-p_{2}|<\delta_{5}$, then
\begin{align}\label{equivalentsystesigb10}
&12T \int_{0}^{T} E\left[\left\|  \int_{ \mathbb{R}^{n}\times[0,1]} (A(p_{1},q)-A(p_{2},q))
F(t,p_{1}, q,z,\widetilde{z}_{p_{1}}(t))\mu_{t,q}(dz)dq \right\|^{2}\right]dt\notag  \\
\leqslant &   \frac{\epsilon}{3T C_{2}^{2}\left(1+2\sup_{p \in[0,1]}E\left[\left\|\widetilde{z}_{p}\right\|_{*,T}^{2}\right]\right)} \int_{0}^{T} E\left[\int_{ \mathbb{R}^{n}\times[0,1]}  \left\|
F(t,p_{1}, q,z,\widetilde{z}_{p_{1}}(t)) \right\|^{2}\mu_{t,q}(dz)dq\right]dt\notag  \\
\leqslant &  \frac{\epsilon}{ T  \left(1+2\sup_{p \in[0,1]}E\left[\left\|\widetilde{z}_{p}\right\|_{*,T}^{2}\right]\right)}
 \int_{0}^{T}  \Bigg(1+E\left[  \|\widetilde{z}_{p}(t)
 \|^{2 }\right] +\bigg(\sup_{q \in {[0,1]}}
\int_{\mathcal{C}_{T}^{n}}\|z\|_{*,T}^{2}\mu_{q}(dz)\bigg)
\Bigg)dt\notag\\
<& \epsilon.
\end{align}
By  Assumption \ref{assumption1} (ii) and H{\"o}lder inequality, we have
\begin{align}\label{equivalentsystesigb11}
& 12T \int_{0}^{T}E\left[\left\|  \int_{ \mathbb{R}^{n}\times[0,1]} A(p_{2},q)
\left(F(t,p_{1}, q,z,\widetilde{z}_{p_{1}}(t))-F(t,p_{1}, q,z,\widetilde{z}_{p_{2}}(t))\right)\mu_{t,q}(dz)dq \right\|^{2}\right]dt\notag\\
\leqslant & 12T \int_{0}^{T}E\left[\int_{ \mathbb{R}^{n}\times[0,1]}\left\|
 F(t,p_{1}, q,z,\widetilde{z}_{p_{1}}(t))-F(t,p_{1}, q,z,\widetilde{z}_{p_{2}}(t))  \right\|^{2}\mu_{t,q}(dz)dq\right]dt\notag\\
\leqslant & 12T \sigma_{4}^{2} \int_{0}^{T} E\left[ \left\|
 \widetilde{z}_{p_{1}}(t) - \widetilde{z}_{p_{2}}(t)  \right\|^{2} \right]dt\notag\\
 \leqslant & 12T \sigma_{4}^{2} \int_{0}^{T}E\left[ \left\|
 \widetilde{z}_{p_{1}}  - \widetilde{z}_{p_{2}}  \right\|_{*,t}^{2} \right]dt.
\end{align}
By  Assumption \ref{assumption1} (ii), we know that for any $\epsilon>0,$ there exists $\delta_{6}>0$ such that
\begin{align}
& \sup\limits_{|p_{1}-p_{2}|<\delta_{6},\ q\in [0,1], \ t \in [0,T],\ z \in \mathbb{R}^{n}} E\big[ \|F(t,p_{1}, q,z,\widetilde{z}_{p_{2}}(t))-F(t,p_{2}, q,z,\widetilde{z}_{p_{2}}(t))\|^2\big]\notag\\
 <& \frac{\epsilon}{12 T^2 \big( 2\sigma_{5} \sup_{p\in [0,1]} E\big[\|\widetilde{z}_{p}\|^2_{*,T}\big]+\sigma_{6}\big)}.\notag
\end{align}
Then, by  H{\"o}lder inequality, we have
\begin{align}\label{equivalentsystesigb12}
& 12T\sup_{|p_{1}-p_{2}|<\delta_{6}}\int_{0}^{T} E\Bigg[\bigg\|  \int_{ \mathbb{R}^{n}\times[0,1]} A(p_{2},q)
\big(F(t,p_{1}, q,z,\widetilde{z}_{p_{2}}(t))\notag\\
& -F(t,p_{2}, q,z,\widetilde{z}_{p_{2}}(t))\big)\mu_{t,q}(dz)dq \bigg\|^{2}\Bigg]dt
<   \epsilon.
\end{align}
By Assumption \ref{assumption001}, we know that for any $\epsilon>0$, there exists $\delta_{1}>0$, such that
\begin{align}\label{equivalentsystesigb2}
  4\sup_{|p_{1}-p_{2}|<\delta_{1}} \left( W_{2}(\mathcal{L}(\widetilde{z}_{p_{1}}(0)),\mathcal{L}(\widetilde{z}_{p_{2}}(0)))\right)^{2}  <\epsilon.
\end{align}
 By Assumption \ref{assumption1} (iii), we know that for any $\epsilon>0$, there exists $\delta_{3}>0$, such that
\begin{align}\label{equivalentsystesigb41}
 \sup_{|p_{1}-p_{2}|<\delta_{3}} \left( W_{2,T}\left(\mathcal{L}(\widetilde{\eta}_{p_{1}}), \mathcal{L}(\widetilde{\eta}_{p_{2}})\right)\right)^{2}  <\min\left\{\frac{\epsilon}{8T^{2}\sigma_{1}},\ \frac{\epsilon}{32T\sigma_{1}}\right\}.
\end{align}
By (\ref{equivalentsystesigb1})-(\ref{equivalentsystesigb12}), we know that, for any $\epsilon>0$, there exists $\delta=\min\{\delta_{i}, \ i=1,\ldots,6\}$, such that if $|p_{1}-p_{2}|<\delta$, then
\begin{align}
  &\sup_{|p_{1}-p_{2}|<\delta}E\left[\left\|\widetilde{z}_{p_{1}}-\widetilde{z}_{p_{2}} \right\|_{*,T}^{2}\right] \notag\\
  \leqslant & \left( 8T\sigma_{1}+32 \sigma_{1}+  12T \sigma_{4}^{2}\right)\int_{0}^{T}\sup_{|p_{1}-p_{2}|<\delta}E\left[ \left\|
 \widetilde{z}_{p_{1}}  - \widetilde{z}_{p_{2}}  \right\|_{*,t}^{2} \right]dt+ 4\epsilon\notag\\
 & + 4 \sup_{|p_{1}-p_{2}|<\delta} E\left[\left\|\widetilde{z}_{p_{1}}(0)-\widetilde{z}_{p_{2}}(0) \right\|^{2}\right] + \left(8T^{2}\sigma_{1}+ 32 \sigma_{1} T \right) \sup_{|p_{1}-p_{2}|<\delta} E\Big[\big\|  \widetilde{\eta}_{p_{1}}  - \widetilde{\eta}_{p_{2}}\big\|_{*,T}^{2}\Big].\notag
\end{align}
This together with  Gr{\"o}nwall's inequality leads to
\begin{align}
 & \sup_{|p_{1}-p_{2}|<\delta} E\left[\left\|\widetilde{z}_{p_{1}}-\widetilde{z}_{p_{2}} \right\|_{*,T}^{2}\right]\notag\\
  \leqslant& \bigg(4\epsilon +4\sup_{|p_{1}-p_{2}|<\delta} E\left[\left\|\widetilde{z}_{p_{1}}(0)-\widetilde{z}_{p_{2}}(0) \right\|^{2}\right]\notag\\
   &+ \left(8T^{2}\sigma_{1}+ 32 \sigma_{1} T \right)\sup_{|p_{1}-p_{2}|<\delta} E\Big[\big\|  \widetilde{\eta}_{p_{1}}  - \widetilde{\eta}_{p_{2}}\big\|_{*,T}^{2}\Big] \bigg) e^{  8T^{2}\sigma_{1}+32 \sigma_{1}T+  12T^{2} \sigma_{4}^{2}}. \notag
\end{align}
Then, by (\ref{defofWassersteindistanceofed}),   (\ref{defofWassersteindistance}), (\ref{equivalentsystesigb2}) and (\ref{equivalentsystesigb41}), we have
\begin{align}
 & W_ {2,T}^2\left(\mu_{p_1},\mu_{p_2}\right)\notag\\
  \leqslant& \bigg(4\epsilon +4\sup_{|p_{1}-p_{2}|<\delta} \left( W_{2}(\mathcal{L}(\widetilde{z}_{p_{1}}(0)),\mathcal{L}(\widetilde{z}_{p_{2}}(0)))\right)^{2}\notag\\
 &+ \left(8T^{2}\sigma_{1}+ 32 \sigma_{1} T \right)\sup_{|p_{1}-p_{2}|<\delta } \left( W_{2,T}\left(\mathcal{L}(\widetilde{\eta}_{p_{1}}), \mathcal{L}(\widetilde{\eta}_{p_{2}})\right)\right)^{2}  \bigg) e^{  8T^{2}\sigma_{1}+32 \sigma_{1}T+  12T^{2} \sigma_{4}^{2}}\notag\\
 \leqslant & 7\epsilon e^{  8T^{2}\sigma_{1}+32 \sigma_{1}T+  12T^{2} \sigma_{4}^{2}}, \notag
\end{align}
that is, $ \{\mu_{p},\ p\in[0,1] \}$ is uniformly continuous  w.r.t. $p$. This together with $$W_ {2}^2\left(\mu_{p_1,t},\mu_{p_2,t}\right) \leqslant W_ {2,T}^2\left(\mu_{p_1},\mu_{p_2}\right)$$ gives the desired assertions.
$\hfill\blacksquare$

\vskip 2mm
\emph{Proof of Lemma \ref{difflisandao}:}
At first, we prove (\ref{NTOINFZ}).
 For any $p \in \big(\frac{i-1}{N},\frac{i}{N}\big]$, Denote $$\Psi_{i,p}(T)=E\bigg[ \sup_{t \in [0,T]} \big\|z_{i}^{N}(t)-z_{p}(t) \big\|^{2} \bigg].$$ By (\ref{randmckvla}), (\ref{finiteparticles}), $C_{r}$ inequality,  H{\"o}lder inequality and Theorem 3.6 in \cite{Frideman}, we have
 \begin{align}
 &\Psi_{i,p}( T)\notag\\
 \leqslant & \epsilon_{1}(N,p)+4T
 \int_{0}^{T}E\bigg[\Big\|G\Big(s,\frac{i}{N},\eta_{\frac{i}{N}}(s),
z_{i}^{N}(s)\Big)-G (s,p,\eta_{p}(s),z_{p}(s) )\Big\|^{2}\bigg]ds\notag\\
 &+4T\int_{0}^{T}E\Bigg[\bigg\|\frac{1}{N}\sum_{j=1}^{N}
\bigg(A^{N}\left(\frac{i}{N},\frac{j}{N}\right) F\left(s,\frac{i}{N}, \frac{j}{N},z_{j}^{N}(s),z_{i}^{N}(s)\right)\bigg)\notag\\
&
-\int_{[0,1]\times \mathbb{R}^{n}}A(p,q)
F(s,p, q,z,z_{p}(s))\mu_{s,q}(dz)dq\bigg\|^{2}\Bigg]ds\notag\\
&+16E\int_{0}^{T}\bigg[\Big\|H\Big(s,\frac{i}{N},\eta_{\frac{i}{N}}(s),
z_{i}^{N}(s)\Big)  -H (s,p,\eta_{p}(s),z_{p}(s) )\Big\|^{2}\bigg]ds\notag\\
=:&l_{1}(T)+l_{2}(T)+l_{3}(T)+l_{4}(T),\label{differencetwoparticls}
 \end{align}
where $\epsilon_{1}(N,p)=4  E
\big[\|z_{\frac{i}{N}}(0)-z_{p}(0) \|^{2}\big]$ and $w_p$ is independent of $p$.
For the second term on the r.h.s. of (\ref{differencetwoparticls}), by Assumption \ref{assumption1} (i)  and $C_{r}$ inequality, we have
\begin{align}
& l_{2}(T)\notag\\
\leqslant & 8T \int_{0}^{T}E\Bigg[\bigg\|G\bigg(s,\frac{i}{N},\eta_{\frac{i}{N}}(s),
z_{i}^{N}(s)\bigg)
-G\big(s,p,\eta_{\frac{i}{N}}(s), z_{i}^{N}(s)\big)\bigg\|^{2}\Bigg]ds\notag\\
& +8T \int_{0}^{T}E\Big[\big\|G\big(s,p,\eta_{\frac{i}{N}}(s),
z_{i}^{N}(s)\big)-G\big(s,p,\eta_{p}(s),z_{p}(s)\big)\big\|^{2}\Big]ds\notag\\
\leqslant & 8T\bigg(\epsilon_{2}(T,N,p)+ \sigma_{1} \epsilon_{3}(T,N,p) + \sigma_{1}\int_{0}^{T}E\left[ \|
z_{i}^{N}(s)-z_{p}(s) \|^{2}\right]ds\bigg)\notag\\
\leqslant & 8T\bigg(  \epsilon_{2}(T,N,p) +  \sigma_{1} \epsilon_{3}(T,N,p) + \sigma_{1} \int_{0}^{T} \Psi_{i,p}( t)dt\bigg),\label{differencetwoparticls2}
\end{align}
where  $\epsilon_{3}(T,N,p)=\int_{0}^{T}    E\big[ \|\eta_{\frac{i}{N}}(s) - \eta_{p}(s) \|^{2} \big] ds $ and $$\epsilon_{2}(T, N,p)  = \int_{0}^{T}E\bigg[ \Big\|G\big(s,p,\eta_{\frac{i}{N}}(s),
z_{i}^{N}(s)\big) -G\big(s, \frac{i}{N},
 \eta_{\frac{i}{N}}(s),
z_{i}^{N}(s)\big) \Big\|^{2}\bigg]ds.$$
  For the fourth term on the r.h.s. of (\ref{differencetwoparticls}), similar to the proof of the above inequality and  by Assumption \ref{assumption1} (i)  and $C_{r}$ inequality, we have
\begin{align}
  l_{4}(T)
\leqslant 32 \bigg( \epsilon_{4}(T,N,p)+ \sigma_{1} \epsilon_{3}(T,N,p) + \sigma_{1} \int_{0}^{T}  \Psi_{i,p}( t)dt\bigg), \label{differencetwoparticls3}
\end{align}
where
 $\epsilon_{4}(T,N,p)=\int_{0}^{T}E\big[ \|H (s,\frac{i}{N},\eta_{\frac{i}{N}}(s),
z_{i}^{N}(s) )-H (s, p,\eta_{\frac{i}{N}}(s),
z_{i}^{N}(s) ) \|^{2}\big]ds$.
 For the third term on the r.h.s. of (\ref{differencetwoparticls}), by $C_{r}$ inequality, we have
\begin{align}
 l_{3}(T)\leqslant l_{5}(T)+l_{6}(T)+l_{7}(T),\label{differencetwoparticls4}
\end{align}
where
\begin{align*}
  l_{5}(T)=& 12T\int_{0}^{T}E\Bigg[ \bigg\|\sum _{j=1}^{N}
\int_{ \big(\frac{j-1}{N},\frac{j}{N}\big]}A^{N}\big(p,q\big)
 \Big(F \Big(s, \frac{i}{N},\frac{j}{N},z_{j}^{N}(s),z_{i}^{N}(s) \Big)\notag\\
 &-F \Big(s,p, q,z_{j}^{N}(s), z_{i}^{N}(s) \Big) \Big) dq \bigg\|^{2}\Bigg]ds,\notag\\
 l_{6}(T)=& 12T \int_{0}^{T}E\Bigg[ \bigg\|\sum_{j=1}^{N}
\int_{ \mathbb{R}^{n}\times\big(\frac{j-1}{N},\frac{j}{N}\big]}A^{N}\big(p,q\big)
 \Big(F \Big(s,p,q, z_{j}^{N}(s),z_{i}^{N}(s) \Big)\notag\\ & -F (s,p,  q,z,z_{p}(s)) \Big) \mu_{s,q}(dz)dq \bigg\|^{2}\Bigg]ds,\notag\\
 l_{7}(T)=&12T \int_{0}^{T}E\Bigg[ \bigg\|
\int_{ \mathbb{R}^{n}\times(0,1]}\left(A^{N}(p,q)-A(p,q)
\right) F(s,p, q,  z, z_{p}(s))\mu_{s,q}(dz) dq \bigg\|^{2}\Bigg]ds.
\end{align*}
%
 For the first term on the r.h.s. of the above inequality, by H{\"o}lder inequality, we have
\begin{align}
l_{5}(T)
\leqslant   12T\epsilon_{5}(T,N,p),\label{differencetwoparticls41}
\end{align}
where $\epsilon_{5}(T,N,p)=\int_{0}^{T}\sum\limits_{j=1}^{N} \int_{ \big(\frac{j-1}{N},\frac{j}{N}\big]}E\big[\|
F(s,\frac{i}{N}, \frac{j}{N},
z_{j}^{N}(s),  z_{i}^{N}(s))-F(s,p, q,z_{j}^{N}(s),z_{i}^{N}(s))\|^{2}\big]dqds$.
 For the second term on the r.h.s. of (\ref{differencetwoparticls4}), by H{\"o}lder inequality and Assumption \ref{assumption1} (ii), we have
\begin{align}
 l_{6}(T)
\leqslant & 12T\int_{0}^{T}\sum\limits_{j=1}^{N}\int_{\big(\frac{j-1}{N},\frac{j}{N}\big]}
E\bigg[\int_{ \mathbb{R}^{n}}\big\|F\big(s,p,q,z_{j}^{N}(s),z_{i}^{N}(s)\big)\notag\\
& -F(s,p, q,z,z_{p}(s))\big\|^{2}\mu_{s,q}(dz)\bigg]dqds\notag\\
\leqslant & 24T\sigma_{4}^{2}\bigg( \int_{0}^{T} \Psi_{i,p}(t)dt+   \int_{0}^{T} \sum_{j=1}^{N} \int_{\big(\frac{j-1}{N},\frac{j}{N}\big]} \Psi_{j,q}(t)dqdt\bigg).\label{differencetwoparticls42}
\end{align}
Fix $M \in (0, \infty)$ and define $F_{M}(s,p,q,x,y)=F(s,p,q,x,y)\mathbb{I}_{\{\|x\|\leqslant M,\|y\|\leqslant M \}}$, where $\mathbb{I}_{\{\|x\|\leqslant M,\|y\|\leqslant M \}}$ equals $1$ if  $\|x\|\leqslant M$ and $\|y\|\leqslant M$, and $0$ otherwise. Note that by Assumption \ref{assumption1} (ii),  $F_{M}$ is Lipschitz continuous w.r.t. $(x,y)$ and the Lipschitz constant is uniform for all $t$, $p$ and $q$. Then, by Corollary 2 in \cite{M. H. Schultz},  there exist  $m=m(M)$ and $\widetilde{F}_{M}(s,p,q,x,y)=\sum_{k=1}^{m}F_{1}(s,p,q,k,m)a_{k}(x)  c_{k}(y)\\
\mathbb{I}_{\{\|x\|\leqslant M,\|y\|\leqslant M \}}$ such that, for any $\ x,\ y \in \mathbb{R}^{n},\ s \in [0,T], \ p,\ q \in [0,1],$
\begin{equation}
 \|\widetilde{F}_{M}(s,p,q,x,y)-F_{M}(s,p,q,x,y) \| \leqslant \frac{1}{M}, \label{FMANDF}
\end{equation}
where $a_{k}$ and $c_{k}$ are the polynomials of $x$ and $y$, respectively, and $F_{1}(s,p,q,k,m)$ is a function of $s,\ p,\ q,\ k,\ m$.
By $C_{r}$ inequality, we have
\begin{align}
&l_{7}(T)\notag\\
\leqslant & 36T\int_{0}^{T}E\Bigg[\bigg\|
\int_{ \mathbb{R}^{n}\times (0,1 ]}\left(A^{N}(p,q)-A(p,q)
\right)
\big(F(s,p,q,z,z_{p}(s)) -F_{M}(s,p, q,z,z_{p}(s))\big)\notag\\
 &\mu_{s,q}(dz)dq\bigg\|^{2}\Bigg]ds +36T\int_{0}^{T}E\Bigg[\bigg\|
\int_{ \mathbb{R}^{n}\times (0,1 ]}\big(A^{N}(p,q) -A(p,q)
\big) \big(F_{M}(s,p,
 q,z,z_{p}(s))\notag\\
& -\widetilde{F}_{M}(s,p, q,z,z_{p}(s))\big)\mu_{s,q}(dz)dq\bigg\|^{2}\Bigg]ds\notag\\
&+36T\int_{0}^{T}E\Bigg[\bigg\|
\int_{ \mathbb{R}^{n}\times (0,1]}\left(A^{N}(p,q)-A(p,q)
\right)
\widetilde{F}_{M}(s,p, q,z,z_{p}(s))\mu_{s,q}(dz)dq\bigg\|^{2}\Bigg]ds\notag\\
=&:l_{8}(T)+l_{9}(T)+ l_{10}(T).\label{differencetwoparticls441}
\end{align}
For the first term on the r.h.s. of (\ref{differencetwoparticls441}), by Assumption \ref{assumption1} (ii), $C_{r}$ inequality,   Lyapunov inequality and  Chebyshev inequality, we have
\begin{align}
& l_{8}(T) \notag\\
\leqslant & 36T\int_{0}^{T}E\Bigg[\bigg(
\int_{ \mathbb{R}^{n}\times (0,1]}\|F(s,p, q,z,z_{p}(s)) -F_{M}(s,p, q,z,z_{p}(s))\|\mu_{s,q}(dz)dq\bigg)^{2}\Bigg]ds\notag\\
\leqslant & 36TC_{2}^{2}\int_{0}^{T}E\Bigg[\bigg\|\int_{ \mathbb{R}^{n}\times (0,1]}\left(1+\|z\|+\|z_{p}(s)\|\right)
 \big(\mathbb{I}_{\{\|z\|>M\}}
+\mathbb{I}_{\{\|z_{p}(s)\|>M\}}\big)
\mu_{s,q}(dz)dq\bigg\|^{2}\Bigg]ds\notag\\
\leqslant &36TC_{2}^{2}\int_{0}^{T}E\Bigg[\bigg\|\int_{(0,1]}
E\left[\mathbb{I}_{\{\|z_{q}(s)\|>M\}}\right]dq
+\mathbb{I}_{\{\|z_{p}(s)\|>M\}}
+\int_{(0,1]}E\left[\|z_{q}(s)\|\mathbb{I}_{\{\|z_{q}(s)\|>M\}}
\right]dq\notag\\
&+\mathbb{I}_{\{\|z_{p}(s)\|>M\}}\int_{(0,1]}
E\left[\|z_{q}(s)\|\right]dq +\|z_{p}(s)\|   \bigg(\int_{(0,1]}E\left[\mathbb{I}_{\{\|z_{q}(s)\|>M\}}\right]dq+ \mathbb{I}_{\{\|z_{p}(s)\|>M\}} \bigg)\bigg\|^{2}\Bigg]ds\notag\\
\leqslant & 432TC_{2}^{2}\bigg(\sup_{p\in(0,1]}\int_{0}^{T}E
\big[\mathbb{I}_{\{\|z_{p}(s)\|>M\}}\big]ds+
\sup_{p\in(0,1]}\int_{0}^{T}E\big[ \|z_{p}(s) \|^{2}
\mathbb{I}_{\{\|z_{p}(s)\|>M\}}
\big]ds  \notag\\
& + \int_{0}^{T} \sup_{p\in(0,1]} E\big[\mathbb{I}_{\{\|z_{p}(s)\|>M\}}\big]
 \sup_{p\in(0,1]}E\big[ \|z_{p}(s) \|^{2}
\big]ds\bigg)\notag\\
\leqslant & 432TC_{2}^{2}\bigg(\frac{1}{M^{2}}\sup_{p\in(0,1]}\int_{0}^{T}E\big[ \|z_{p}(s) \|^{2}
\big]ds +\frac{1}{M^{2}}\sup_{p\in(0,1]}\int_{0}^{T}\left(E\big[ \|z_{p}(s) \|^{2}
\big]\right)^{2}ds\notag\\
&+ \sup_{p\in(0,1]}\int_{0}^{T}\left(E\big[ \|z_{p}(s) \|^{2+\upsilon}
\big]\right)^{\frac{2}{2+\upsilon}} \left( E\big[\mathbb{I}_{\{\|z_{p}(s)\|>M\}}
\big]\right)^{\frac{\upsilon}{2+\upsilon}}ds\bigg)\notag\\
\leqslant & 432TC_{2}^{2}\bigg(\frac{1}{M^{2}}\sup_{p\in(0,1]}\int_{0}^{T}\left(E\big[ \|z_{p}(s) \|^{2+\upsilon}
\big]\right)^{\frac{2}{2+\upsilon}}ds\notag\\
&+ M^{-\frac{2\upsilon}{(2+\upsilon)^2}} \int_{0}^{T}\bigg(
\sup_{p\in(0,1]}\left(E\big[ \|z_{p}(s) \|^{2+\upsilon}
\big]\right)^{\frac{2}{2+\upsilon}}  \sup_{p\in(0,1]} \left(E\big[ \|z_{p}(s) \|^{2+\upsilon}
\big]\right)^{\frac{2\upsilon}{(2+\upsilon)^{2}}}\bigg)ds\notag\\
&+\frac{1 }{M^{2}}\int_{0}^{T}\sup_{p\in(0,1]} \left(E\big[ \|z_{p}(s) \|^{2+\upsilon}
\big]\right)^{\frac{4}{2+\upsilon}}ds\bigg)\notag\\
\leqslant & 432T^{2}C_{2}^{2}\bigg(\frac{1}{M^{2}}\sup_{p\in(0,1]}
\left(B_{p}(T)\right)^{\frac{2}{2+\upsilon}}  + M^{-\frac{2\upsilon}{(2+\upsilon)^2}} \sup_{p\in(0,1]}\big(B_{p}(T)
\big)^{\frac{2}{2+\upsilon}} \sup_{p\in(0,1]}\left(B_{p}(T)\right)^{\frac{2\upsilon}{(2+\upsilon)^{2}}}
\notag\\
& +\frac{1}{M^{2}}\sup_{p\in(0,1]}\left(B_{p}(T)\right)^{\frac{4}{2+\upsilon}}\bigg)
=  L_{3}(M,T),\label{differencetwoparticls4411}
\end{align}
where $B_{p}(T)=E\big[\sup_{s\in [0,T]} \|z_{p}(s) \|^{2+\upsilon}
\big]$.
For the second term on the r.h.s. of (\ref{differencetwoparticls441}), by (\ref{FMANDF})  and
H{\"o}lder inequality, we have
\begin{align}
 l_{9}(T)
\leqslant & 36T\int_{0}^{T}E\bigg[\int_{ \mathbb{R}^{n}\times (0,1 ]} \|F_{M}(s,p, q,z,z_{p}(s)) -\widetilde{F}_{M}(s,p, q,z,z_{p}(s)) \|^{2}\mu_{s,q}(dz)dq\bigg]ds\notag\\
\leqslant & \frac{36T^2}{M^{2}}.\label{differencetwoparticls4412}
\end{align}
By Corollary 2 in \cite{M. H. Schultz} and Assumption \ref{assumption1} (ii), there exists a constant $K(T,m)>0$, such that $$ \sup_{p,q \in [0,1],\ s \in [0,T], \ k=1,2,\ldots,m}\|F_{1}(s,p,q,k,m)\|\leqslant K(T,m).$$
For the third term on the r.h.s. of (\ref{differencetwoparticls441}), by  H{\"o}lder
 inequality and noting that $a_{k}$ and  $c_{k}$ are the polynomials independent of $p$ and $q$, we know that there exists $C_{M,T}$ such that
\begin{align}
&l_{10}(T)\notag\\
\leqslant &36Tm\sum_{k=1}^{m}\int_{[0,T]\times [0,1]}E\Bigg[\bigg\|
\int_{ \mathbb{R}^{n}\times(0,1]}\big(A^{N}(p,q) -A(p,q)
\big)F_{1}(s,p,q,k,m)a_{k}(z)\mathbb{I}_{\{\|z\|\leqslant M\}}c_{k}(z_{p}(s))\notag\\
& \times \mathbb{I}_{\{\|z_{p}(s)\|\leqslant M\}}
\mu_{s,q}(dz)dq\bigg\|^{2}\Bigg]dsdp\notag\\
\leqslant &36T^{2}mK^{2}(T,m)\int_{[0,1]}\sup_{s\in[0,T]}\sum_{k=1}^{m} E\big[
\|\mathbb{I}_{\{\|z_{p}(s)\|\leqslant M\}}  c_{k}(z_{p}(s))
\|^{2}\big]\bigg\|\int_{(0,1]}\big(A^{N}\big(p,q\big)-A(p,q)
\big) \notag\\
& \times \int_{ \mathbb{R}^{n}}a_{k}(z)\mathbb{I}_{\{\|z\|\leqslant M\}} \mu_{s,q}(dz)dq\bigg\|^{2}\notag\\
\leqslant &  C_{M,T}\left\|A^{N}-A\right\|_{\infty \to 1}^{2}.\label{specoal}
\end{align}
By (\ref{differencetwoparticls})-(\ref{differencetwoparticls42}) and (\ref{differencetwoparticls441})-(\ref{specoal}), we have
\begin{align}
&\int_{[0,1]}E\bigg[\sup_{t \in [0,T]}\|\bar{z}_{p}^{N}(t)-z_{p}(t)\|^{2}\bigg]dp\notag\\ 
\leqslant &  P_{1}(M,N,T) + P_{2}(T) \int_{0}^{T}\int_{[0,1]}E\bigg[\sup_{s\in [0,t]} \|
\bar{z}_{p}^{N}(s)-z_{p}(s) \|^{2}\bigg]dpdt,\notag
\end{align}
where $P_{1}(M,N,T)= \epsilon_{1}(N)+8T\epsilon_{2}(T,N)+ 8T\sigma_{1} \epsilon_{3}(T,N)+32 \epsilon_{4}(T,N)
 +32\sigma_{1}\epsilon_{3}(T,N)
+12T\\ \epsilon_{5}(T,N)+L_{3}(M,T)+\frac{36T^{2}}{M^{2}}
 + C_{M,T}\|A^{N}-A\|_{\infty \to 1}^{2} $, $P_{2}(T)=  8T\sigma_{1} +32\sigma_{1}+48T  \sigma^{2}_{4} $,\ $\epsilon_{1}(N)=\\ \int_{[0,1]}\epsilon_{1}(N,p)dp,$\ $\epsilon_{i}(T,N)=\int_{[0,1]}\epsilon_{i}(T, N,p) dp,\ i=2,\ldots,5$. 
Then, by  Gr{\"o}nwall's inequality, we  have
\begin{align}
\int_{[0,1]}E\bigg[\sup_{t \in [0,T]} \|\bar{z}_{p}^{N}(t)-z_{p}(t) \|^{2}\bigg]dp
\leqslant   e^{P_{2}(T)T }P_{1}(M,N,T).\label{large2}
\end{align}
By    Assumption \ref{assumption01}  and Lemma 8.11 in \cite{L. Lovasz}, we have
\begin{align}
\lim_{N\to \infty}\|A^{N}-A\|_{\infty \to 1}^{2}=0.\label{fidsts1}
  \end{align}
  By Assumption \ref{assumption001}  and $W_{2} \Big(\delta_{z_{\frac{i}{N}}(0)},\delta_{z_{p}(0)} \Big)= \|z_{\frac{i}{N}}(0)-z_{p}(0)  \|$, we have
\begin{align}
\lim_{N\to \infty}\epsilon_{1}(N)  =0.\label{fidsts2}
 \end{align}
 By Assumption \ref{assumption1} and Theorem \ref{existencegeneral}, we have
 \begin{align}
\lim_{N\to \infty}\epsilon_{i}(T,N)=0, \ i=2,\ldots,5.\label{fidsts3}
 \end{align}
By Theorem \ref{existencegeneral}, we have
\begin{align}
\lim_{M\to \infty}L_{3}(M,T)=0.\label{fidsts6}
 \end{align}
 Then, letting $N$ and $M$ tend to infinity and  by (\ref{large2})-(\ref{fidsts6}),  we have  (\ref{NTOINFZ}).

 Then, we prove (\ref{convergewasserstein1}). At first, we claim that
 \begin{align}\label{convergewasserstein2}
     \lim_{N\to \infty}\frac{1}{N}\sum_{i=1}^{N}E\left[ \left\|z_{i}^{N}-z_{\frac{i}{N}} \right\|_{*,T}^{2}\right]=0.
 \end{align}
  Suppose that the claim holds.  By  (\ref{w1wassersteinlip}), we have
\begin{align}
&W_{1,T}\bigg(\frac{1}{N}\sum_{i=1}^{N}\delta_{z_{i}^{N}},\int_{[0,1]}
\mu_{p}dp\bigg)\notag\\
=&W_{1,T}\bigg(\int_{[0,1]}\mu_{p}^{N}dp, \int_{[0,1]}
\mu_{p}dp\bigg)\notag\\
= &\sup_{ f \in \mathcal{C}_{L}}  \int_{[0,1]}\bigg( \int_{\mathcal{C}_{T}^{n}}  f(z) (\mu_{p}^{N}(dz)    -\mu_{p}(dz))\bigg)dp\notag\\
\leqslant  &\int_{[0,1]}\sup_{f \in \mathcal{C}_{L}}  \int_{\mathcal{C}_{T}^{n}} f(z) (\mu_{p}^{N}(dz) -
\mu_{p}(dz) )dp\notag\\
=& \int_{[0,1]}W_{1,T}\left(\mu_{p}^{N}, \mu_{p}\right)dp,\notag
\end{align}
 where $\mu_{p}^{N}=\delta_{z_{i}^{N}}, \ p\in   \big(\frac{i-1}{N},\frac{i}{N}\big],\ i=1,\ldots,N$ and $\mu_{0}^{N}=\delta_{z_{0}}$.  Then, by  the triangle inequality of $W_{1,T}$, we have
 \begin{align}\label{wasser1}
    &E\bigg[W_{1,T}\Big(\frac{1}{N}\sum_{i=1}^{N}\delta_{z_{i}^{N}},\int_{[0,1]}
\mu_{p}dp\Big)\bigg] \notag\\
\leqslant  & \int_{[0,1]}E\left[W_{1,T}\left(\mu_{p}^{N}, \bar{\mu}_{p}^{N}\right)\right]dp +\int_{[0,1]}E\left[W_{1,T}\left(\bar{\mu}_{p}^{N},
\widetilde{\mu}_{p}^{N}\right)\right]dp+\int_{[0,1]}W_{1,T}\left(\widetilde{\mu}_{p}^{N}, \mu_{p}\right)dp,
 \end{align}
 where  $\bar{\mu}_{p}^{N}=\delta_{z_{\frac{i}{N}}},\ \widetilde{\mu}_{p}^{N}=\mu_{\frac{i}{N}}, \ p\in \big(\frac{i-1}{N},\frac{i}{N}\big],\ i=1,\ldots,N$, $\bar{\mu}_{0}^{N}=\delta_{z_{0}}$, $\widetilde{\mu}_{0}^{N}=\mu_{0}$.
 Note that $\delta_{ \big(z_{i}^{N},  z_{\frac{i}{N}} \big)}$ is a coupling of $\delta_{ z_{i}^{N}}$ and $\delta_{z_{\frac{i}{N}}}$. Then,  similar to the proof of Theorem 3.1, by   (\ref{defofWassersteindistance}), Lyapunov inequality and   $C_{r}$ inequality, we have 
 \begin{align}
  & \bigg(\int_{[0,1]}E\big[W_{1,T}\big(\mu_{p}^{N}, \bar{\mu}_{p}^{N}\big)\big]dp\bigg)^{2}\notag\\
  \leqslant &  \bigg(\int_{[0,1]} E\big[W_{2,T}(\mu_{p}^{N}, \bar{\mu}_{p}^{N})\big] dp \bigg)^{2}
  \leqslant \frac{1}{N}\sum\limits_{i=1}^{N}E\big[W_{2,T}^{2} (\delta_{z_{i}^{N}},  \delta_{z_{\frac{i}{N}}} )\big]
 \leqslant   \frac{1}{N}\sum\limits_{i=1}^{N}E\Big[\big\|z_{i}^{N}-z_{\frac{i}{N}}\big\|^{2}_{*,T}
 \Big].\notag
 \end{align}
 This together with (\ref{convergewasserstein2}) gives
 \begin{align}
    \lim_{N \to \infty} \int_{[0,1]}E\left[W_{1,T}\left(\mu_{p}^{N}, \bar{\mu}_{p}^{N}\right)\right]dp=0.  \label{wasser2}
 \end{align}
 Note that, for $i \neq j,\ i,\ j=1,2,...,N$, $z_{\frac{i}{N}}$ is independent of $z_{\frac{j}{N}}$ by the independence of $\{(z_{p}(0),\ \eta_{p},\\ w_{p}),\ p\in [0,1] \}$. Similarly to the estimation of \eqref{wasser2}, we have 
\begin{align}
 \bigg(\int_{[0,1]}E\left[W_{1,T}\left(\bar{\mu}_{p}^{N},\widetilde{\mu}_{p}^{N}\right)\right]dp \bigg)^2
\leqslant  \frac{2}{N^2}\sum_{i=1}^{N}  E\Big[ \big\|z_{\frac{i}{N}} \big\|_{*,T}^{2} \Big].\notag
\end{align}
By Theorem \ref{existencegeneral}, we have $\sup\limits_{1\leqslant i \leqslant N, \ N\in \mathbb{N}^{+}}  E\big[ \|z_{\frac{i}{N}} \|_{*,T}^{2}\big]<\infty.$ This together with the above inequality gives
\begin{align}\label{facaq}
  \lim_{N\to\infty} \int_{[0,1]}E\left[W_{1,T}\left(\bar{\mu}_{p}^{N},
\widetilde{\mu}_{p}^{N}\right)\right]dp =0.
\end{align}
By (\ref{defofWassersteindistance}), Lyapunov inequality  and H{\"o}lder inequality, we have 
  \begin{align}
 &\bigg(\int_{[0,1]}W_{1,T}\big(\widetilde{\mu}_{p}^{N}, \mu_{p}\big)dp\bigg)^{2}\notag\\
   \leqslant&  \bigg(\int_{[0,1]}W_{2,T} (\widetilde{\mu}_{p}^{N},\mu_{p} )dp\bigg)^{2}\notag\\
   \leqslant  &   \int_{[0,1]} W_{2,T}^{2} (\widetilde{\mu}_{p}^{N}, \mu_{p} )dp
  =   \sum_{i=1}^{N}\int_{ \big(\frac{i-1}{N},\frac{i}{N}\big]} W_{2,T}^{2} (\mu_{\frac{i}{N}},   \mu_{p} )dp,
  \end{align}
 which together with Lemma \ref{labelconti}  gives
  \begin{align}\label{lianxu}
   \lim_{N\to\infty}  \int_{[0,1]}W_{1,T} (\widetilde{\mu}_{p}^{N}, \mu_{p} )dp=0.
 \end{align}
This together with (\ref{wasser1})-(\ref{facaq}) gives (\ref{convergewasserstein1}).

 Finally, we prove claim (\ref{convergewasserstein2}).
 By  (\ref{randmckvla}) and (\ref{finiteparticles}), similar to the proof of (\ref{differencetwoparticls}), we have
 \begin{align}\label{fini1}
  &E\bigg[\left\|z_{i}^{N}-z_{\frac{i}{N}} \right\|_{*,T}^{2} \bigg]\notag\\
 \leqslant & 3T
 \int_{0}^{T}E\bigg[\Big\|G\Big(s,\frac{i}{N},\eta_{\frac{i}{N}}(s),
z_{i}^{N}(s)\Big)-G\Big(s,\frac{i}{N},\eta_{\frac{i}{N}}(s),z_{\frac{i}{N}}(s) \Big)\Big\|^{2}\bigg]ds\notag\\
&+12E\int_{0}^{T}\bigg[\Big\|H\Big(s,\frac{i}{N},\eta_{\frac{i}{N}}(s),
z_{i}^{N}(s)\Big)-H \Big(s,\frac{i}{N},\eta_{\frac{i}{N}}(s),z_{\frac{i}{N}}(s) \Big)\Big\|^{2}\bigg]ds
\notag\\
& +3T\int_{0}^{T}E\Bigg[\bigg\|\frac{1}{N}\sum_{j=1}^{N}
\bigg(A^{N}\left(\frac{i}{N},\frac{j}{N}\right) F\left(s,\frac{i}{N}, \frac{j}{N},z_{j}^{N}(s),z_{i}^{N}(s)\right)\bigg)
\notag\\
&-\int_{[0,1]\times \mathbb{R}^{n}}A\Big(\frac{i}{N},q\Big)
F\Big(s,\frac{i}{N}, q,z,z_{\frac{i}{N}}(s)\Big)\mu_{s,q}(dz)dq\bigg\|^{2}\Bigg]ds.
 \end{align}
By Assumption \ref{assumption1} (i), we have
\begin{align}\label{fini2}
  &
 3T\int_{0}^{T}E\bigg[\Big\|G\Big(s,\frac{i}{N},\eta_{\frac{i}{N}}(s),
z_{i}^{N}(s)\Big)-G \Big(s,\frac{i}{N},\eta_{\frac{i}{N}}(s),z_{\frac{i}{N}}(s) \Big)\Big\|^{2}\bigg]ds\notag\\
&+12E\int_{0}^{T}\bigg[\Big\|H\Big(s,\frac{i}{N},\eta_{\frac{i}{N}}(s),
z_{i}^{N}(s)\Big)-H \Big(s,\frac{i}{N},\eta_{\frac{i}{N}}(s),z_{\frac{i}{N}}(s) \Big)\Big\|^{2}\bigg]ds\notag\\
\leqslant & (3T+12)\sigma_{1}  \int_{0}^{T} E\Big[\big\|
z_{i}^{N}(s) -z_{\frac{i}{N}}(s) \big\|^{2}\Big]ds\notag\\
\leqslant &  (3T+12)\sigma_{1}  \int_{0}^{T} E\Big[\big\|
z_{i}^{N} -z_{\frac{i}{N}}  \big\|_{*,t}^{2}\Big]dt.
\end{align}
By $C_{r}$ inequality, we have
\begin{align}\label{fini3}
  &3T\int_{0}^{T}E\Bigg[\bigg\|\frac{1}{N}\sum_{j=1}^{N}
\bigg(A^{N}\left(\frac{i}{N},\frac{j}{N}\right) F\left(s,\frac{i}{N}, \frac{j}{N},z_{j}^{N}(s),z_{i}^{N}(s)\right)\bigg)
\notag\\
&-\int_{[0,1]\times \mathbb{R}^{n}}A\Big(\frac{i}{N},q\Big)
F\left(s,\frac{i}{N}, q,z,z_{\frac{i}{N}}(s)\right)\mu_{s,q}(dz)dq\bigg\|^{2}\Bigg]ds\notag\\
\leqslant & 15T\int_{0}^{T}E\Bigg[\bigg\| \sum_{j=1}^{N}\int_{\big(\frac{j-1}{N}, \frac{j}{N}\big]}
 A^{N}\left(\frac{i}{N},\frac{j}{N}\right) \bigg(F\left(s,\frac{i}{N}, \frac{j}{N},z_{j}^{N}(s),z_{i}^{N}(s)\right)\notag\\
&-F\left(s,\frac{i}{N}, q,z_{j}^{N}(s), z_{i}^{N}(s)\right)\bigg)dq \bigg\|^{2}\Bigg]ds\notag\\
&+ 15T\int_{0}^{T} E\Bigg[\bigg\| \sum_{j=1}^{N}\int_{\big(\frac{j-1}{N}, \frac{j}{N}\big]}
 A^{N}\left(\frac{i}{N},\frac{j}{N}\right) \bigg(F\left(s,\frac{i}{N}, q,z_{j}^{N}(s),z_{i}^{N}(s)\right)\notag\\
&-F\left(s,\frac{i}{N}, q,z_{ \frac{j}{N}}(s),z_{\frac{i}{N}}(s)\right)\bigg)dq \bigg\|^{2}\Bigg]ds\notag\\
&+ 15T\int_{0}^{T} E\Bigg[\bigg\| \sum_{j=1}^{N}\int_{\big(\frac{j-1}{N}, \frac{j}{N}\big]\times\mathbb{R}^{n}}
 A^{N}\left(\frac{i}{N},\frac{j}{N}\right) \bigg(F\left(s,\frac{i}{N}, q,z_{ \frac{j}{N}}(s),z_{\frac{i}{N}}(s)\right)\notag\\
 &-F\left(s,\frac{i}{N}, q,z,z_{\frac{i}{N}}(s)\right) \bigg)\mu_{s, \frac{j}{N}}(dz)dq \bigg\|^{2}\Bigg]ds\notag\\
 &+ 15T\int_{0}^{T} E\Bigg[\bigg\| \sum_{j=1}^{N}\int_{\big(\frac{j-1}{N}, \frac{j}{N}\big]\times\mathbb{R}^{n}}
 A^{N}\left(\frac{i}{N},\frac{j}{N}\right) F\Big(s,\frac{i}{N}, q,z,z_{\frac{i}{N}}(s)\Big) \big(\mu_{s, \frac{j}{N}}(dz)\notag\\
  &-\mu_{s,q}(dz)\big)dq \bigg\|^{2}\Bigg]ds+ 15T\int_{0}^{T} E\Bigg[\bigg\| \sum_{j=1}^{N}\int_{\big(\frac{j-1}{N}, \frac{j}{N}\big]\times\mathbb{R}^{n}}
 \left(A^{N}\left(\frac{i}{N},\frac{j}{N}\right)-A\left(\frac{i}{N},q\right)\right) \notag\\
 &\times F\Big(s,\frac{i}{N}, q,z,z_{\frac{i}{N}}(s)\Big) \mu_{s,q}(dz) dq \bigg\|^{2}\Bigg]ds.
\end{align}
 By H{\"o}lder inequality, we have
 \begin{align}\label{fini4}
 &  15T\int_{0}^{T}E\Bigg[\bigg\| \sum_{j=1}^{N}\int_{\big(\frac{j-1}{N}, \frac{j}{N}\big]}
 A^{N}\left(\frac{i}{N},\frac{j}{N}\right) \bigg(F\left(s,\frac{i}{N}, \frac{j}{N},z_{j}^{N}(s),z_{i}^{N}(s)\right)\notag\\
&-F\left(s,\frac{i}{N}, q,z_{j}^{N}(s),z_{i}^{N}(s)\right)\bigg)dq \bigg\|^{2}\Bigg]ds\notag\\
\leqslant & 15T \epsilon_{5}\left(T,N,\frac{i}{N}\right),
 \end{align}
 where $\epsilon_{5}\left(T,N,\frac{i}{N}\right)=\int_{0}^{T}\sum\limits_{j=1}^{N}\int_{ \big(\frac{j-1}{N},\frac{j}{N}\big]}E\big[\|
F(s,\frac{i}{N}, \frac{j}{N},
z_{j}^{N}(s),  z_{i}^{N}(s))-F\big(s,\frac{i}{N}, q,z_{j}^{N}(s),z_{i}^{N}(s)\big)\|^{2}\big]  dqds$.
 By Assumption \ref{assumption1} (ii) and H{\"o}lder inequality, we have
 \begin{align}\label{fini5}
   & 15T\int_{0}^{T} E\Bigg[\bigg\| \sum_{j=1}^{N}\int_{\big(\frac{j-1}{N}, \frac{j}{N}\big]}
 A^{N}\left(\frac{i}{N},\frac{j}{N}\right) \bigg(F\left(s,\frac{i}{N}, q,z_{j}^{N}(s),z_{i}^{N}(s)\right)\notag\\
&-F\left(s,\frac{i}{N}, q,z_{ \frac{j}{N}}(s),z_{\frac{i}{N}}(s)\right)\bigg)dq \bigg\|^{2}\Bigg]ds\notag\\
\leqslant & 15T \sigma_{4}^{2}\bigg( \frac{1}{N}\sum_{j=1}^{N} \int_{0}^{T} E\Big[\big\|z_{j}^{N}(s)- z_{ \frac{j}{N}}(s)\big\|^{2}\Big]ds+\int_{0}^{T} E\Big[\big\|z_{i}^{N}(s)- z_{ \frac{i}{N}}(s)\big\|^{2}\Big]ds\bigg)\notag\\
\leqslant & 15T \sigma_{4}^{2}\bigg( \frac{1}{N}\sum_{j=1}^{N} \int_{0}^{T} E\Big[\big\|z_{j}^{N}  - z_{ \frac{j}{N}} \big\|_{*,t}^{2} \Big]dt+\int_{0}^{T} E\Big[\big\|z_{i}^{N} - z_{ \frac{i}{N}} \big\|_{*,t}^{2}\Big]dt\bigg).
 \end{align}
For $l\neq i,\ j$, by the independence of $\left\{z_{\frac{i}{N}}, \ i=1,\ldots,N\right\}$, we have $E\Big[\Big(\int_{\big(\frac{j-1}{N}, \frac{j}{N}\big]\times\mathbb{R}^{n}}
 A^{N}\big(\frac{i}{N},\frac{j}{N}\big) \big(F\big(s,\\ \frac{i}{N}, q,z_{ \frac{j}{N}}(s),z_{\frac{i}{N}}(s)\big) -F\big(s,\frac{i}{N}, q,z,z_{\frac{i}{N}}(s)\big) \big)  \mu_{s, \frac{j}{N}}(dz)dq\Big)^{\mathsf{T}}\Big(\int_{\big(\frac{l-1}{N}, \frac{l}{N}\big]\times\mathbb{R}^{n}}
 A^{N}\left(\frac{i}{N},\frac{l}{N}\right) \big(F\big(s,\frac{i}{N}, q, z_{ \frac{l}{N}}(s), \\ z_{\frac{i}{N}}(s)\big)-F\big(s,\frac{i}{N}, q,z,z_{\frac{i}{N}}(s)\big) \big)\mu_{s, \frac{l}{N}}(dz)dq\Big)\Big]=0.$ Then, we have
 \begin{align}\label{fini6}
    &15T\int_{0}^{T} E\Bigg[\bigg\| \sum_{j=1}^{N}\int_{\big(\frac{j-1}{N}, \frac{j}{N}\big]\times\mathbb{R}^{n}}
 A^{N}\left(\frac{i}{N},\frac{j}{N}\right) \bigg(F\left(s,\frac{i}{N}, q,z_{ \frac{j}{N}}(s),z_{\frac{i}{N}}(s)\right)\notag\\
 &-F\left(s,\frac{i}{N}, q,z,z_{\frac{i}{N}}(s)\right) \bigg)\mu_{s, \frac{j}{N}}(dz)dq \bigg\|^{2}\Bigg]ds\notag\\
 =& 15T\int_{0}^{T}  \sum_{j=1}^{N}E\Bigg[\bigg\|\int_{\big(\frac{j-1}{N}, \frac{j}{N}\big]\times\mathbb{R}^{n}}
 A^{N}\left(\frac{i}{N},\frac{j}{N}\right) \bigg(F\left(s,\frac{i}{N}, q,z_{ \frac{j}{N}}(s),z_{\frac{i}{N}}(s)\right)\notag\\
 &-F\left(s,\frac{i}{N}, q,z,z_{\frac{i}{N}}(s)\right) \bigg)\mu_{s, \frac{j}{N}}(dz)dq \bigg\|^{2}\Bigg]ds+15T\int_{0}^{T} \sum_{j=1}^{N}E\Bigg[\bigg(\int_{\big(\frac{j-1}{N}, \frac{j}{N}\big]\times\mathbb{R}^{n}}
 A^{N}\left(\frac{i}{N},\frac{j}{N}\right) \notag\\
 & \times \bigg(F\left(s,\frac{i}{N}, q,z_{ \frac{j}{N}}(s),z_{\frac{i}{N}}(s)\right)-F\left(s,\frac{i}{N}, q,z,z_{\frac{i}{N}}(s)\right) \bigg)\mu_{s, \frac{j}{N}}(dz)dq \bigg)^{\mathsf{T}} \notag\\
 &\times\bigg(\int_{\big(\frac{i-1}{N}, \frac{i}{N}\big]\times\mathbb{R}^{n}}
 A^{N}\left(\frac{i}{N},\frac{i}{N}\right) \bigg(F\left(s,\frac{i}{N}, q,z_{ \frac{i}{N}}(s),z_{\frac{i}{N}}(s)\right)\notag\\
 &-F\left(s,\frac{i}{N}, q,z,z_{\frac{i}{N}}(s)\right) \bigg)\mu_{s, \frac{i}{N}}(dz)dq \bigg)\Bigg]ds.
 \end{align}
 By Assumption \ref{assumption1} (ii) and H{\"o}lder inequality, we have
 \begin{align}\label{fini7}
 & 15T\int_{0}^{T}  \sum_{j=1}^{N}E\Bigg[\bigg\|\int_{\big(\frac{j-1}{N}, \frac{j}{N}\big]\times\mathbb{R}^{n}}
 A^{N}\left(\frac{i}{N},\frac{j}{N}\right) \bigg(F\left(s,\frac{i}{N}, q,z_{ \frac{j}{N}}(s),z_{\frac{i}{N}}(s)\right)\notag\\
 &-F\left(s,\frac{i}{N}, q,z,z_{\frac{i}{N}}(s)\right) \bigg)\mu_{s, \frac{j}{N}}(dz)dq \bigg\|^{2}\Bigg]ds\notag\\
 \leqslant & \frac{15T}{N}  \sum_{j=1}^{N} \int_{0}^{T}E\Bigg[\int_{\big(\frac{j-1}{N}, \frac{j}{N}\big]\times\mathbb{R}^{n}}\bigg\|F\left(s,\frac{i}{N}, q,z_{ \frac{j}{N}}(s),z_{\frac{i}{N}}(s)\right)\notag\\
 &-F\left(s,\frac{i}{N}, q,z,z_{\frac{i}{N}}(s)\right)  \bigg\|^{2}\mu_{s, \frac{j}{N}}(dz)dq\Bigg]ds\notag\\
 \leqslant & \frac{1}{N^{2} } 15T\sigma_{4}^{2} \sum_{j=1}^{N}\int_{0}^{T} E\Bigg[\int_{ \mathbb{R}^{n}}\big\| z_{ \frac{j}{N}}(s) -z \big\|^{2}\mu_{s, \frac{j}{N}}(dz)\Bigg]ds\notag\\
 \leqslant & \frac{1}{N } 30 T^{2}\sigma_{4}^{2}    \sup_{p\in [0,1],\ t \in [0,T]}E\Big[ \big\| z_{p}(t) \big\|^{2} \Big].
 \end{align}
 Similar to the proof of the above inequality and by $C_{r}$ inequality, we have
 \begin{align}\label{fini8}
  &15T \sum_{j=1}^{N}\int_{0}^{T}E\Bigg[\bigg(\int_{\big(\frac{j-1}{N}, \frac{j}{N}\big]\times\mathbb{R}^{n}}
 A^{N}\left(\frac{i}{N},\frac{j}{N}\right)  \bigg(F\left(s,\frac{i}{N}, q,z_{ \frac{j}{N}}(s),z_{\frac{i}{N}}(s)\right)\notag\\
 &  -F\left(s,\frac{i}{N}, q,z,z_{\frac{i}{N}}(s)\right) \bigg)\mu_{s, \frac{j}{N}}(dz)dq \bigg)^{\mathsf{T}} \bigg(\int_{\big(\frac{i-1}{N}, \frac{i}{N}\big]\times\mathbb{R}^{n}}
 A^{N}\left(\frac{i}{N},\frac{i}{N}\right)\notag\\
 &\times \bigg(F\left(s,\frac{i}{N}, q,z_{ \frac{i}{N}}(s),z_{\frac{i}{N}}(s)\right)-F\left(s,\frac{i}{N}, q,z,z_{\frac{i}{N}}(s)\right) \bigg)\mu_{s, \frac{i}{N}}(dz)dq \bigg)\notag\\
 \leqslant & 30T \sum_{j=1}^{N}\int_{0}^{T}E\Bigg[\bigg\|\int_{\big(\frac{j-1}{N}, \frac{j}{N}\big]\times\mathbb{R}^{n}}
 A^{N}\left(\frac{i}{N},\frac{j}{N}\right)  \bigg(F\left(s,\frac{i}{N}, q,z_{ \frac{j}{N}}(s),z_{\frac{i}{N}}(s)\right)\notag\\
 &  -F\left(s,\frac{i}{N}, q,z,z_{\frac{i}{N}}(s)\right) \bigg)\mu_{s, \frac{j}{N}}(dz)dq \bigg\|^{2}\Bigg]+ 30T N \int_{0}^{T}E\Bigg[\bigg\| \bigg(\int_{\big(\frac{i-1}{N}, \frac{i}{N}\big]\times\mathbb{R}^{n}}
 A^{N}\left(\frac{i}{N},\frac{i}{N}\right)\notag\\
 &\times \bigg(F\left(s,\frac{i}{N}, q,z_{ \frac{i}{N}}(s),z_{\frac{i}{N}}(s)\right)-F\left(s,\frac{i}{N}, q,z,z_{\frac{i}{N}}(s)\right) \bigg)\mu_{s, \frac{i}{N}}(dz)dq  \bigg\|^{2}\Bigg]\notag\\
 \leqslant & \frac{1}{N } 120 T^{2}\sigma_{4}^{2}    \sup_{p\in [0,1],\ t \in [0,T]}E\Big[ \big\| z_{p}(t) \big\|^{2} \Big].
 \end{align}
 By Remarks 6.5-6.6 in \cite{Villani}, we have $W_{2}(\mu, \nu) \geqslant \sup\limits_{f:\ f\ \text{is\ 1-Lipschitz} }\big|\int_{\mathbb{R}^{n}}f(z) \mu(d z)-\int_{\mathbb{R}^{n}} f(z) \nu(d z)\big|,$
where $ \mu, \ \nu \in \mathscr{P}(\mathbb{R}^{n})$.
 Then, by  Assumption \ref{assumption1} (ii) and (\ref{defofWassersteindistance}),  we have
\begin{align}\label{fini9}
  &15T\int_{0}^{T} E\Bigg[\bigg\| \sum_{j=1}^{N}\int_{\big(\frac{j-1}{N}, \frac{j}{N}\big]\times\mathbb{R}^{n}}
 A^{N}\left(\frac{i}{N},\frac{j}{N}\right) F\Big(s,\frac{i}{N}, q,z,z_{\frac{i}{N}}(s)\Big) \big(\mu_{s, \frac{j}{N}}(dz)-\mu_{s,q}(dz)\big)dq \bigg\|^{2}\Bigg]ds\notag\\
 \leqslant & 15T\int_{0}^{T} \sum_{j=1}^{N} \int_{\big(\frac{j-1}{N}, \frac{j}{N}\big]} E\Bigg[\bigg\| \int_{\mathbb{R}^{n}} F\Big(s,\frac{i}{N}, q,z,z_{\frac{i}{N}}(s)\Big) \big(\mu_{s, \frac{j}{N}}(dz)-\mu_{s,q}(dz)\big) \bigg\|^{2}\Bigg]dqds\notag\\
 \leqslant & 15Tn\sigma_{4}^{2} \int_{0}^{T} \sum_{j=1}^{N} \int_{\big(\frac{j-1}{N}, \frac{j}{N}\big]} E\left[  W_{2}^{2}\left(\mu_{s, \frac{j}{N}},\mu_{s,q}\right)  \right]dqds.
\end{align}
 Similar to the proof of (\ref{differencetwoparticls441})-(\ref{specoal}), we have
 \begin{align}
 & 15T\int_{[0,T]\times [0,1]}E\Bigg[\bigg\| \sum_{j=1}^{N}\int_{\big(\frac{j-1}{N}, \frac{j}{N}\big]\times\mathbb{R}^{n}}
 \left(A^{N}\left(\frac{i}{N},\frac{j}{N}\right)-A\left(\frac{i}{N},q\right)\right)  \notag\\
 &\times F\Big(s,\frac{i}{N}, q,z,z_{\frac{i}{N}}(s)\Big) \mu_{s,q}(dz) dq \bigg\|^{2}\Bigg]dsdp\notag\\
 \leqslant & \frac{3}{2}\left(L_{3}(M,T)+\frac{36T^2}{M^{2}}+C_{M,T}\left\|A^{N}-A\right\|_{\infty \to 1}^{2}\right).\notag
 \end{align}
 By the above inequality  and (\ref{fini1})-(\ref{fini9}), we have
 \begin{align}
  &\frac{1}{N}\sum_{i=1}^{N} E\bigg[ \big\|z_{i}^{N}-z_{\frac{i}{N}} \big\|_{*,T}^{2} \bigg]\notag\\
  \leqslant & \left(3T\sigma_{1} +12\sigma_{1} +15T n \sigma_{4}^{2}+15T \sigma_{4}^{2}\right)  \int_{0}^{T}\frac{1}{N}\sum_{i=1}^{N}  E\Big[\big\|
z_{i}^{N} -z_{\frac{i}{N}}  \big\|_{*,t}^{2}\Big]dt\notag\\
&+ \frac{15T }{N}\sum_{i=1}^{N} \epsilon_{5}\left(T,N,\frac{i}{N}\right)+\frac{1}{N } 150 T^{2}\sigma_{4}^{2}    \sup_{p\in [0,1],\ t \in [0,T]}E\Big[ \big\| z_{p}(t) \big\|^{2} \Big]+  \frac{3}{2}\bigg(L_{3}(M,T)\notag\\
&+\frac{36T^2}{M^{2}}+C_{M,T}\left\|A^{N}-A\right\|_{\infty \to 1}^{2}\bigg)+15Tn\sigma_{4}^{2} \int_{0}^{T} \sum_{j=1}^{N} \int_{\big(\frac{j-1}{N}, \frac{j}{N}\big]} E\left[  W_{2}^{2}\left(\mu_{s, \frac{j}{N}},\mu_{s,q}\right)  \right]dqds.\notag
 \end{align}
 This together with Gr{\"o}nwall's inequality gives
 \begin{align}\label{fini10}
   &\frac{1}{N}\sum_{i=1}^{N} E\bigg[ \big\|z_{i}^{N}-z_{\frac{i}{N}} \big\|_{*,T}^{2} \bigg]\notag\\
   \leqslant&
   e^{T \left(3T\sigma_{1} +12\sigma_{1} +15T n \sigma_{4}^{2}+15T \sigma_{4}^{2}\right)} \Bigg( 15T \frac{1}{N}\sum_{i=1}^{N} \epsilon_{5}\left(T,N,\frac{i}{N}\right)+\frac{1}{N } 150T^{2}\sigma_{4}^{2}  \notag\\
&\times  \sup_{p\in [0,1],\ t \in [0,T]}E\Big[ \big\| z_{p}(t) \big\|^{2} \Big]+  \frac{3}{2}\Big(L_{3}(M,T)+\frac{36T^2}{M^{2}}+C_{M,T}\left\|A^{N}-A\right\|_{\infty \to 1}^{2}\Big)\notag\\
&+15Tn\sigma_{4}^{2} \int_{0}^{T} \sum_{j=1}^{N} \int_{\big(\frac{j-1}{N}, \frac{j}{N}\big]} E\left[  W_{2}^{2}\left(\mu_{s, \frac{j}{N}},\mu_{s,q}\right)  \right]dqds\Bigg).
    \end{align}
 By Lemma \ref{labelconti}, we have
 \begin{align}\label{fini11}
  \lim_{N\to \infty} 15Tn\sigma_{4}^{2} \int_{0}^{T} \sum_{j=1}^{N} \int_{\big(\frac{j-1}{N}, \frac{j}{N}\big]} E\left[  W_{2}^{2}\left(\mu_{s, \frac{j}{N}},\mu_{s,q}\right)  \right]dqds=0.
 \end{align}
  By Assumption \ref{assumption1} (ii)  and Theorem  \ref{existencegeneral}, we have
  \begin{align}\label{fini12}
    \lim_{N\to \infty} 15T \frac{1}{N}\sum_{i=1}^{N} \epsilon_{5}\left(T,N,\frac{i}{N}\right)=0.
  \end{align}
 By Theorem \ref{existencegeneral}, we have
 \begin{align}
  \lim_{N\to \infty} \frac{1}{N } 150 T^{2}\sigma_{4}^{2}    \sup_{p\in [0,1],\ t \in [0,T]}E\Big[ \big\| z_{p}(t) \big\|^{2} \Big]=0.\notag
 \end{align}
  Letting $N$ and $M$ tend to infinity and by the above equality, (\ref{fidsts1}), (\ref{fidsts6}) and (\ref{fini10})-(\ref{fini12}), we have (\ref{convergewasserstein2}).
$\hfill\blacksquare$

\vskip 1.5mm
\emph{Proof of Theorem \ref{finitetographon}:}
 By $C_{r}$ inequality,  we have
\begin{align}
&\int_{[0,1]}E\bigg[\sup_{t \in [0,T]}\left\|\hat{z}_{p}^{N,k}(t)-z_{p}(t)\right\|^{2}\bigg]dp\notag\\
 \leqslant & 2\int_{[0,1]}E\bigg[\sup_{t \in [0,T]}\left\|\hat{z}_{p}^{N,k}(t)-\bar{z}^{N}_{p}(t)\right\|^{2}\bigg]dp  +2\int_{[0,1]}E\bigg[\sup_{t \in [0,T]}\left\|\bar{z}^{N}_{p}(t)-z_{p}(t)\right\|^{2}\bigg]dp.\notag
\end{align}
This together with Lemmas \ref{dsffinite}-\ref{difflisandao} leads to (\ref{NTOINFTYGRS}).

By the triangle inequality of $W_{1,T}$, we have
\begin{align}\label{NTOINFTYwass1}
  & E\bigg[W_{1,T}\bigg(\frac{1}{N}\sum_{i=1}^{N}\delta_{z_{i}^{N,k}},\int_{[0,1]}
\mu_{p}dp\bigg)\bigg]\notag\\
\leqslant & E\bigg[W_{1,T}\bigg(\frac{1}{N}\sum_{i=1}^{N}\delta_{z_{i}^{N,k}},
\frac{1}{N}\sum_{i=1}^{N}\delta_{z_{i}^{N}}\bigg)
\bigg] +E\bigg[W_{1,T}\bigg(\frac{1}{N}\sum_{i=1}^{N}\delta_{z_{i}^{N}},\int_{[0,1]}
\mu_{p}dp\bigg)\bigg].
\end{align}
Then,   by (\ref{defofWassersteindistance}) and Lyapunov inequality, we have
\begin{align}
\label{W1Taddaddadd1}
 E\Big[W_{1,T}\big(\delta_{z_{i}^{N,k}},  \delta_{z_{i}^{N}}\big)
\Big]\leq&
E\Big[W_{2,T} \big(\delta_{z_{i}^{N,k}}, \delta_{z_{i}^{N}}\big)
\Big]
\leq \Big(E\big[W_{2,T}^{2}\big( \delta_{z_{i}^{N,k}},\delta_{z_{i}^{N}}\big)
\big]\Big)^{\frac{1}{2}}.
\end{align}
Noting that $\delta_{(z_{i}^{N,k},z_{i}^{N})}$ is a coupling of $\delta_{z_{i}^{N,k}}$ and $\delta_{z_{i}^{N}}$, by (\ref{defofWassersteindistance}), we have
\begin{align}
 W_{2,T}^{2}\big( \delta_{z_{i}^{N,k}},\delta_{z_{i}^{N}}\big)
\leqslant   \int_{\mathcal{C}_{T}^{n} \times \mathcal{C}_{T}^{n}} \|x-y \|_{*, T}^2  \delta_{(z_{i}^{N,k},z_{i}^{N})}(dx,dy)
 = \|z_{i}^{N,k}-  z_{i}^{N} \|_{*,T}^{2}.\label{dirac}
\end{align}
This together with $C_{r}$ inequality:  $\frac{1}{N}\sum\limits_{i=1}^{N}a_i\leq \Big(\frac{1}{N}\sum\limits_{i=1}^{N}a_i^2\Big)^{\frac{1}{2}}$ for $a_i\geq0$, $i=1,2,...,N$, (\ref{w1wassersteinlip}) and (\ref{W1Taddaddadd1}) leads to
\begin{align}
 & E\Big[W_{1,T}\Big(\frac{1}{N}\sum _{i=1}^{N}\delta_{z_{i}^{N,k}},\frac{1}{N}
\sum_{i=1}^{N}\delta_{z_{i}^{N}}\Big)
\Big]\notag\\
 =& E\bigg[\sup_{f \in \mathcal{C}_{L}}\Big( \frac{1}{N}\sum \limits_{i=1}^{N} \int_{\mathcal{C}_{T}^{n}} f(z) (\delta_{z_{i}^{N,k}}(d z)
 -\delta_{z_{i}^{N}}(d z))\Big)\bigg]\notag\\
\leqslant&   \frac{1}{N}\sum\limits_{i=1}^{N} E\Big[W_{1,T}\left(\delta_{z_{i}^{N,k}},  \delta_{z_{i}^{N}}\right)
\Big]\notag\\
 \leqslant&
 \Bigg( \frac{1}{N} \sum\limits_{i=1}^{N}E\bigg[W_{2,T}^{2}\left( \delta_{z_{i}^{N,k}},\delta_{z_{i}^{N}}\right)
\bigg]\Bigg)^{\frac{1}{2}}\notag\\
\leqslant &
\Big(\frac{1}{N} \sum_{i=1}^{N}E\big[  \|z_{i}^{N,k}-  z_{i}^{N} \|_{*,T}^{2}\big] \Big)^{\frac{1}{2}}\notag\\
 =&
\bigg(\int_{[0,1]}E\bigg[\sup\limits_{t \in [0,T]} \|\hat{z}_{p}^{N,k}(t)-\bar{z}^{N}_{p}(t) \|^{2}\bigg]
dp\bigg)^{\frac{1}{2}}.\label{17}
\end{align}
%
This together with  (\ref{NTOINFTYwass1}) and Lemmas \ref{dsffinite}-\ref{difflisandao} gives (\ref{NTOINFTYwass}).
\hfill$\square$

\begin{lemma}(\cite{Kallenberg})\label{infimeas}
 (infinite product measures) For any probability spaces $\left(S_i,\ \mathcal{S}_i,\ \mu_i\right),\ i \in \Lambda$, there exist some independent random elements $\xi_i$ in $S_i$ with
$$
\mathcal{L}\left(\xi_i\right)=\mu_i,\   i \in \Lambda.
$$
\end{lemma}

\end{document}